\documentclass[10pt]{article}

\usepackage{amsmath}

\usepackage{array}

\usepackage{appendix}

\usepackage{tocloft}                   % Adds Part structure to table of contents

\usepackage{graphicx}

\usepackage{amsfonts}

\usepackage{amssymb}

\usepackage{mathrsfs}

\usepackage{yfonts}

\usepackage{euscript}

\usepackage{centernot}                 % Permits me to trivially make a centred not version of any object

\usepackage{ifsym}                     % Gives a nicer sharp

\usepackage{upgreek}

\usepackage{mathtools}

\usepackage{color}

\usepackage{slantsc}
\usepackage{calligra}

\usepackage{bbold}          % Gives mathbb acting on lower case and, hopefully, numbers...

\usepackage[T1]{fontenc}

\usepackage{epsf}

\usepackage{latexsym}

\usepackage{tipa}

\usepackage{makeidx}

\makeindex

\def\b{\begin{equation}}

\def\e{\begin{equation}}

\def\be{\begin{equation}}              % Longer older ones kept for rext import compatibility.

\def\ee{\end{equation}}

\def\beq{\begin{equation}}

\def\eeq{\end{equation}}

\def\bea{\begin{eqnarray}}

\def\eea{\end{eqnarray}}

\def\m{\mbox{ }}

\def\mma {\m , \m \m }

\def\!{\hspace{-1.6667em}}

\def\n{\noindent}

\def\u{\underline}

\def\w{\widetilde}

\def\s{\stackrel}

\def\slTheta{\mathit{\Theta}}                     % triangleland azimuth variable. A continuous deformation.

\def\slPhi{\mathit{\Phi}}                         % relative angles in general (suffixed) and relative angle in triangleland in particular (unsuffixed). Also a SIC variable.

\def\biq{\mbox{\boldmath$q$}}

\def\biN{\mbox{\boldmath$N$}}

\def\lbiN{\mbox{\Large\boldmath$N$}}

\def\sbiM{\mbox{\scriptsize\boldmath$M$}}

\def\birho{\mbox{\boldmath$\rho$}}

\def\bchi{\mbox{\boldmath$\chi$}}                                   % Mass-weighted relative Jacobi configuration space vector

\def\brho{\birho}                                   % Mass-weighted relative Jacobi configuration space vector

\def\mB{\mbox{B}}  

\def\mC{\mbox{C}}                        % Collinear configuration/ equator in triangleland, can be higher-d regions in other models

\def\mD{\mbox{D}}                        % Double collision/ points in triangleland, can be higher-d regions in other models.

\def\mG{\mbox{G}}

\def\mH{\mbox{H}} 

\def\mK{\mbox{K}}

\def\mM{\mbox{M}}                        % Mergers

\def\mO{\mbox{O}}

\def\mP{\mbox{P}}

\def\mR{\mbox{R}}                        % Region, regular triangle

\def\mS{\mbox{S}}                        % Region within $\bupSigma$, spurious points in triangleland: C $\cap$ M

\def\mU{\mbox{U}}                        % Region within generalized hypersurface $\Upsilon$

\def\me{\mbox{e}}

\def\ml{\mbox{l}}

\def\mo{\mbox{o}}

\def\mp{\mbox{p}}

\def\urho{{\u{\rho}}}

\def\bK{\mbox{\bf K}}

\def\bL{\mbox{\bf L}}

\def\br{\mbox{\bf r}}

\def\fs{\mbox{\sffamily s}}

\def\sa{\mbox{\scriptsize a}}

\def\sd{\mbox{\scriptsize d}}

\def\se{\mbox{\scriptsize e}}

\def\sh{\mbox{\scriptsize h}} 

\def\si{\mbox{\scriptsize i}}

\def\sm{\mbox{\scriptsize m}}

\def\sn{\mbox{\scriptsize n}} 

\def\so{\mbox{\scriptsize o}} 

\def\sp{\mbox{\scriptsize p}}

\def\sr{\mbox{\scriptsize r}}

\def\sss{\mbox{\scriptsize s}}  %TO AVOID ARXIV changing \ss to German double s.

\def\st{\mbox{\scriptsize t}}

\def\sx{\mbox{\scriptsize x}}

\def\sA{\mbox{\scriptsize A}}

\def\sC{\mbox{\scriptsize C}}

\def\sD{\mbox{\scriptsize D}}

\def\sF{\mbox{\scriptsize F}}

\def\sM{\mbox{\scriptsize M}}

\def\sP{\mbox{\scriptsize P}}

\def\sR{\mbox{\scriptsize R}}

\def\sS{\mbox{\scriptsize S}}

\def\sT{\mbox{\scriptsize T}}

\def\sfP{\mbox{\sffamily{\scriptsize P}}}      % index for primary constraints

\def\sfQ{\mbox{\sffamily{\scriptsize Q}}}      % index currently used for spacetime observables.

\def\sfT{\mbox{\sffamily{\scriptsize T}}}      % index for Lie algebra generators, and for true dynamical Dof's if needed.

\def\sfX{\mbox{\sffamily{\scriptsize X}}}      % index for complement to general cyclic coordinate (in Routhian reduction), except where it is given a SIC meaning.

\def\sumi2{\sum\mbox{}_{\mbox{}_{\mbox{\scriptsize $i$=1}}}^2}

\def\sumi3{\sum\mbox{}_{\mbox{}_{\mbox{\scriptsize $i$=1}}}^3}

\def\sumABcycles3{\sum\mbox{}_{\mbox{}_{\mbox{\scriptsize cycles $A,B$=1}}}^{3}}

\def\sumCDcycles3{\sum\mbox{}_{\mbox{}_{\mbox{\scriptsize cycles $C,D$=1}}}^{3}}

\def\sumj3{\sum\mbox{}_{\mbox{}_{\mbox{\scriptsize $j$=1}}}^3}

\def\sumk3{\sum\mbox{}_{\mbox{}_{\mbox{\scriptsize $k$=1}}}^3}

\def\prodiA1{\prod\mbox{}_{\mbox{}_{\mbox{\scriptsize $i$=1}}}^{A - 1}}

\def\bigtimes{\mbox{\Large $\times$}}

\def\sumpn{\sum\mbox{}_{\mbox{}_{\mbox{\scriptsize $p$ = 1}}}^{n - 1}}

\def\d{\textrm{d}}                                                  % ordinary derivative

\def\es{\m = \m}

\def\:={\m := \m}

\def\=:{\m =: \m}

\def\geqs{\m \geq \m}

\def\FrT{\mathfrak{T}}                                         % A time-line. 

\def\FrC{\mbox{$\mathfrak{C}$}}                                
                                                       
\def\FrX{\mathfrak{X}}                                         % A set; 
\def\sFrR{\mbox{\scriptsize $\mathfrak{R}$}}                   % Region symbol
\def\FrS{\mbox{\Large $\mathfrak{s}$}}                         % For the general space, roughly meaning 'in general equipped set'.  
\def\sFrS{\mbox{\large$\mathfrak{s}$}}                         % smaller version of the same

\def\lFrg{\mbox{\Large$\mathfrak{g}$}}                         % Irrelevant group, Lie group.
\def\FrB{\mbox{$\mathfrak{B}$}}                                % Base space of a fibre bundle. 

\def\FrT{\mbox{\boldmath$\mathfrak{T}$}}                       % Used for tangent space and then cotangent space is $^*$ of this. 
\def\Hilb{\mbox{{\boldmath$\mathfrak{H}$}ilb}}                 % Hilbert space
\def\FrQ{\mbox{\Large $\mathfrak{q}$}}                               % Configuration space
\def\sFrQ{\mbox{\large $\mathfrak{q}$}}                              % smaller version 
\def\bFrL{\mbox{\boldmath$\mathfrak{L}$}}                            % The space of light degrees of freedom

\def\Phase{\mbox{{\boldmath$\mathfrak{P}$}hase}}                     % Phase space.

\def\bFrR{\mbox{\boldmath$\mathfrak{R}$}}                            % First letter of RigPhase, also used for Riem etc.  Is also, by itself, a ring.
\def\Rig-Phase{\bFrR\mbox{ig-}\Phase}                                % Rigged Phase Space
                                                                													   
\def\lFrr{\mbox{\Large $\mathfrak{r}$}}                              % Relative space

\def\sFrr{\mbox{\large $\mathfrak{r}$}}  

\def\FrP{\mbox{\Large $\mathfrak{p}$}}                                 % Preshape space.     
\def\sFrP{\mbox{\large $\mathfrak{p}$}}                                % smaller version

\def\FrR{\mbox{\boldmath$\mathfrak{R}$}}                             % Relational space

\def\sFrR{\mbox{\scriptsize\boldmath$\mathfrak{R}$}}                 % Relational space
\def\bFrR{\mbox{\boldmath$\mathfrak{R}$}}                            % Used in regularity structure symbol

\def\bFrR{\mbox{\boldmath$\mathfrak{R}$}}                            % Used in incipient regularity structure symbol

\def\diam{\delta iam}                                                % Diameter of an (N, 1) configuration

\def\1mat{\u{\u{1}}}                                                 % unit-entry matrix

\def\Leib{\bFrL\mbox{eib}}                                           % Leibniz space.

\def\Positive-Modespace{\mbox{{\boldmath$\mathfrak{M}$}odespace$^+$}}% Positive modespace

\def\POSITIVE-MODESPACE{\mbox{{\boldmath$\mathfrak{M}$}ODESPACE$^+$}}% Positive modespace alongside scalar field matter inhomogeneous modes.

\def\FrO{\mbox{$\mathfrak{O}$}}                                      % Individual gauge orbits.

\def\Top{\FrT\mo\mp}

\def\Rel{\FrR\me\ml}

\def\lattice{\mbox{\bf\Large$\mathfrak{L}$}}                                      % A lattice; $\lattice_{\bsFrc}$ is the lattice of first-class constraint algebraic structures, 
\def\Kin-Hilb{\mbox{{\boldmath$\mathfrak{K}$}in-\Hilb}}                     % Dynamical Hilbert space 

\def\Mid-Hilb{\mbox{{\boldmath$\mathfrak{M}$}id-\Hilb}}                     % Dynamical Hilbert space 

\def\Dyn-Hilb{\mbox{{\boldmath$\mathfrak{D}$}yn-\Hilb}}                     % Dynamical Hilbert space 

\def\5Star{\mbox{\Large$\star$}}              % Rectified time derviative actually used

\begin{document}

\begin{titlepage}

\begin{center}

\Large \lbiN{\bf -Body Problem:} \normalsize

\vspace{0.1in}

\large{\bf Minimal \biN's for Qualitative Nontrivialities}

\vspace{0.2in}

{\normalsize \bf Edward Anderson$^*$}

\vspace{.2in}

\end{center}

\begin{abstract}

We review the $N$-Body Problem in arbitrary dimension $d$ at the kinematical level, with modelling Background Independence in mind. 
In particular, we give a structural analysis of its reduced configuration spaces, 
decomposing this subject matter into basic Topology, Geometry, Group Theory, Linear Algebra, Graph Theory and Order Theory.  
At the metric level, these configuration spaces of shapes form basic geometric series for 1- and 2-$d$: 
$\mathbb{S}^{N - 2}$ and $\mathbb{CP}^{N - 2}$, though there are no more such series for $d \geq 3$.  
$d \geq 3$ also sees an onset of stratification. 
Casson's diagonal, for which $N = d + 1$, plays a critical role which we explain in simple Linear Algebra terms; these are moreover topologically spheres.
$N = d$ and $N = d + 2$ have further significance as well, the latter as regards a counting notion of genericity of the isotropy groups and kinematical orbits realized.  
These observations twin the $N$ = (3, 4, 5) progression in qualitative complexity that almost all $N$-Body Problem work for concrete $N$ concentrates on 
with the much better-known  $N$ = (2, 3, 4) progression in 2-$d$: from intervals to triangles to quadrilaterals: a large source of intuitions and mathematical analogies.  
We furthermore provide an Order-Theoretic genericity criterion, which is almost always bounded by the double-slope $N = 2 d + 1$ line, 
though an accidental relation pushes $N$ up by 1 to 8 in 3-$d$.  
We finally consider rubber shapes, for which the configuration spaces are graphs: much simpler than stratified manifolds, 
and yet containing quite a few of metric-level shapes' qualitative features. 
This singles out $N = 5$ and 6 in 1-$d$ and $N = 6$ and 8 in $\geq$ 2-$d$ for the onsets of various graph-theoretical nontrivialities.  

\end{abstract}

\n {\bf PACS}: 04.20.Cv , 02.40.Yy , 02.70.Ns . 	

\m 

\n {\bf Physics keywords}: $N$-Body Problem, Configuration Spaces, Background Independence, Topological and Geometrical Methods in Theoretical Physics.

\m

\n {\bf Mathematics keywords}: Shape Theory, Applied Topology, Applied Geometry, Orbit Spaces, Shape Statistics, Interplay between Linear Algebra and Graph Theory. 

\vspace{0.1in}
  
\n $^*$ Dr.E.Anderson.Maths.Physics@protonmail.com

\vspace{0.1in}

\end{titlepage}

%===================================================================================================================================================================================
%===================================================================================================================================================================================
\section{Introduction}\label{Intro-I}
%==================================================================================================================================================================================
%==================================================================================================================================================================================

The $N$-Body Problem exhibits major qualitative leaps in complexity between each rung of the following ladder.  

\m

\n A) The most familar 1-body problem and the 2-body problem \cite{Goldstein} 
which reduces to analogous mathematics by passing to centre of mass frame and discarding the centre of mass's position.  

\m 

\n B) The 3-body problem, classical in the Celestial Mechanics \cite{Euler, Lagrange, Dziobek, Wintner, McGehee, Marchal} and Molecular Physics \cite{Dragt, Iwai87, LR97} literatures. 

\m 

\n C) The 4-body problem, considered in e.g.\ \cite{MM75, Palmore, ABC96, LR97, LMRAC98, Albouy-1}.  

\m 

\n D) The 5-body problem; little specific-$N$ analytic work has made it this far \cite{Xia, ML00, Albouy}.  

\m 

\n  This ladder is moreover usually considered in the following context.  

\m 

\n{\bf Modelling assumption 1} The `underlying absolute space model' is $\mathbb{R}^d$: flat and topologically unidentified.  

\m 

\n{\bf Modelling assumption 2} In most of these works, the spatial dimension is $d = 3$.   

\m 

\n{\bf Modelling assumption 3} The Euclidean and/or similarity group are to be treated as physically-irrelevant automorphisms.  
The Euclidean group $Eucl(d)$ consists of translations and rotations, whereas the similarity group $Sim(d)$ includes dilations as well 
(these are the continuous versions of these groups, the full versions in each case also containing reflections).  

\m

\n Our first exploratory strategy is to let $d$ be arbitrary;  
this is fairly well-known to give systematic progressions of reduced configuration spaces for $d = 1$ and 2. 
The 1-$d$ case of this rests on its similarity shape spaces being spheres $\mathbb{S}^{N - 2}$, or its corresponding Euclidean scaled shape spaces being flat $\mathbb{R}^{N - 1}$. 
The 2-$d$ case rests on its similarity shape spaces being complex-projective spaces $\mathbb{CP}^{N - 2}$; this was already known to Smale at the topological level \cite{Smale70}, 
whereas Kendall established furthermore \cite{Kendall84} that these are equipped with the standard Fubini--Study metric in a natural manner.  
By similarity shapes, we mean $N$-point configurations (constellations) in $\mathbb{R}^d$ quotiented by $Sim(d)$.  
Shape Theory in this sense of similarity shapes has been further developed by Kendall, 
motivated by setting up a theory of Shape Statistics \cite{Kendall89, Small, Kendall, FileR, Bhatta, I, II, III, QuadI, PE16}. 
By Euclidean scaled shapes, we mean $N$-point constellations in $\mathbb{R}^d$ quotiented by $Eucl(d)$. 
Shape-and-Scale Theory in this sense is more commonly considered in the Mechanics and Molecular Physics literatures \cite{Iwai87, LR95, LR97, ML00}, 
though Shape Theory has been considered in this context as well \cite{M02, M05, FileR, M15, AMech}.
Both are known under the alias of `internal spaces' in the Molecular Physics literature.
For fixed $d \geq 3$, moreover, such systematic progressions are absent. 

\m 

\n Kendall also brought attention to \cite{Kendall} Casson's result that there is a third topological-level progression along the diagonal $(d, N) = (d, d + 1)$.
This is a first indication of structural insights arising from letting both $N$ and $d$ concurrently vary.
This is moreover a {\it basis diagonal} in terms of relative position vectors.   
The current article both provides this simple Linear Algebra interpretation and also works out its consequences, rendering this part of the subject rather more transparent.  

\m 

\n We further motivate the current article by noting the {\it Relational Aufbau Principle} \cite{I} -- 
starting out from small $N$, $d$ and automorphism group $\lFrg$ being quotiented out and build up. 
This is proving to be very useful, because smaller $d$, $N$ and $\lFrg$ re-enter the study of larger such, in particular as submanifolds, strata and significant subgroups.  
In this way, understanding the smaller models becomes a prerequisite in working out the topology and geometry of the larger models. 
In the current article, we work up $N$, $d$ and $\lFrg$ [as far as $Sim(d)$]. 

\m  

\n We consider mininal-$N$ features of constellation space on the carrier space $\mathbb{R}^d$ in Sec 2, 
of the relative space obtained from this by quotienting out the translations in Sec 3, 
and of Kendall's preshape space, for which the dilations are quotiented out as well, in Sec 4.    
We next outline constellation space and relative angle space on the carrier space $\mathbb{S}^1$ in Sec 5, 
and discrete quotients of configuration spaces in Sec 6.  
Sec 7 outlines very recent work \cite{Top-Shapes, Top-Shapes-2} on rubber shapes; 
here there are just three coarser universality classes for carrier spaces which are connected, Hausdorff and without boundary: 
all the $\mathbb{C}^d$ for $d \geq 2$ work out the same but each of $\mathbb{R}$ and $\mathbb{S}^1$ are extra cases.    
We return to to carrier space $\mathbb{R}^d$ in Sec 8, now  quotienting out rotations as well, 
at the level of dimension-counting and active versus (partially) inactive group actions. 
This gives the basisland, alias Casson, diagonal \cite{Kendall} in the $(n, d)$ grid (for $n := N - 1$), 
which splits the rest of this grid into nonspanninglands (below the diagonal) and (linearly) dependentlands above it.  
Sec 9 considers minimal $N$'s for various significant qualitative features to appear in flat-space polygonal shapes. 

\vspace{10in}

%%%%%%%%%%%%%%%%%%%%%%%%%%%%%%%%%%%%%%%%%%%%%%%%%%%%%%%%%%%  R E L A T I O N A L   T H E O R Y   P R O P E R  %%%%%%%%%%%%%%%%%%%%%%%%%%%%%%%%%%%%%%%%%%%%%%%%%%%%%%%%%%%%%%%%%%%%%%
%
\n We next embark on Relational Theory proper -- comprising both Shape Theory and Shape-and-Scale Theory --
with Sec 10's outline of minimal $N$ for various nontrivialities of topological relational space graphs corresponding to the rubber configurations.
$N = 5$ and 6 for $\mathbb{R}$ and $N = 6$ and 8 for $\mathbb{S}^1$ and $\geq$ 2-$d$ are singled out here by the onsets of various further nontrivialities.  
Metric-level relational configurations' relational spaces are then covered in Sec 11 at the topological level and Sec 12 at the metric level. 
In particular, $d = 1$, $d = 2$ and the basis diagonal form topological series, the first two of which remain metric-level series as well.
Also nonspanninglands entries for a given $N$'s dimensional equality is accompanied by topological equivalence, 
with the first diagonal line of nonspanninglands providing minimal representatives.  
Symmetry, uniformity, Lagrangian and Jacobian structure's qualitatively distinct small-$N$ features are outlined in Sec 13, 
and relational space isometry groups in Sec 14.   

\m 

\n Stratification -- a key feature in the Relational Theory of geometrical shapes -- is outlined in Sec 15. 
While this does not occur for 1- or 2-$d$ Similarity Shape Theory, it does in all subsequent dimensions. 
Its sole exemplar in 2-$d$ Euclidean Scaled Shape Theory is the maximal collision, 
a cause of many difficulties some of which subsequently affect other manifestations of stratification in larger models.
Paper II moreover discusses stratification becoming qualitatively harder to handle in affine and projective shape theories, 
conjecturing that this bears relation to the quotiented-out group being non-compact in a non-trivially resolvable manner. 

\m 

\n We next consider isotropy groups and kinematical orbit spaces in Sec 16; 
the realization of this was noted to be generic for the 5-body Problem by Mitchell and Littlejohn \cite{ML00}.  
We here point out, firstly, that this notion of genericity extends to $N = d + 2$, thus giving further significance to the minimal dependentlands.
With the opposite side's $N = d$ being the first model for a given $d$ on which the rotation group does not act fully, the 

\n($d$, $d$ + 1, $d$ + 2) triple of values of $N$.  
For $d = 3$, this triple of $N$ values is (3, 4, 5), accounting for many increasing qualitative complexities in ascending from 3- to 4- and 5-body problems, 
and moreover in a manner which loses a major source of qualitative distinctions among the $N \geq 5$.  
Our Linear Algebra conceptualization moreover twins this triple with (2, 3, 4) in 2-$d$. 
By this, some qualitative complications in 3-$d$'s passage from 4-to-5-body problems  
have far more geometrically familiar analogues in how intervals, triangles and quadrilaterals manifest increasing complexity.  
This moreover points to $N = 6$ in 4-$d$ as a particular next research frontier.

\m 

\n Secondly, we observe another sense of genericity. 
On the one hand, Mitchell and Littlejohn's is realization of full count of isotropy subgroups, so we term it {\it C-genericity}.
On the other hand, our new criterion is the further realization of these in the form of the generic lattice of isotropy subgroups, so we term it {\it O-genericity}, 
as our new genericity criterion thus has Order-Theoretic \cite{Lattice1, Lattice1} roots.  
This generally obeys a $N = 2 \, d + 1$ bound: double the Casson diagonal's slope in the $(d, N)$ grid.  
In 3-$d$, however, a Lie group accidental relation pushes $N$ up by 1 to 8.    

\m

\n We conclude in Sec 17 with interplays between the counting, rubber, topological, geometrical and stratificational effects.  
N.B.\ the current article concerns solely kinematics: no dynamics or work with specific potentials are included. 
Much of what is covered, however, is known to have dynamical consequences, as Appendix A outlines; Appendix B outlines large-$N$ treatments. 
Appendix C outlines Stiefel spaces, Grassmann spaces and a generalization, since all of these occur in the current article. 

\m 

\n{\bf Application 1} is to models of \cite{BB82, B03, FORD, Cones, FileR, QuadI, APoT1, ASphe, AMech, AConfig, ABook, I, II, III, A-Monopoles, Forth} 
Background Independence \cite{A64, A67, Giu06}. 
This has relevance to the Absolute versus Relational Motion Debate -- which dates back at least as far as Leibniz versus Newton \cite{Newton, L, M, DoD, Buckets, ABook, Generic} --
and has furthermore now been treated at the quantum level                                                                       \cite{AF, +Tri, FileR, QuadI, ABook}.
To models of the dynamical structure of GR \cite{B94I, RWR, ABook}, 
and to the Problem of Time \cite{Battelle, DeWitt67, K92, I93, B94I, Kiefer04, APoT1, ABook} that follows from the previous items in this paragraph.   

\m

\n{\bf Application 2} is to recent considerable expansion in scope of Kendall-type Shape Theories and Shape Statistics \cite{Bhatta, PE16}, 
which has renewed interest in the underlying theory. 
This has long been known to involve topology and geometry, though I have more recently shown it involves Graph Theory as well \cite{I, II, III, Top-Shapes, Top-Shapes-2}. 

\m  

\n{\bf Application 3} The Background Independence and Problem of Time application point can now be strengthened, by use of 
{\sl arbitrary group relationalism} \cite{AMech, ABook}, {\sl comparative study of absolute space models} \cite{ABook}, 
merely topological relationalism \cite{I89, Top-Shapes} and topology change \cite{GH92}.

\m 

\n Applications 2 and 3 are addressed in Paper II \cite{Minimal-N-2} of the current series, 
with 3 amounting to a continuation of the Aufbau Principle in $\lFrg$'s application into the Affine, Projective and Conformal Shape Theory cases. 
The current article and Paper II serves to replace the {\sl somewhat} useful adage that {\sl 3-Body Problem insights are necessary to model many aspects of Background Independence} 
with a detailed feature-by-feature account of {\sl when 4, 5, 6 and 8 Body Problems are required}, 
and of {\sl which of these features depend more generally on} $d$, $N - d$ {\sl or other} $(d, N)$ {\sl interplay}.   
 
\vspace{10in}
 
%==================================================================================================================================================================
%==================================================================================================================================================================
\section{$\mathbb{R}^d$ constellation space}\label{Q}
%==================================================================================================================================================================
%==================================================================================================================================================================

\n{\bf Definition 1} The {\it carrier space} $\FrC^d$, 
                   alias {\it absolute space} in the case of modelling physical space to be an at-least-provisional model for the structure of space.

\m 

\n{\bf Example 1} The most usually considered carrier space is flat Euclidean space $\mathbb{R}^d$.

\m 
 
\n{\bf Definition 2} A {\it constellation} is a collection of $N$ points $q^{a \, I}$ on a given carrier space, 
where $a$ is a spatial     index running from 1 to $d$ 
  and $I$      is a point label index running from 1 to $N$.  
We use $\biq$ as an index-free notation for this. 

\m 

\n{\bf Remark 1} In some physical applications, the points model material particles (classical, and taken to be of negligible extent).
Because of this, we subsequently refer to constellations as consisting of {\it points-or-particles}. 
The current article furthermore considers just the case of equal masses in any detail.  

\m 

\n{\bf Remark 2} Constellations can include coincident points, or collisions of particles; we subsequently refer to these special configurations as {\it coincidences-or-collisions}. 

\m 

\n{\bf Remark 3} We argue that $d$ is to be viewed as an independent variable and $N$ as a dependent variable, by which 
\be 
N = N(d) 
\ee
is entertained.
This is based on the premise of considering whichever number $N$ of points-or-particle on a given carrier space.
This means that the conventional presentation of $d$ and $N$ is to be a $(d, N)$ grid, with horizontal axis labelled $d$ and vertical axis labelled $N$.   
This covers the upper-right quadrant, since 
\be
N \mma d \m \geq \m 0  \m .
\ee
\n{\bf Definition 3} {\it Constellation space} $\FrQ(\FrC^d, N)$ is the space of all $N$-point-or-particle constellations on a fixed carrier space $\FrC^d$. 
%
%FFFFFFFFFFFFFFFFFFFFFFFFFFFFFFFFFFFFFFFFFFFFFFFFFFFFFF  C O O R D I N A T E   S Y S T E M S   F O R   3   P A R T I C L E S  FFFFFffFFFFFFFFFFFFFFFFFFFFFFFFFFFFFFFFFFFFFFFFFFFFFFFFFFFFF
{            \begin{figure}[!ht]
\centering
\includegraphics[width=1.0\textwidth]{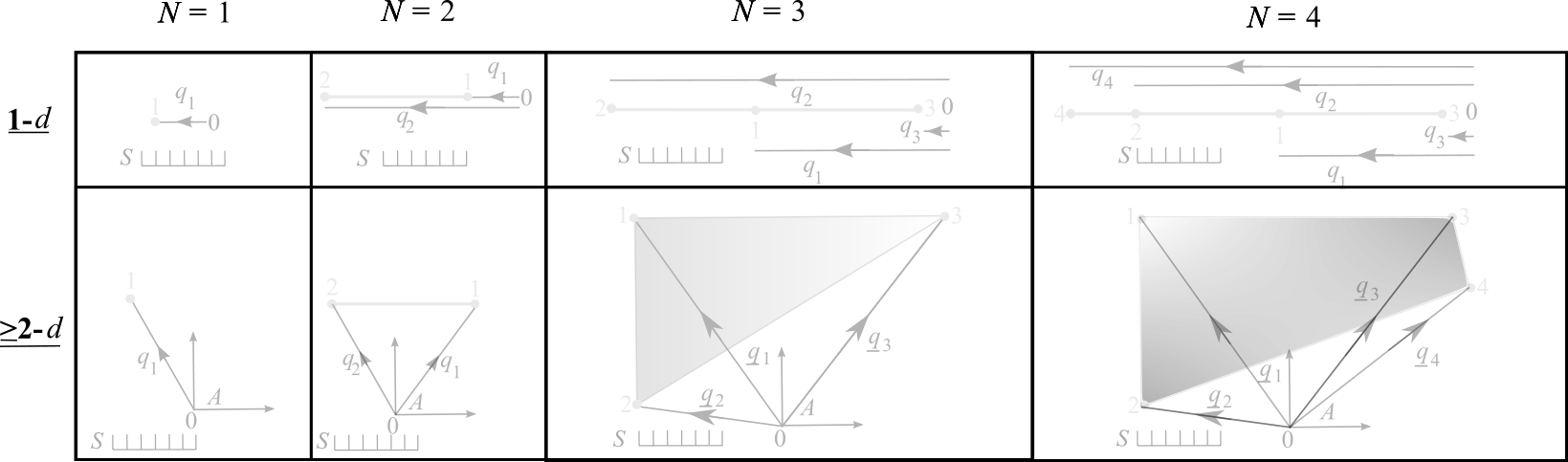}
\caption[Text der im Bilderverzeichnis auftaucht]{        \footnotesize{Position coordinates relative to a fixed absolute scale $S$, 
                                                                                                                absolute origin $0$, 
																										      and absolute axes $A$ (in $d \geqs 2$).
This article uses grey to distinguish out metric-level configurations from the corresponding configuration spaces (black), 
topological-level configurations (pastel blue) and their configuration spaces (bright blue).} }
\label{Position-Coordinates}\end{figure}            }
%FFFFFFFFFFFFFFFFFFFFFFFFFFFFFFFFFFFFFFFFFFFFFFFFFFFFFFFFFFFFFFFFFFFFFFFFFFFFFFFFFFFFFFFFFFFFFFFFFFFFFFFFFFFFFFFFFFFFFFFFFFFFFFFFFFFFFFFFFFFFFFFFFFFFFFFFFFFFFFFFFFFFFFFFFFFFFFFFFFFFFFFFF

\m

\n{\bf Proposition 1} Constellation space takes the product space form 
\be 
\FrQ(\FrC^d, N)  \es  \bigtimes_{i = 1}^N \FrC^d     \m , 
\label{Prod}
\ee 
of dimension 
\be
\mbox{dim}(\FrQ(\FrC^d, N))  =  N \, d               \m  .
\label{dim-Q}
\ee 
\n{\bf Remark 4} For the 1-Body Problem, $q^i = q^{i1}$ is all. 
Thus, 
\be 
N = 1 \m \mbox{ makes no distinction between configuration space and carrier space } : \m \FrQ(\FrC^d, 1) = \FrC^d   \m . 
\ee
$$
N \geq 2 \m \mbox{ is required to support a notion of configuration space that is distinct from carrier space }: 
$$
\be
\FrQ(\FrC^d, N)        \es       \bigtimes_{I = 1}^N \FrC^d  
                  \m  \neq \m    \FrC^d                       
				  \m \mbox{ for } \m  N \geqs 2                                     \m .  
\ee
\n{\bf Definition 4} We denote configuration space dimension by $k$.
If the configuration space is the most reduced for a given modelling situation, this is the corresponding number of degrees of freedom. 

\m 

\n{\bf Combinatorial Classification} A model is 

\m 

\n 1) {\it trivial} if it has $k = 0$ degrees of freedom, 

\m 

\n 2) {\it nontrivial}           if it has $k \geq  1$, and 

\m

\n 3) {\it minimally nontrivial} if it has exactly $k = 1$. 

\m 

\n{\bf Remark 5} Having 1 degree of freedom does not moreover suffice for those modelling situations in 
which degrees of freedom have to evolve with respect to other degrees of freedom. 
These are modelling situations in which absolute times such as Newton's have no meaning or existence: : a criterion of `Temporal Relationalism' \cite{ABook}.
The minimal model is now of one degree of freedom evolving with respect to another. 

\m  

\n 4) A model is {\it relationally trivial}  if it has $k \leq 1$ degrees of freedom, 

\m 

\n 5) {\it relationally trivial} if it has $k \geq 2$, and 

\m 

\n 6) {\it minimally relationally nontrivial}  if it has exactly $k = 2$. 

\m  

\n{\bf Remark 6} With $k = 0$ being relationally trivial as well, in the rest of this article, we refer to $k = 1$ as relationally trivial 
rather than its pre-relational alias minimally nontrivial.  

\m 

\n{\bf Example 1} For constellation space triviality,  
\be 
\mbox{dim($\FrQ(d, N)$) = 0  \m for \m $N \, d = 0$ \m $\Rightarrow$ \m $N = 0$ \m or \m $d = 0$  }         \m .   
\ee
For constellation space minimal nontriviality and relational triviality, 
\be
\mbox{dim($\FrQ(d, N)$) = 1 \m for \m  $N \, d = 1$ \m $\Rightarrow$ $(d, N) = (1, 1)$ }                    \m . 
\ee 
For constellation space minimal relational nontriviality, 
\be 
\mbox{dim($\FrQ(d, N)$) = 2 \m for \m  $N \, d = 2$ \m $\Rightarrow$ $(d, N) = (2, 1)$ \m or \m $(1, 2)$ }  \m .    
\ee 
\n{\bf Definition 5} The $p$th {\it moment} of $\u{q}_I$ is 
\be 
\u{Q}_I^p = m_I\u{q}^p_I                                             \m . 
\ee
The $p$th {\it scalar moment} is 
\be 
S_I^p = m_I |q_I|^p                                                  \m . 
\ee
The $I$th partial moment of inertia is 
\be 
I_I = S_I^2                                                          \m .
\ee 
The {\it inertia quadric} is
\be 
I = \sum_{I = 1}^N I_I = M_{IJ} q_I q_J = ||\biq||_{\sbiM}\mbox{}^2  \m , 
\ee 
for $M_{IJ} := \mbox{diag}(m_I)$ and $||\mbox{ }||_{\sbiM}$ the corresponding norm.

\m 

\n{\bf Definition 6} 
\be
\u{\chi}_I := \sqrt{\mu_I}\u{q}_{I}   \m 
\ee
are mass-weighted coordinates; we also denote these with indices suppressed as $\bchi$.   
Then 
\be 
I = \sum_{I = 1}^N|\chi_I|^2 = |\bchi|^2_{\mathbb{R}^{N \, d}} \m , 
\ee 
so we can interpret 
\be 
\chi_I = \sqrt{\mu_I} q_I \m \mbox{ as } \m \sqrt{I_I} \m : \m \m \mbox{the square root of the $I$th partial moment of inertia} \m .  
\ee
\n{\bf Remark 7} It is moreover straightforward to show that $\FrQ(d, N) = \mathbb{R}^{N \, d}$ additionally carries the standard flat metric 
with the $\bchi$ in the role of standard Cartesian coordinates: 
\be 
\d s^2_{\sFrQ(d, N)}  \es  \sum_{a = 1}^d \sum_{I = 1}^N   \d \chi^{a \, I \, \,  2}        \m .  
\ee
\n{\bf Definition 7} A {\it maximal coincidence-or-collision} is one in which all the points-or-particles are at a single location.  

\m 

\n{\bf Remark 8} For maximal coincidences-or-collisions O, $I = 0$.  
This transcends to  generic configuration G = O in $N = 1$ and binary coincidence-or-collision B alias 2 = O for $N = 2$.  

\m 

\n{\bf Definition 8}, $d = 0$'s configuration spaces are {\it maximallands}: 
all points-or-particles have to be piled up into the maximal coincidence-or-collision here; this is irrespective of $N \geq 1$.

\m

\n{\bf Remark 9} Maximalland constellationspace takes the form 
\be 
\FrQ(0, N)  \es  \prod_{i = 1}^N \FrC^0  \es  \prod_{i = 1}^N\{\mbox{pt}\}  
                        \es  \{\mbox{pt}\}
\ee 

\n{\bf Definition 9} We term the $d = 1$ models {\it N-stop metrolands} since their configurations look like underground train lines. 

\m  
   
\n                   We term the $d = 2$ models {\it N-a-gonlands} (collectively {\it polygonlands})     since their configurations are planar $N$-sided polygons. 

\m 
	
\n{\bf Remark 10} We shall moreover subsequently find technical reasons to accord distinction to metrolands and polygonlands.  	

\m 

\n{\bf Remark 11} If $d = 0$ as well, this is a maximalland as well as a pointland; in this case, moreover, maximal ceases to be a coincidence-or-collision. 
\be 
N \geq 2 \m \mbox{ is required to support notions of coincidence-or-collison}  \m .  
\ee

\m 

\n{\bf Definition 10} $N = 1$'s configuration spaces are {\it pointlands}: just one realized point-or-particle regardless of dimension. 

\m 

\n{\bf Remark 12} For $N \geq 1$, pointlands take the form 
\be 
\FrQ(d, 1) = \{ \mbox{pt} \} \m .  
\ee 
\n{\bf Definition 11} $N = 0$'s configurations are {\it emptylands}. 

\m 

\n If $d = 0$ as well, we have {\it empty pointland}: a single-point topology which does not model any {\sl realized} point-or-particle. 

\m 

\n{\bf Definition 12} A {\it partial coincidence-or-collision} is one in which not all the points-or-particles are at a single location.  

\m 

\n{\bf Remark 13}
\be 
N \geq 3 \m \mbox{ is required to have distinction between maximal and partial coincidences-or-collisions}  \m .  
\ee 
This is since for $N = 2$, the binary collision $2$ alias $\mB$ is both maximal and the only possible collision.  

\m 

\n{\bf Remark 14} Maximal coincidences-or-collisions $\mO$ are significant due to dominating both dynamical behaviour and technical intractability. 
So, on the one hand for some mathematical purposes they are excised.  
But, on the other hand, some such excisions are either physically contentious or requiring justification outside of the remit of classical nonrelativistic point-particle models.  

\m

\n{\bf Remark 15} Some of the reasons for excluding the maximal coincidence-or-collison $\mO$ in $N$-Body Problem studies moreover remain absent for $N = 2$ as well.
In this way, binary good behaviour turns out to trump maximal bad behaviour in this $N = 2$ case in which these two notions conflate. 

\vspace{10in}

%==================================================================================================================================================================
%==================================================================================================================================================================
\section{$\mathbb{R}^d$ relative space}\label{r}
%==================================================================================================================================================================
%==================================================================================================================================================================

\n{\bf Remark 1} Relational Theory furthermore takes some group of automorphisms $\lFrg$ of $\FrC^d$ -- or $\FrQ(\FrC^d, N)$ by its product space structure (\ref{Prod}) -- 
and regards these as irrelevant to the modelling in question. 
This includes e.g.\ quotienting out the Euclidean group of translations and rotations in a bid to free one's modelling from $\mathbb{R}^d$ absolute space. 
This is an example of quotienting out an isometry group, since 
\be
Isom(\mathbb{R}^d) = Eucl(d)                                         \m .  
\ee 
\n{\bf Remark 2} In the current section, however, we only quotient out $Tr(d)$. 
This corresponds to freeing our modelling from having an absolute origin, $0$.

\m 

\n{\bf Remark 3}
\be
Tr(d) = \mathbb{R}^d                                                 \m 
\ee
as a manifold. 
[On $\mathbb{R}^d$ we can continue transpating forever in $d$ mutually perpendicular directions, the corresponding basis vectors for which generate $Tr(d)$.]  
Thus 
\be
\mbox{dim}(Tr(d)) = d                                                                   \m .  
\label{dim-Tr}
\ee
\n{\bf Definition 1} Quotienting $Tr(d)$ out from $\FrQ(d, N)$ gives {\it relative space} :
\be
\lFrr(d, N)  \:=  \frac{\FrQ(d, N)}{Tr(d)}                                                  \m . 
\ee
\n{\bf Proposition 1} 
\be
\mbox{dim}\left(\frac{A}{B}\right)  \es  \mbox{dim}(A) - \mbox{dim}(B)                       
\label{A-B}
\ee
is the na\"{\i}ve count for the dimension of a quotient.  

\m

\n{\bf Remark 4} While `inactivity' and `strata' caveats qualifying this na\"{\i}vety appear in Secs \ref{Counts} and \ref{Strata} respectively, 
the current section and all sections up to where each caveat is introduced are not affected by these caveats.  

\m

\n{\bf Corollary 1}
\be
\mbox{dim}(\lFrr(d, N))  =  \mbox{dim}(\FrQ(d, N)) - \mbox{dim}(Tr(d)) 
                        =  \mbox{dim}(\mathbb{R}^{N \, d}) - \mbox{dim}(\mathbb{R}^d) 
  					    =  N \, d - d = (N - 1) d 
					    =  n \, d
\ee 
for 
\be 
n  :=  N - 1                                                                            \m . 
\label{n-def}
\ee 
\n{\bf Remark 5} Consequently, relative space triviality occurs for   
\be 
\mbox{dim($\lFrr(d, N)$) = 0  \mma  for \m $n \, d = 0$ \m  $\Rightarrow$ \m $d = 0$ \m \mbox{ or } \m $n = 0$ \m so \m $N = 1$ }   \m .   
\ee
In the first case, for $d = 0$, there is no nontrivial $Tr(d)$ to eliminate: 
\be 
Tr(0) = id                \m ,   
\ee 
so   
\be
\lFrr(0, N) = \FrQ(0, N)  \m . 
\ee  
In the second case, 
\be
\mbox{ for $N = 1$ \mma removing \m $Tr(d)$ \m leaves no degrees of freedom }  \m .  
\ee
\n{\bf Remark 6} From this point onward, working with an $(d, n)$ rather than $(d, N)$ grid starts to be convenient. 
Its range is $\mathbb{N}_0 \times \mathbb{N}_0$.    

\m 

\n{\bf Remark 7} Between the current article and its sequel, there are many degree of freedom minimality counts, so we phrase these systematically in terms of discrete 
equations to be solved in non-negative integers $\mathbb{N}_0$.  

\m 

\n{\bf Remark 8} Relative space minimal nontriviality and relational triviality occur for 
\be
\mbox{dim($\lFrr(d, N)$) = 1  \m for \m  $n \, d = 1$ \m $\Rightarrow$ \m $(d, n) = (1, 1)$ \m $\Rightarrow$ \m $(d, N) = (1, 2)$ }  \m . 
\ee 
\n{\bf Remark 9} Relative space minimal relational nontriviality occurs for 
\be 
\mbox{dim($\lFrr(d, N)$) = 2 \m for \m  $n \, d = 2$ \mma $\Rightarrow$ \m $(d, n) = (1, 2)$ \m or \m (2, 1) \m $\Rightarrow$ \m  $(d, N) = (1, 3)$ \m or \m $(2, 2)$ }  \m .  
\ee
\n{\bf Definition 2} The {\it minimal nontrivial unit} and the {\it minimal relationally nontrivial unit} are the smallest-$N$ configurations that a given model possesses 
with at least 1 and at least 2 degrees of freedom respectively.  
If exactly $k = 1, 2$ are respectively attainable; tis definition collapses to cases 3) and 6) of our combinatorial classification, 
though for some models, only some $l > 1$ and $m > 2$ are available as solutions in $\mathbb{N}_0$.  

\m

\n{\bf Remark 10} For some models, the cases of precisely 1 and/or 2 degrees of freedom are not realized. 
Thus the minimal (relationally) nontrivial unit may realize more than the minimal (relationally) nontrivial number of degrees of freedom.  
In Sec 2 and 3's examples, the minimal values coincide with the minimal units.  
In such cases, we refer to the models realizing the minimal relational nontrivial unit as {\it MNRUlands}.  

\m 

\n{\bf Remark 11} By Remark 5, a translation-invariant single-particle universe model is bereft of dynamical content. 

\m 

\n $N \geq 2$ however retains information after translations have been quotiented out. 

\m 

\n For $d = 1$, this is minimally nontrivial: a single degree of freedom in the form of a relative point-or-particle separation. 

\m 

\n For $d = 2$, this is minimally nonrelational: two degrees of freedom in the form of a relative separation 2-vector. 

\m 

\n For $d \geq 3$, this is a minimally relationally nontrivial unit that is in excess of minimally nonrelational: 
the    $d \geq 3$  degrees of freedom of a relative separation spatial vector. 

\m

\n{\bf Definition 3} Centre of mass position $\u{R}$ is defined as a point about which the total mass moment equates to the sum of first moments about all points-or-particles,  
\be 
M \u{R}  \es  \sum_{I = 1}^N m_i \u{q}_i                                  \m , 
\ee
where 
\be 
M        \es  \sum_{I = 1}^N  m_I                                             
\ee 
is the total mass. 
So
\be 
\u{R}  \es   \langle \u{q}^{I} \rangle  
       \es  \frac{1}{M} \, \sum_{I = 1}^1 m_I \u{q}^{I}                   \m .
\ee
\n{\bf Remark 12} Relational space is moreover all of computationally, topologically and geometrically trivial, in the sense that 
\be
\lFrr(d, N)  \:=  \frac{\FrQ(d, N)}{Tr(d)} 
            \es  \frac{\mathbb{R}^{N \, d}}{\mathbb{R}^d} 
			\es  \mathbb{R}^{n \, d}                                      \m .  
\ee
We shall see below that this corresponds to the quotienting's admitting the standard basic interpretation of {\sl passing to centre of mass coordinates}.  
For $N = 1$, 
\be
\langle  \u{q}^{1} \rangle  \es  \frac{1}{1} \, \sum_{I = 1}^1 \u{q}^{I} 
                            \es                                \u{q}^{1}   \m . 
\ee
For $N = 2$ -- minimal for a nontrivially realized centre of mass --
\be
\langle  \u{q}^{I} \rangle  \es  \frac{1}{2} \, \sum_{I = 1}^2 \u{q}^{I} 
                            \es  \frac{1}{2} ( \u{q}^{1} + \u{q}^{2} )    \m , 
\ee
which has the interpretation of being the midpoint between the two points-or-particles.  

\m

\n{\bf Structure 1} {\it Relative separation vectors} -- differences between position vectors (Fig \ref{Relative-Lag-Jac-Coords}) --
\be 
\u{r}^{IJ} := \u{q}^J - \u{q}^I 
\ee 
occur here because such differences are {\sl invariants} corresponding to the $Tr(d)$ group. 
We use the notation $\br$ to suppress both $a$ and $IJ$ indices.  
Relative separation vectors are also widely known as {\it relative Lagrange coordinates}. 
In this way, part of Remark 6's content can be rephrased as  
\be
N \geqs 2 \m \mbox{ is required for relative Lagrange coordinates to be defined } .  
\ee
{\it Relative (Lagrange) separations} themselves are the corresponding magnitudes,
\be
r^{IJ}  :=  ||\u{r}^{IJ}||  \m .  
\ee
The general translation invariant takes the form 
\be 
f(r^{IJ})                   \m ; 
\ee  
aside from being well-known, in this case, such results follow more generally from solving observables PDEs \cite{ABeables}. 

%FFFFFFFFFFFFFFFFFFFFFFFFFFFFFFFFFFFFFFFFFFFFFFFFFFFFFF  C O O R D I N A T E   S Y S T E M S   F O R   3   P A R T I C L E S  FFFFFffFFFFFFFFFFFFFFFFFFFFFFFFFFFFFFFFFFFFFFFFFFFFFFFFFFFFF
{            \begin{figure}[!ht]
\centering
\includegraphics[width=1.0\textwidth]{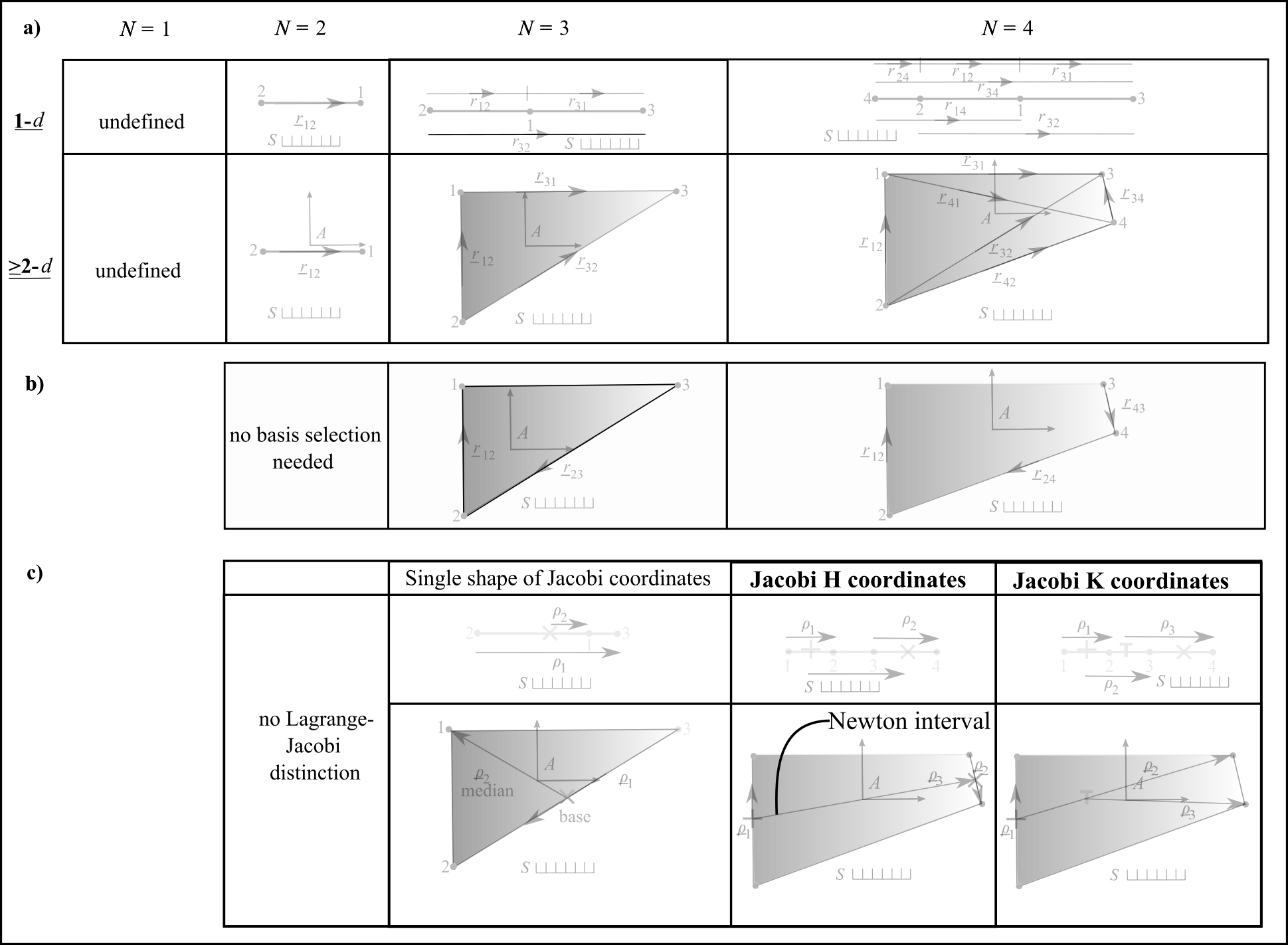}
\caption[Text der im Bilderverzeichnis auftaucht]{        \footnotesize{a) Relative separation vectors.   
Note that these are still defined with respect to a fixed absolute scale $S$, and absolute axes $A$ (in $d \geqs 2$).  

\m 

\n b) Some basis choices amongst these. 

\m 

\n c) Relative Jacobi coordinates.
We use the centre of mass notation $\sfX$ for double and $\sfT$ for triple.
} }
\label{Relative-Lag-Jac-Coords}\end{figure}            }
%FFFFFFFFFFFFFFFFFFFFFFFFFFFFFFFFFFFFFFFFFFFFFFFFFFFFFFFFFFFFFFFFFFFFFFFFFFFFFFFFFFFFFFFFFFFFFFFFFFFFFFFFFFFFFFFFFFFFFFFFFFFFFFFFFFFFFFFFFFFFFFFFFFFFFFFFFFFFFFFFFFFFFFFFFFFFFFFFFFFFFFFFF

\m

\n{\bf Remark 13} 
\be
N \geq 2 \m  \mbox{ is required to support a nontrivial notion of centre of mass } . 
\ee
This is in the sense that for $N = 1$, centre of mass just coincides with the point-or-particle itself, 
                  whereas for $N = 2$ it is midway between the two points-or-particles. 

\m 

\n{\bf Remark 14} The following Linear Algebra features enter at the $N = 3$ level. 
\n 1) 
\be
\mbox{$N \geq 3$ \m is required for not all relative Lagrange separation vectors \m $\u{r}_{IJ}$ \m to be linearly independent}  \m .
\ee   
\n A first counter to this is to pick a basis among the $\u{r}_{IJ}$. 
Observing the number of elements in this basis, we can moreover re-issue definition (\ref{n-def}), now imbued with conceptual meaning: 
\be 
n := N - 1 \m  \mbox{ is the number of independent relative separation vectors supported by \m $N$}                             \m . 
\label{n-def-2}
\ee
\n 2) Working with such a basis however imparts nondiagonality upon the inertia quadric
\be
I  \es  \frac{1}{M}\sum_{I}\sum_{> \m J} m_I m_J r_{IJ}^2                                                                      \m .
\ee
E.g.\ for $N = 3$, and equal masses, taking $\u{r}_{12}$ and  $\u{r}_{23}$ as basis, 
\be
I  =   {r_{12}}^2 + {r_{23}}^2 + ({r_{12}} + {r_{23}})^2 
  \es  \frac{2}{3} \, (  {r_{12}}^2 + {r_{23}}^2 + \u{r}_{12} \cdot \u{r}_{23}  ) 
   =   \frac{1}{3}(  {r_{12}} \m r_{13} )\left( \s{\mbox{2 \m 1}}{1 \, \m \, 2} \right) \left(  \s{\mbox{$r_{12}$}}{r_{13}}   \right)   \m .
\ee 
Throughout the $N \geq 3$ cases for which this occurs, this can be remedied \cite{Marchal, I} by applying diagonalization. 
The resulting coordinates are widely known as {\it Jacobi coordinates}. 
It follows that
\be
N \geqs 3 \m \mbox{ is required to have relative Jacobi to Lagrange coordinate distinction }                                 \m . 
\ee
since for $N = 2$, the single relative Lagrange coordinate already has diagonal status. 

\m 

\n{\bf Remark 15} Examining the Jacobi coordinates from a conceptual point of view moreover identifies them as inter point-or-particle {\sl cluster} separations. 
Thus working in these coordinates amounts to point-or-particle separation's primary status being shared by point-or-particle cluster separations as well. 
In fact, viewing points-or-particles as clusters of 1 point-or-particle, this description can be simplified to according clusters primary status \cite{I}.   
Thus 
\be
N \geqs 3 \m \mbox{ is required to have a distinct notion of cluster separation }                          \m 
\ee
(and subsequent allocation of primary status to clusters in general rather than just to individual points-or-particles or separations between these).
From this Jacobian point of view, (\ref{n-def}) can be further reconceived of as  
\be 
n := N - 1 \m  \mbox{ is the number of independent relative cluster separation vectors supported by \m $N$} \m . 
\ee
The inertia quadric is now 
\be
I = \sum_i \mu_i R_i^2
\ee
for $\mu_i$ the Jacobi masses; note that equal particle masses $m_I$ does not imply equal Jacobi masses $\mu_i$

\m

\n{\bf Remark 16} It is furthermore convenient to work with {\it mass-weighted relative Jacobi coordinates}, 
\be 
\underline{\rho}_i := \sqrt{\mu_i}\u{R}_i \m . 
\ee
We use the notation $\brho$ to suppress both $\alpha$ and $i$ indices.  
Now 
\be 
I = \sum_i |\rho_i|^2 = |\brho|^2   \m .
\ee
Also $\lFrr(d, N) = \mathbb{R}^{n \, d}$ additionally carries the standard flat metric with $\rho^{a  \, i}$ in the role of standard Cartesian coordinates: 
\be 
\d s^2_{\sFrr(d, N)} = \sum_{a \, i = 1}^{n \, d} \d \rho^{a \, i \, \, 2}        \m .  
\ee 
\n{\bf Remark 17} 
\be 
\u{R}_i \m \mbox{ and } \m \u{\rho}_i \mma \mbox{are also bona fide $Tr(d)$ invariants}            \m , 
\ee
since, as linear combinations of differences $\u{r}_{IJ}$, they are indeed of the form $f(\u{r}_{IJ})$. 
\be 
\rho_i = \sqrt{\mu_i} R_i = \sqrt{I_i}   \m : 
\ee  
the $i$th Jacobi partial moment of inertia.  

\m 

\n{\bf Remark 18} For $N = 3$, there are moreover 3 Jacobi coordinate systems -- i.e.\ choice of clusterings -- corresponding to label permutations (Fig 4.a).  
This is a $d$-independent statement. 
%
%FFFFFFFFFFFFFFFFFFFFFFFFFFFFFFFFFFFFFFFFFFFFFFFFFFFFFFFFFFFFFFFFFFFFFFFFFFFFFFFFFFFFFFFFFFFFFFFFFFFFFFFFFFFFFFFFFFFFFFFFFFFFFFFFFFFFFFFFFFFFFFFFFFFFFFFFFFFFFFFFFFFFFFFFFFFFFFFFFFFFFFFFF
{\begin{figure}[ht]
\centering
\includegraphics[width=0.85\textwidth]{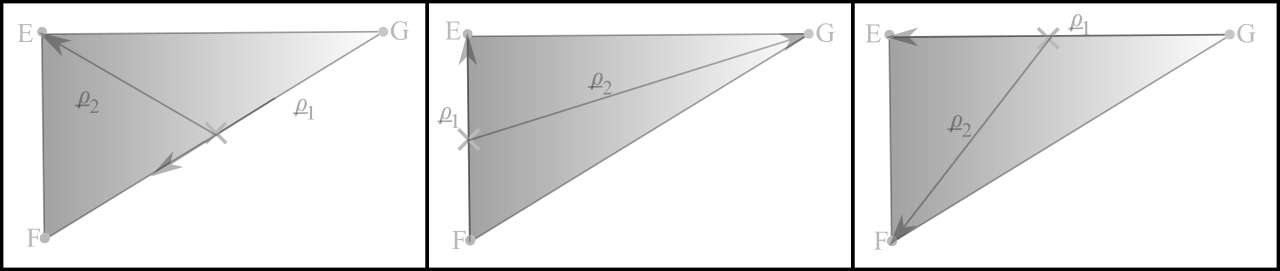}
\caption[Text der im Bilderverzeichnis auftaucht]{\footnotesize{The 3 clustering choices pictured.  
Let us label the 3-Body Problem's points-or-particles by E, F and G;  the choice of cluster are then EF, G, FG, E and GE, F, with Jacobi vectors as depicted.}} 
\label{Cluster-Choices}\end{figure} } 
%FFFFFFFFFFFFFFFFFFFFFFFFFFFFFFFFFFFFFFFFFFFFFFFFFFFFFFFFFFFFFFFFFFFFFFFFFFFFFFFFFFFFFFFFFFFFFFFFFFFFFFFFFFFFFFFFFFFFFFFFFFFFFFFFFFFFFFFFFFFFFFFFFFFFFFFFFFFFFFFFFFFFFFFFFFFFFFFFFFFFFFFFFF

\m

\n{\bf Remark 19} However, 
\be
\mbox{For $N \geq  4$ \m diversity of Jacobi alias clustering coordinate systems goes beyond just label permutations}  \m . 
\ee 
\n{\bf Remark 20}  A first reason for this is that more than one {\it two-part partition} is now possible, imparting cluster shape distinctions. 
I.e.\ 4 can be split as 2 + 2, giving the Jacobi H coordinates shape, or as 3 + 1, giving the Jacobi K coordinates shape (both are in row 2 of \ref{Jacobi-Trees-3-to-6}).
This is in contrast with 3 admitting 2 + 1 alone as such a split. 

\m 

\n{\bf Remark 21} Using $C(p, q)$ to denote `$p$ choose $q$', there are 
\be 
\frac{C(4, 2)}{2} = 3 \m \mbox{ labelling choices of Jacobi H's} 
\ee 
and 
\be 
C(4, 1) = 4 \m \mbox{ labelling choices of } \m 3 + 1 \mbox{ split Jacobi K's}    \m . 
\ee 
The Jacobi K is moreover also the smallest example of cluster {\sl hierarchy}, i.e. (2 + 1) + 1 split, 
by which there is a further multiplicative factor of 3 labellings of 2 + 1 in according the final count of 
\be 
12 \m \mbox{ labelling choices of Jacobi K's}  \m .  
\ee 
\n{\bf Remark 22} A further way of phrasing the Jacobi H versus Jacobi K diversity is that 
\be 
\mbox{$N = 4$ \m vertices supports the first tree graph that is not a path} \m : 
\ee
the {\it claw graph} alias {\it 3-star graph} in Fig \ref{Jacobi-Trees-3-to-6}.
This graph-theoretic point of view on clustering moreover quickly becomes crucial in finding the range of possible Jacobi coordinate systems as $N$ increases 
(also supplied in Fig \ref{Jacobi-Trees-3-to-6} for $N = 5$ and $6$).   
In turn, this permits coordinates adapted to whatever cluster hierarchy, which is of clear physical use, for instance Celestial Mechanics and Molecular Physics.    
%
%FFFFFFFFFFFFFFFFFFFFFFFFFFFFFFFFFFFFFFFFFFFFFFFFFFFFFF  C O O R D I N A T E   S Y S T E M S   F O R   3   P A R T I C L E S  FFFFFffFFFFFFFFFFFFFFFFFFFFFFFFFFFFFFFFFFFFFFFFFFFFFFFFFFFFF
{            \begin{figure}[!ht]
\centering
\includegraphics[width=1.0\textwidth]{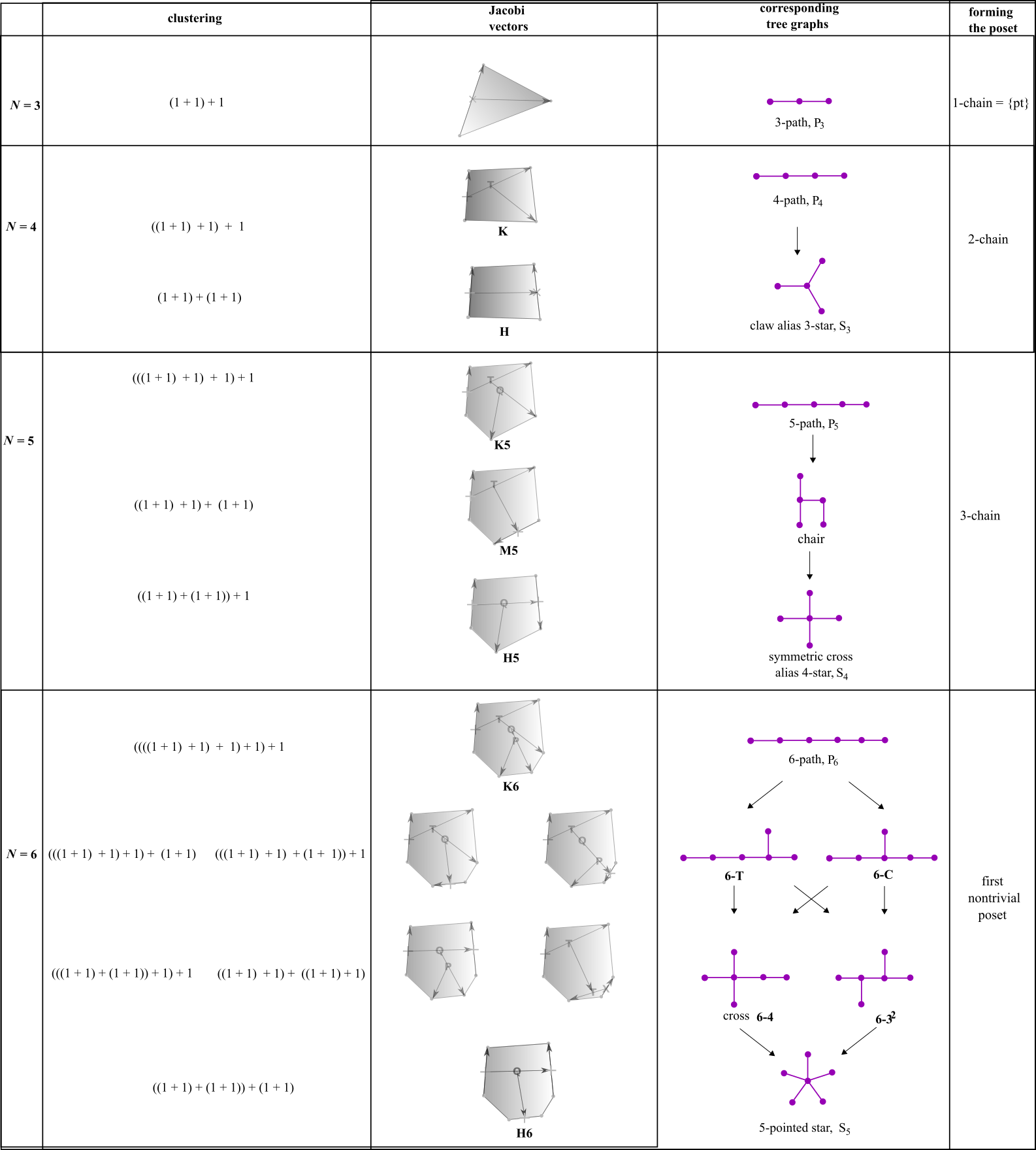}
\caption[Text der im Bilderverzeichnis auftaucht]{        \footnotesize{Correspondence between Jacobi coordinate clusterings and tree graphs. 
We find that the Jacobi K and H generalize to arbitrary $N$ as top and bottom elements KN and HN of the poset of trees ordered by amount of branching. 
Thus other names for these are the path of length $N$, $\mP_N$, and the $n$-pointed star, $\mS_n$.  
For $N = 5$, the poset first has a middle element, which we denote by M5; the corresponding tree graph is the chair. 
For $N = 6$ there are now 4 middle elements, which we distinguish as 
-T: terminal branch, 
-C: central branch, 
-$3^2$: two valency-3 vertices and 
-4: one valency-4 vertex, alias the cross graph.
centre of mass notation: $\sfQ$ for quadruple and $\sfP$ for pentuple. } }
\label{Jacobi-Trees-3-to-6}\end{figure}            }
%FFFFFFFFFFFFFFFFFFFFFFFFFFFFFFFFFFFFFFFFFFFFFFFFFFFFFFFFFFFFFFFFFFFFFFFFFFFFFFFFFFFFFFFFFFFFFFFFFFFFFFFFFFFFFFFFFFFFFFFFFFFFFFFFFFFFFFFFFFFFFFFFFFFFFFFFFFFFFFFFFFFFFFFFFFFFFFFFFFFFFFFFF

\m 

\n{\bf Remark 23} The cluster hierarchies moreover form a poset \cite{Lattice1} under the operation `has more levels of hierarchy'. 
This is in 1 : 1 correspondence with the poset of `how branched' a tree graph is. 
This poset moreover has unique greatest and least elements in each case. 
\be 
\mbox{For $N = 3$ \mma the top is the same as the bottom, so there is only one point in the cluster hierarchy poset.} 
\ee 
\be 
\mbox{$N = 4$ \m is minimal for most and least hierarchical to be distinct} 
\ee
this maps to 
\be
\mbox{$N = 4$ \m is minimal for most and least branched trees to be distinct}   \m .    
\ee 
The least branched trees are of course the $N$-paths $\mP_N$, whereas the most branched trees are the $n$-pointed stars $\mS_{n}$, as per Fig 4.  

\m 

\n Moreover, 
\be 
\mbox{$N = 5$ \m is minimal for the cluster hierarchy poset to have a middle}                                                                            \m , 
\ee 
in this case the {\it chair graph} exhibited within Fig \ref{Jacobi-Trees-3-to-6}.  
This case's hierarchy poset is still just a chain (the 3-chain alias directed 3-path).   

\m 

\n Finally,  
\be 
\mbox{$N = 6$ \m is minimal to have a nontrivial cluster hierarchy poset}                                                                                \m , 
\ee
in the sense of being more than just a chain. 
This is attained through having the central and non-central single branch, forming the poset at the bottom of Fig \ref{Jacobi-Trees-3-to-6}.   
 
\m 

\n{\bf Remark 24} That detailed understading of a system benefitting from using of {\sl all} shapes of cluster hierarchy \cite{Forth} poses a problem even for moderate $N$, 
since the number of trees grows quickly with $N$ (as a larger-$N$ effect, we postpone discussion of this to the Conclusion).  

\m 

\n{\bf Remark 25} See Sec \ref{TSS} and \cite{I, II, III, Top-Shapes, Top-Shapes-2} for further instances of Graph Theory entering Shape Theory. 

\m 

\n \n{\bf Definition 4} {\it Exit channels} are the possible bound states leaving an interaction process. 

\m 

\n{\bf Remark 26} These are tight binaries, ternaries T alias 3... which are {\sl approximately} binary, ternary... collisions.  
In this way, partial collisions moreover enter the physically important theory of exit channels. 

\m 

\n{\bf Example 1} The 3-Body Problem has 3 labellings of B alias 2, and thus of exit channel.  

\m 

\n{\bf Example 2} The 4-Body Problem has 7 two-fragment partial collision: three double binaries $2^2$ and four ternaries.  
\be
\mbox{$N = 3$'s diversity of partial collisions is pure labelling} \m ,  
\ee 
while 
\be 
\mbox{$N = 4$ \m is minimal for exit channel diversity to have further topological content} \m .   
\ee 
This is directly tied to the Jacobi H to K distinction, with the three types of H collapsing to the three labellings of $2^2$ 
and the four ahierarchical labellings of K collapsing to the four labellings of 3.  

\m 

\n{\bf Proposition 1} 
\be
|\mbox{$n$-ary 2-fragment partial collisions}|  \es  |\mbox{$n$-ary 2-fragment exit channels}| 
                                                \es  2^n - 1
\ee
\n{\bf Remark 27} This is the $N$ to $n$ recasting of \cite{ML00}'s mechanics-level result, and is underlied by a well-known basic Combinatorics result. 

\m 

\n{\bf Remark 28} Maximalland relative spaces are 
\be 
\lFrr(0, N) = \emptyset     \m . 
\ee 
Metroland relative spaces are 
\be 
\lFrr(1, N) = \mathbb{R}^n  \m ; 
\ee  
these are not distinctive at the current section's level, but are at the level of relational spaces. 
Pointland relative spaces are  
\be 
\lFrr(d, 1) = \emptyset     \m .
\ee 
\n{\bf Definition 5} We term $(d, 2)$ configurations spaces {\it intervallands}, since they are characterized by a single relative separation.  

\m 

\n{\bf Remark 29} Intervalland relative spaces are of the form 
\be 
\lFrr(d, 2) = \mathbb{R}_0  \m .  
\ee 

\vspace{10in}

%===================================================================================================================================================================================
%===================================================================================================================================================================================
\section{Kendall's preshape space}\label{Preshape}
%===================================================================================================================================================================================
%===================================================================================================================================================================================

\n We next consider the group of dilations, $Dil$.  

\m

\n{\bf Proposition 1} 
\be
Dil = \mathbb{R}^+ \m  \mbox{ independently of dimension}                   \m .  
\ee 
This is clear from the possible values that magnification factors can take [excluding the inversion: i.e.\ making the continuous choice].  

\m  

\n{\bf Corollary 1}
\be
\mbox{dim}(Dil)  =  \mbox{dim}(\mathbb{R}_+) 
                 =  1                                                       \m  .
\ee
\n {\bf Definition 1} Considering the translations and dilations together amounts to considering the {\it dilatational group}   
\be
Dilatat(d) = Tr(d) \rtimes Dil  \m , 
\ee 
where $\rtimes$ denotes semidirect product.  

\m 

\n{\bf Definition 2} If the carrier space scale $S$ is to join the carrier space origin $0$ in having no meaning,  
the incipient constellation space $\FrQ(d, N)$ is quotiented by the dilatational group $Dilatat(d)$ to form Kendall's \cite{Kendall}
\be
(\mbox{\it preshape space}) \mma  \FrP(d, N)  \:=  \frac{  \FrQ(d, N)  }{  Dilatat(d)  } 
                                              \es  \frac{  \FrQ(d, N)  }{  Tr(d) \times Dil  }  
                                              \es  \frac{  \lFrr(d, N)  }{  Dil  } 
                                              \es  \frac{  \mathbb{R}^{n \, d}  }{  \mathbb{R}_+  }  \m .
\label{Shape-Space}
\ee
\n{\bf Corollary 1} Proposition 1 of Sec 3 applies to Dil quotients without caveats as well, giving 
\be
\mbox{dim}(\FrP(d, N)) \es  \mbox{dim}\left(  \frac{    \lFrr(d, N)    }{     Dil       }  \right) 
                       \es  \mbox{dim}\left(  \frac{ \mathbb{R}^{n \, d}  }{ \mathbb{R}_+  }  \right) 
				       \es  \mbox{dim}(  \mathbb{R}^{n \, d}  ) - \mbox{dim}(  \mathbb{R}_+  )
                       \es  n \, d - 1                                                                                                                     \m , 
\ee
\n{\bf Remark 1} Thus 
\be 
\mbox{dim($\FrP(d, N)$) = 0 \m for \m  $n \, d = 1$ \m $\Rightarrow$ \m $(d, N) = (1, 2)$                   : \m preshape space triviality}                         \m ,   
\ee
\be
\mbox{dim($\FrP(d, N)$) = 1 \m for \m  $n \, d = 2$ \m $\Rightarrow$ \m $(d, N) = (1, 3)$ \m or \m $(2, 2)$ : \m preshape space relational triviality}              \m , 
\ee  
$$ 
\mbox{dim($\FrP(d, N)$) = 2 \m for \m  $n \, d = 3$ \m $\Rightarrow$ \m $(d, N) = (1, 3)$ \m or \m (3, 1)} 
$$
\be
\mbox{\m $\Rightarrow$ \m $(d, N) = (1, 4)$ \m or \m $(3, 2)$  : \m minimal preshape space relational nontriviality}   \m . 
\ee 
This is the first occurrence of 4 points-or-particles on a line and of the triangle in the plane.

\m 

\n{\bf Structure 1} The corresponding invariants are now functions of ratios of magnitudes -- or components -- whether of relative Lagrange, 
relative Jacobi or or mass-weighted relative Jacobi quantities 
\be 
f\left(\frac{r_{IJ}}{r_{KL}}\right)          \mma  
f\left(\frac{R_i}{R_j}\right)                \mma 
f\left(\frac{\rho_i}{\rho_i}\right)          \m   . 
\ee 
\n{\bf Structure 2} The {\it normalized mass-weighted Jacobi coordinates}  
\be 
\u{n}_i := \frac{\u{\rho}_i}{\rho}
\ee 
for\footnote{$\rho$ is interpreted in detail in \cite{I}.}  
\be
\rho := \sqrt{I}
\ee 
the square root of the total moment of inertia $I$ are also significant.  
\be 
f(n_i)
\ee
is thus another functional form of general interest. 

\m 

\n{\bf Proposition 1} The preshape spaces are moreover spheres \cite{Kendall84}, both topologically and metrically: 
\be 
\FrP(d, N) = \mathbb{S}^{n \, d - 1}                                                                                         \m , 
\ee 
naturally equipped with standard (hyper)spherical metric 
\be 
\d s^2_{\sFrP(d, N)} = \d s^2_{\sss\sp\sh\se} 
                    = \sumpn \prod\mbox{}_{\mbox{}_{\mbox{\scriptsize $m$ = 1}}}^{p - 1} \mbox{sin}^2\theta_m \d\theta_p^2   \m . 
\label{HS}
\ee 
\n{\bf Remark 3} The first part of this can be seen from normalization corresponding to the on-sphere condition (standardizing its radius to 1): $\rho_i$ flat metric subject to 
\be 
\sum_{i = 1}^n \rho_i^2 \m = \m \rho^2 = I
\ee  
i.e.\ 
\be 
\sum_{i = 1}^n n_i^2 \m = \m \sum_{i \, a  = 1}^{n \, d}  n_{ia}\mbox{}^2 = 1 \m : \m \mbox{ the on-$\mathbb{S}^{n \, d - 1}$-condition } .  
\ee
\n The second part arises via the metric in an obvious choice of basis of ratio coordinates 
\be 
\frac{\rho_p}{\rho_n} \mma  p = 1 \m \mbox{ to } \m n - 1 
\ee 
being recognizable as \cite{FileR} the hypersphere in Beltrami coordinates, followed by the standard conversion from these to the given hyperspherical coordinates $\theta_p$.

\m

\n{\bf Remark 4} Following on from Remark 9 in Sec 2, $N = 1$ admits no normalizable preshapes, 
which require $N \geq 2$ to be realized, as non-coincident pair of points (or non-colliding pair of particles). 
This gives a major reason to exclude the maximal coincidence-or-collision O from the shapes. 

\m

\n{\bf Remark 5} This first part of Remark 3 exemplifies a spherical method, 
whereas the first part of the second a projective method, which is then converted to spherical form.  
In general, spheres benefit from being able to draw on projective as well as spherical methods, `doubling' the amount of techniques and methods available.  

\m

\n{\bf Remark 6} While many applications have meaningful scale, it turns out to be simpler to remove scale, remove rotations and then re-introduce scale 
than to remove rotations directly.  
This is in part because the preshape space is a sphere, and thus well-known and in particular compact.
So we know the resulting quotients are compact \cite{Lee}.
It is also in part because the reintroduction of the scale is merely a cone construct, as we shall detail in Secs 11 and 12.  

\m 

\n{\bf Structure 3} Some special cases are as follows.  
The maximalland preshape spaces are  
\be 
\FrP(0, N) = \emptyset                           \m . 
\ee 
The metroland preshape spaces are  
\be 
\FrP(1, N) = \mathbb{S}^{n - 1}                  \m ; 
\ee 
these are not distinguished at the preshape space level, but are at the shape space level.  

\m 

\n The pointland preshape spaces are 
\be
\FrP(d, 1)  = \emptyset                          \m .   
\ee 
Finally, the intervalland preshape spaces are 
\be
\FrP(d, 2)  = \{\mbox{pt}\} \m \m d \geq 2       \m , 
\ee
with the exception that 
\be
\FrP(1, 2) = C_2                                 \m . 
\ee
Note that             $\mbox{dim}(\{\mbox{pt}\}) = 0 $, 
whereas by convention $\mbox{dim}(\emptyset)     = -1$.   

\vspace{10in}

%===================================================================================================================================================================================
%===================================================================================================================================================================================
\section{Constellations on $\mathbb{S}^1$}\label{S1}
%===================================================================================================================================================================================
%===================================================================================================================================================================================

\n{\bf Remark 1} Configurations here are conveniently described in terms of $N$ angular coordinates $\varphi^I$ (Fig \ref{Angle-Lag-Jac}.a).
%
%FFFFFFFFFFFFFFFFFFFFFFFFFFFFFFFFFFFFFFFFFFFFFFFFFFFFFFFFFFFFFFFFFFFFFFFFFFFFFFFFFFFFFFFFFFFFFFFFFFFFFFFFFFFFFFFFFFFFFFFFFFFFFFFFFFFFFFFFFFFFFFFFFFFFFFFFFFFFFFFFFFFFFFFFFFFFFFFFFFFFFFFFF
{            \begin{figure}[!ht]
\centering
\includegraphics[width=0.75\textwidth]{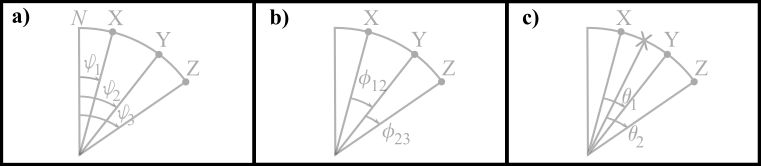}
\caption[Text der im Bilderverzeichnis auftaucht]{        \footnotesize{a) Absolute angles with respect to a preferred axis $N$. 
b) Relative (Lagrange) angles and c) Relative Jacobi angles both cease to refer to $N$. } }
\label{Angle-Lag-Jac} \end{figure}          }
%FFFFFFFFFFFFFFFFFFFFFFFFFFFFFFFFFFFFFFFFFFFFFFFFFFFFFFFFFFFFFFFFFFFFFFFFFFFFFFFFFFFFFFFFFFFFFFFFFFFFFFFFFFFFFFFFFFFFFFFFFFFFFFFFFFFFFFFFFFFFFFFFFFFFFFFFFFFFFFFFFFFFFFFFFFFFFFFFFFFFFFFFF

\m 

\n{\bf Structure 1} The corresponding constellation space is 
\be 
\FrQ(\mathbb{S}^1, N)  \es  \bigtimes_{i = 1}^N \mathbb{S}^1 
                       \es  \mathbb{T}^N                            \m : \m \m \mbox{the $N$-torus} \m .  
\ee 
\n{\bf Structure 2} Quotienting out the sole $SO(2)$ rotation is straightforward in this case -- a compactified version of removing a translation.  

\m 

\n{\bf Remark 2} {\it Relative angles} are differences between absolute angles (Fig \ref{Angle-Lag-Jac}.b), in the present context  
\be
\theta_{IJ} := \varphi_J - \varphi_I \m. 
\ee 
arise here as invariants corresponding to the $SO(2) = U(1)$ group.  
These can be thought of as compactified analogues of relative Lagrange separations on $\mathbb{R}$.  

\m 

\n{\bf Definition 1} The {\it centre of angle} is 
\be
\Phi = \frac{1}{N}\sum_{I = 1}^N \varphi_I \m ; 
\ee 
this is for the equal-masses unit-radius circle case.
More generally -- in the distinct-masses, fixed arbitrary circle radius case -- this would be a {\it centre of inertia}. 
This is the circle point-or-particle model's analogue of a centre of mass.
This is the total version; clearly centre of angle can also be defined by averaging over a partial subset of the angles.  

\m 

\n{\bf Definition 2} The {\it angular inertial quadric} is 
\be 
I_{\sA}(\varphi^I)   \es   \sum_{I}{\varphi_I}^2 \m . 
\ee
\n{\bf Remark 3} This can be re-expressed in terms of relative angles as 
\be 
I_{\sA}(\theta_{IJ}) \:= \frac{1}{N} \, \sum_{I}\sum_{> \m J}{\theta_{IJ}}^2  \m .
\ee 
\n{\bf Structure 3} The corresponding {\it angular} alias {\it circular relative spaces} are   
\be 
\lFrr(\mathbb{S}^1, N)  \:=  \frac{\FrQ(\mathbb{S}^1), N}{SO(2)} 
                       \es  \frac{\bigtimes_{i = 1}^N \mathbb{S}^1 }{\mathbb{S}^1} 
					   \es  \bigtimes_{i = 1}^n \mathbb{S}^1 
					   \es  \mathbb{T}^n                                            \m : \m \mbox{ the $n$-dimensional torus }  \m ,  
\ee 
where $n$ admits the somewhat new interpretation 
\be
\mbox{(independent circular relative angle number) } \m  n  :=  N - 1                                                           \m .  
\ee
\n{\bf Remark 4} The various minimal counts are exactly the same as for $\mathbb{R}$ subject to $Tr(1)$.

\m 

\n{\bf Remark 5} The following Linear Algebra features enter at the $N = 3$ level. 

\m 

\n 1) 
\be
\mbox{$N \geq 3$  \m  is required for not all relative angles \m $\theta_{IJ}$ \m to be linearly independent}  \m .
\ee   
\n Again, a first counter to this is to pick a basis among the $\theta_{IJ}$. 
Observing the number of elements in this basis,  
\be 
n := N - 1 \m  \mbox{ is the number of independent relative angles supported by $N$} \m . 
\label{n-def-3}
\ee
\n 2) Again, working with such a basis however imparts nondiagonality upon the angular inertia quadric.   

\m 

\n For $N \geq 3$ this can again be remedied by applying diagonalization. 
The resulting coordinates are inter angle {\sl cluster} separations $\slTheta_i$ (Fig \ref{Angle-Lag-Jac}.c), 
in direct parallel to Jacobi coordinates as regards their coefficients and the clustering structure they correspond to;  to date, they are far less well-known.
\be
N \geqs 3 \m \mbox{ is required to have a distinct notion of angle cluster separation}                          \m  . 
\ee
\n{\bf Remark 6} It is furtherly convenient to work with {\it mass-weighted relative cluster separation angles}, 
\be 
\Theta_i := \sqrt{\mu_i}\slTheta_i                                                                      \m . 
\ee
These mass weightings are also in 1 : 1 correspondence with the familiar case of Jacobi coordinates.    

\m 

\n{\bf Structure 4} It is straightforward to show that $\lFrr(\mathbb{S}^1, N) = \mathbb{T}^{n}$ additionally carries the standard flat-torus metric  
with $\theta^{a  \, i}$ in the role of standard angular coordinates: 
\be 
\d s^2_{\sFrr(\mathbb{S}^1, N)}  \es  \sum_{i = 1}^{n} \d \Theta^{i \, \, 2}                            \m .  
\ee 
\n{\bf Remark 7} Counts, H and K cluster shape differences, correspondence to tree graphs and poset structure thereupon are as per the standard Jacobi coordinates. 

\m 

\n{\bf Remark 8} The invariants in this $\mathbb{S}^1$ relational theory are of the general form  
\be 
f(\theta_{IJ}) \m \mbox{ or } \m f(\Theta_i) \m .  
\ee 

\vspace{10in}

%===================================================================================================================================================================================
%===================================================================================================================================================================================
\section{Discrete quotients}\label{Gamma}
%===================================================================================================================================================================================
%===================================================================================================================================================================================

\n{\bf Modelling ambiguity 1} Whether to identify mirror images.  

\m 

\n{\bf Modelling ambiguity 2} Whether points-or-particles are distinguishably or indistinguishably labelled. 

\m 

\n{\bf Remark 1} $N = 1$ does not however support meaningful particle labels, since an unlabelled particle is equivalent to a particle labelled by the unlabel. 
Thus 
\be
N \geq 2 \m \mbox{ is required to have meaningful label distinguishability }    \m .   
\ee
This starts with two particles being labelled EF being distinct from two both labelled E (or, equivalently, not labelled at all).  

\m 

\n{\bf Remark 2} A first quartet of discrete group operations are then $\Gamma = id$, $C_2$-ref (acting reflectively), $S_N$ and $S_N \times C_2$; 
for $N = 2$, $S_2 = C_2$ as well, but now acting as $C_2$-label.  

\m

\n{\bf Remark 3} 
\be 
N \geq 3 \m \mbox{ is required to have partial label distinguishability }       \m :
\ee 
a further feature necessitating $S_N$'s subgroups to be considered as well. 
(The first proper subgroup of $S_N$ occurs for $N = 3$: $A_3 = C_3$.)  
One is now to consider the bounded lattice of subgroups of $\mathbb{S}_N \times \mathbb{C}_2$.   
The unit here is the whole group, the zero the trivial group, the join is intersection and the meet is the group jointly generated.  

\m 

\n{\bf Remark 4} 
\be 
\mbox{$N \geq 3$ \m is required to have a nontrivial lattice of discrete quotients}  \m .  
\ee
\n{\bf Remark 5} Another conceptually and technically useful way \cite{A-Monopoles} of viewing this extension concerns the lattice of distinguishable group actions 
of the subgroups of $S_N \times C_2$.
We shall see in Sec 8 that moreover for some $(d, N)$, $S_N$ suffices; 
more accurately, it is the {\sl lattice of distinct discrete subgroup actions} of $S_N \times C_2$ or $S_N$ on a given configuration space that is realized; 
see \cite{A-Monopoles} for further details.    

\m 

\n{\bf Remark 6} If mirror images are distinct (a valid choice in 1-$d$), Example 1 and 2 of Sec 2's configurational counts are all doubled. 

\m 

\n{\bf Remark 7} The quotients by these themselves then form a corresponding upside-down lattice.  
The bounded lattice unit here is the original configuration space without discrete quotienting, 
whereas the bounded lattice zero is the most discretely quotiented configuration space, 
which we term {\it Leibniz space} due to its minimal implementation of Leibniz's `Identity of Indiscernibles' \cite{L}.   
% 
%FFFFFFFFFFFFFFFFFFFFFFFFFFFFFFFFFFFFFFFFFFFFFFFFFFFFFFFFFFFFFFFFFFFFFFFFFFFFFFFFFFFFFFFFFFFFFFFFFFFFFFFFFFFFFFFFFFFFFFFFFFFFFFFFFFFFFFFFFFFFFFFFFFFFFFFFFFFFFFFFFFFFFFFFFFFFFFFFFFFFFFFFF
{            \begin{figure}[!ht]
\centering
\includegraphics[width=0.75\textwidth]{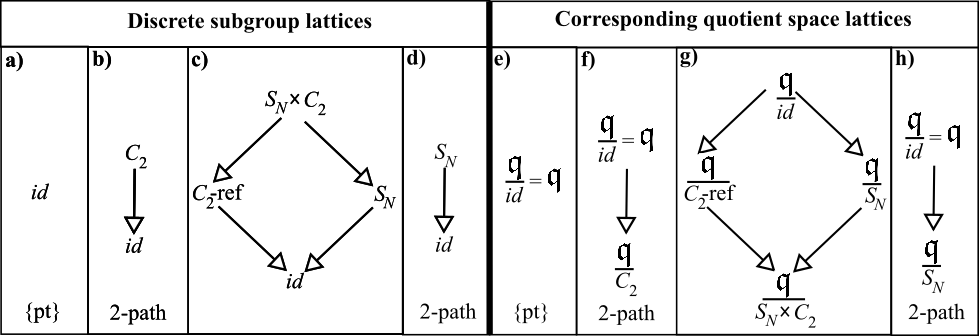}
\caption[Text der im Bilderverzeichnis auftaucht]{        \footnotesize{Discrete subgroup lattices, and the corresponding quotient space lattices, as feature in the current article. 
c) and g) are for those $(d, N)$ for which mirror image identification is optional, whereas d) and h) are for those for which this is not optional.  
} }
\label{Specific-Lattices} \end{figure}          }
%FFFFFFFFFFFFFFFFFFFFFFFFFFFFFFFFFFFFFFFFFFFFFFFFFFFFFFFFFFFFFFFFFFFFFFFFFFFFFFFFFFFFFFFFFFFFFFFFFFFFFFFFFFFFFFFFFFFFFFFFFFFFFFFFFFFFFFFFFFFFFFFFFFFFFFFFFFFFFFFFFFFFFFFFFFFFFFFFFFFFFFFFF

\m 

\n{\bf Structure 1} The mirror-images identified preshape space is
\be 
\FrO\FrP(d, N) = \mathbb{RP}^{n \, d - 1}  \m .
\ee 
In particular, 
\be 
\FrO\FrP(1, N) = \mathbb{RP}^{n  - 1}  \m .
\ee

%===================================================================================================================================================================================
%===================================================================================================================================================================================
\section{Rubber Relationalism}\label{Co}
%===================================================================================================================================================================================
%===================================================================================================================================================================================

\n{\bf Definition 1} A {\it rubber shape} alias {\it topological notion of shape} is the topological content of the $N$-point constellation on some carrier space $\FrC^d$.  

\m 

\n{\bf Remark 1} Rubber shapes continue to distinguish between the types of coincidence-or-collision.
The configurations here are the topological types of coincidence-or-collison and the topological type or types of generic configuration exhibiting no coincidences-or-collisions.  
See Fig \ref{1-to-4} for examples. 
Coincidences-or-collisions being an important part of Geometry and Dynamics 
further justifies their study in isolation by the creation of Rubber Relationalism \cite{Top-Shapes, Top-Shapes-2}.    
%
%FFFFFFFFFFFFFFFFFFFFFFFFFFFFFFFFFFFFFFFFFFFFFFFFFFFFFFFFFFFFFFFFFFFFFFFFFFFFFFFFFFFFFFFFFFFFFFFFFFFFFFFFFFFFFFFFFFFFFFFFFFFFFFFFFFFFFFFFFFFFFFFFFFFFFFFFFFFFFFFFFFFFFFFFFFFFFFFFFFFFFFFFF
{            \begin{figure}[!ht]
\centering
\includegraphics[width=0.8\textwidth]{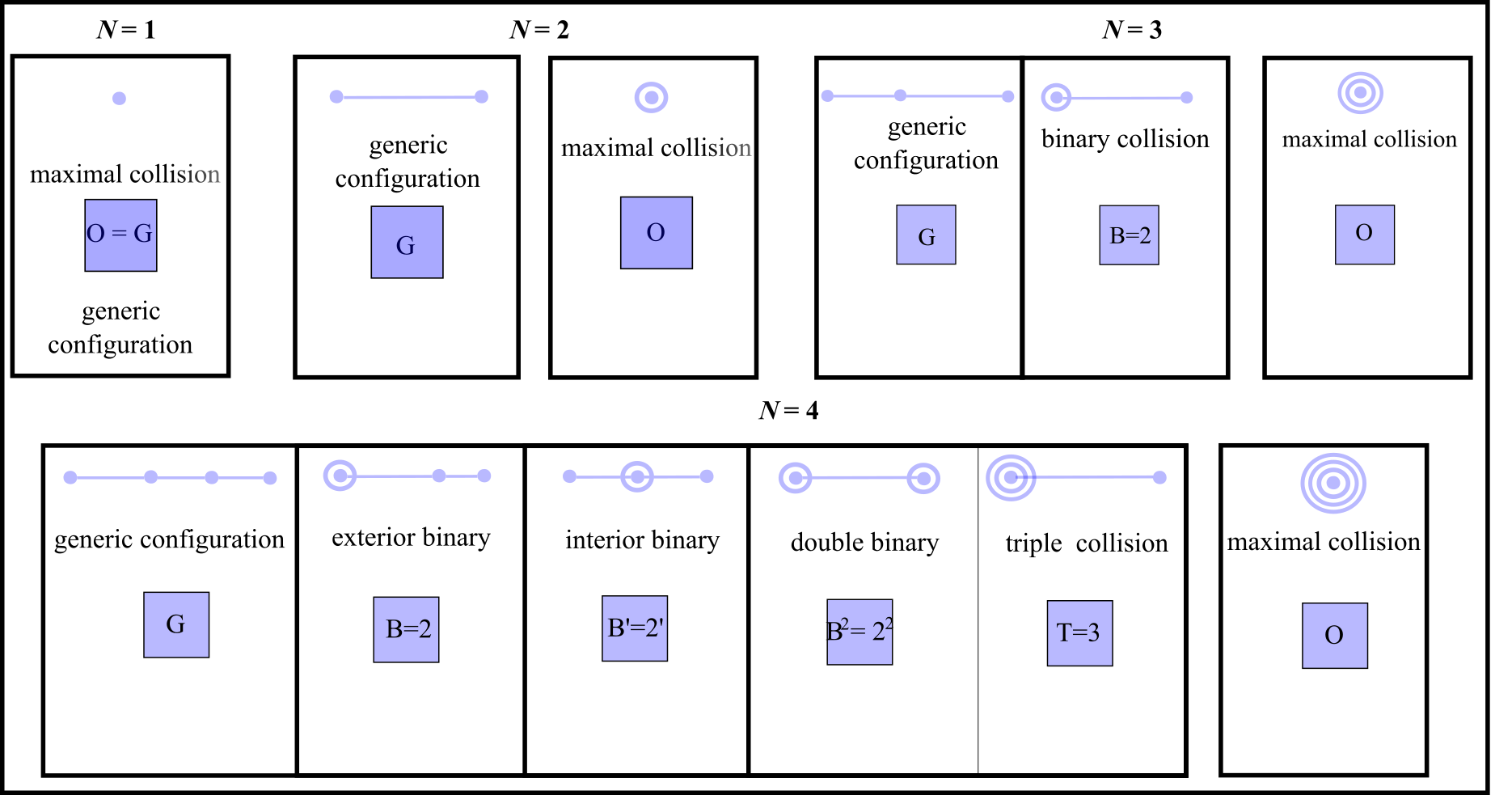}
\caption[Text der im Bilderverzeichnis auftaucht]{        \footnotesize{Types of collision for $N = 1, 2, 3$ and $4$.}}
\label{1-to-4} \end{figure}          }
%FFFFFFFFFFFFFFFFFFFFFFFFFFFFFFFFFFFFFFFFFFFFFFFFFFFFFFFFFFFFFFFFFFFFFFFFFFFFFFFFFFFFFFFFFFFFFFFFFFFFFFFFFFFFFFFFFFFFFFFFFFFFFFFFFFFFFFFFFFFFFFFFFFFFFFFFFFFFFFFFFFFFFFFFFFFFFFFFFFFFFFF

\m 

\n{\bf Definition 1}  A {\it rubber relational theory} is a quadruple   
\be
( \, \FrC^d, \, N, \, S, \Gamma \, )         \m .
\ee 
$\FrC^d$ and $N$ having been introduced in Sec 2 and $\Gamma$ in Sec 6, it remains for us to introduce $S$.  
This is binary-valued: $s$ if the theory has scale or $\emptyset$ if it does not. 
  
\m

\n{\bf Definition 2} {\it Rubber Relational space} is the quotient space
\be  
\Rel(\FrC^d, N; S, \Gamma)  \es  \frac{\FrQ(\FrC^d, N; S)}{  \Gamma  }                                            \m .  
\ee
If $S = s$, the rubber relational space notion specializes to {\it rubber shape-and-scale space} 
\be 
\FrR(d, N; \Gamma)  \:=  \Rel(d, N; s, \Gamma)  \m , 
\ee
whereas if $S = \emptyset$, it specializes to {\it rubber shape space}  
\be 
\FrS(d, N; \Gamma)  \:=  \Rel(d, N; \emptyset, \Gamma)   \m . 
\ee

\m

\n{\bf Remark 2} Under the modelling assumption that $\FrC^d$ is a connected Hausdorff manifold without boundary, 
it turns out that there are just three classes of rubber shape theories \cite{Top-Shapes, Top-Shapes-2}. 
In particular, all such $d \geq 2$ models all work out the same regardless of what the carrier space $\FrC^d$ is. 
The other two distinct cases are $\mathbb{R}$ \cite{Top-Shapes} and $\mathbb{S}^1$ \cite{Top-Shapes-2}.  

\m 

\n This universality is moreover subject to the caveat that some $\FrC^d$ support the removal of scale while others do not. 
Rubber Relationalism thus includes both Rubber Shape Theory and Rubber Scaled-Shape Theory.

\m 

\n The underlying reason why 1-$d$ is different is because removing a point disconnects the real line but not any space with $d \geq 2$ 
or the circle (which is turned into an interval). 
This enters consideration in Rubber Relationalism via collisions realizing such excised points. 
For $\mathbb{S}^1$, this enters consideration upon removing {\sl two} points, since the second point disconnects the interval created by removing the first point.  

\m

\n{\bf Remark 3} Rubber Relationalism admits various simplified cases as per the rest of this section.   

\m  

\n{\bf Proposition 1} For $d \geq 2$, rubber shapes-and-scales are in 1 : 1 correspondence with partitions.  
So in the mirror-image identified and indistinguishable case, 
\be 
|\Top\mbox{-}\Leib_{\sFrR}|  =     p(N)   \m , 
\ee 
and
\be 
|\Top\mbox{-}\Leib_{\sFrS}|  =  p(N) - 1  \m ,    
\ee
for $p(N)$ the well-studied partition number \cite{p(N)}.  
(This parallels Proposition 1 of Sec 3.) 

\m 

\n{\bf Proposition 2} For $N = 1$ to $3$, the three classes coincide (modulo the $\mathbb{S}^1$ class not realizing the mirror-images-distinct shape class).   

\m 

\n{\bf Remark 12} The cases including and excluding scale are different though the former appending the maximal collision to the latter.  
This is stucturally the first place that the maximal $\mO$ makes a difference in the $N$-Body Problem theory, 
but it does not present a difficulty on this occasion: it is just a vertex like any other partition is. 

\m 

\n{\bf Remark 13} 
\be
\mbox{$N = 4$ \m  gives the first departure \cite{Top-Shapes} of rubber Leibniz configurations from partitions, for \m $\FrC^d = \mathbb{R}$}  \m .
\ee
This is due to the $2$ to $2^{\prime}$ distinction among binary coincidence-or-collisions, as depicted in Fig \ref{Top-Leib-Order}.b). 
Such distinction occurs for all subsequent $N$ as well. 

\m 

\n{\bf Remark 14} 
\be 
\mbox{$N = 6$ \m marks the first departure of \m $\FrC^d = \mathbb{S}^1$ \m  rubber Leibniz configurations from partitions} \m . 
\ee
This is due to the $2^2$ to $2^{2 \, \prime}$ distinction among double-binary coincidence-or-collisions, as depicted in Fig \ref{Top-Leib-Order}.c). 
Such distinction occurs for all subsequent $N$ as well. 

\m 

\n{\bf Remark 15} The above two departures can be viewed as successive fine grainings of partitions: 
a $\mathbb{S}^1$ distinction fine-graining, 
followed by a further $\mathbb{R}$-specific fine-graining. 

\m 

\n{\bf Remark 16} Consequently 
\be 
r(N)  \m \geq \m  s(N) 
      \m \geq \m  p(N)   \m , 
\ee
for $r(N)$ the number of distinct rubber Leibniz configurations on $\mathbb{R}$ 
and $s(N)$ the number of distinct rubber Leibniz configurations on $\mathbb{S}^1$;  
the first few values of these are given in Fig \ref{Top-Leib-Order}.a).  
%
%FFFFFFFFFFFFFFFFFFFFFFFFFFFFFFFFFFFFFFFFFFFFFFFFFFFFFFFFFFFFFFFFFFFFFFFFFFFFFFFFFFFFFFFFFFFFFFFFFFFFFFFFFFFFFFFFFFFFFFFFFFFFFFFFFFFFFFFFFFFFFFFFFFFFFFFFFFFFFFFFFFFFFFFFFFFFFFFFFFFFFFFFF
{            \begin{figure}[!ht]
\centering
\includegraphics[width=0.6\textwidth]{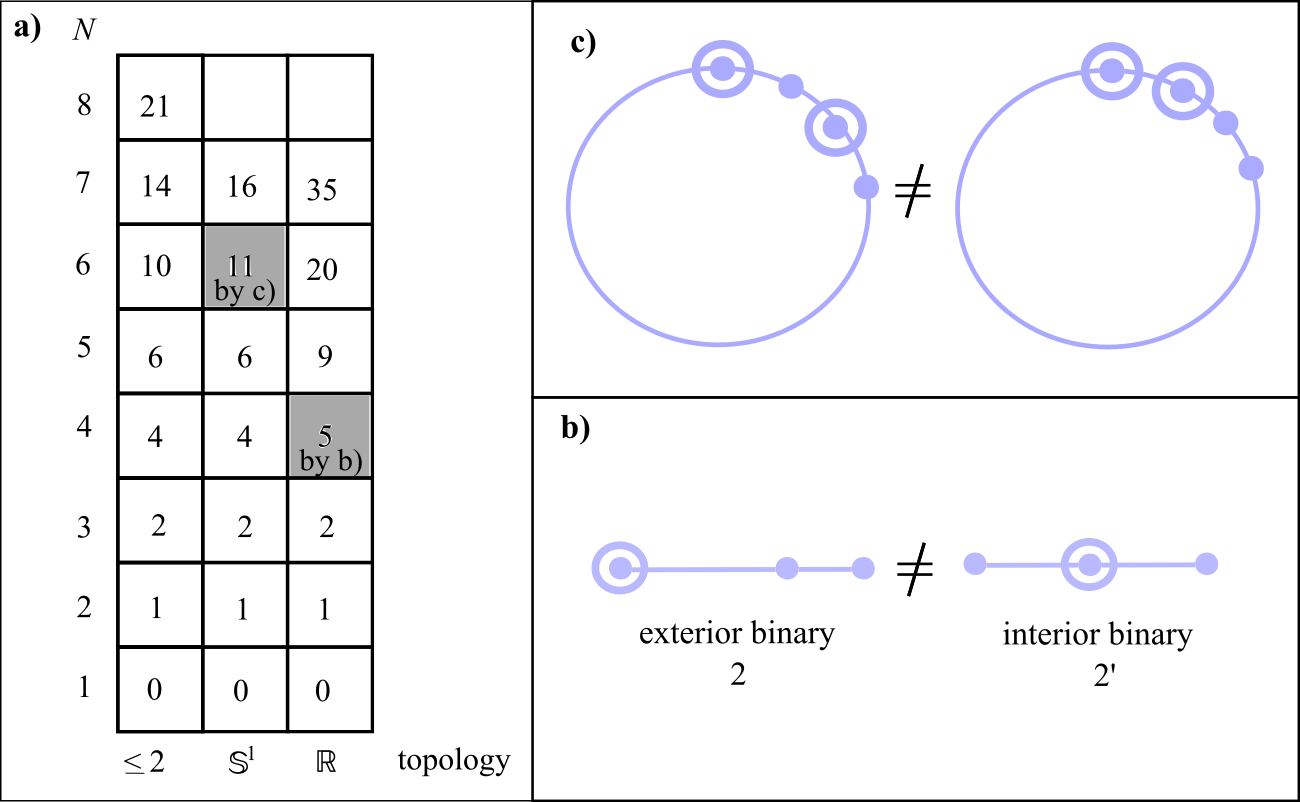}
\caption[Text der im Bilderverzeichnis auftaucht]{        \footnotesize{a) Rubber shape counts = |$\Top\mbox{-}\Leib(\FrC^d)$|, 
with b) and c) indicating $\mathbb{R}$'s first distinction at $N = 4$ and $\mathbb{S}^1$'s at $N = 6$.     
} }
\label{Top-Leib-Order} \end{figure}          }
%FFFFFFFFFFFFFFFFFFFFFFFFFFFFFFFFFFFFFFFFFFFFFFFFFFFFFFFFFFFFFFFFFFFFFFFFFFFFFFFFFFFFFFFFFFFFFFFFFFFFFFFFFFFFFFFFFFFFFFFFFFFFFFFFFFFFFFFFFFFFFFFFFFFFFFFFFFFFFFFFFFFFFFFFFFFFFFFFFFFFFFFFF

\m 

\n{\bf Remark 17} For less quotiented relational spaces, for $N = 2$, there are two topological classes of (non-)shape (Fig \ref{1-to-4}). 
In this case, excluding O still leaves us with a topological theory.

\m

\n Excluding O from $N = 2$, however, leaves one with no topological class distinction; one needs to consider at least $N = 3$ to have this feature. 

\m 

\n{\bf Proposition 3}  
\be
|\Top\mbox{-}{\cal R}| = l(N) = 2^N \m \mbox{ and } \m |\Top\mbox{-}\FrS| = l(N) - 1 = 2^N - 1  \m , 
\ee 
where $l$ stands for `labelled'.

%==================================================================================================================================================================
%==================================================================================================================================================================
\section{Similarity and Euclidean relational theories' dimension counts}\label{Counts}
%==================================================================================================================================================================
%==================================================================================================================================================================

%=========================================================================================================================================================================================
\subsection{Shape and relational spaces}\label{Preamble-1}
%=========================================================================================================================================================================================

\n{\bf Definition 1}  A {\it relational theory} is a quadruple   
\be
( \, \FrC^d, \, N, \, \lFrg, \Gamma \, )         \m .
\ee 
$\FrC^d$ and $N$ having been introduced in Sec 2 and $\Gamma$ in Sec 6, it remains for us to introduce 
\be 
\lFrg := Aut(\langle\FrC^d, \sigma\rangle)
\ee  
as the continuous group of automorphisms acting on $\FrC^d$ for $\sigma$ some level of mathematical structure on $\FrC^d$.

\m

\n{\bf Definition 2} {\it Relational space} is the quotient space
\be  
\Rel(\FrC^d, N; \lFrg)  \es  \frac{\FrQ(\FrC^d, N)}{\lFrg}                                           \m .  
\ee
Including discrete automorphisms as well, 
\be
\Rel(\FrC^d, N; \lFrg, \Gamma)  \es \frac{  \FrQ(\FrC^d, \, N)  }{  \lFrg \circ \Gamma  }  
                                \es \frac{  \Rel(\FrC^d, N; \lFrg)  }{  \Gamma  }                    \m ,
\ee
where $\circ$ is a generic product (of the form $\times$ or $\rtimes$ in all examples in the current article).

\m

\n{\bf Definition 3} For those $\lFrg$ that include a scaling transformation $s$, 
the relational space notion specializes to {\it shape-and-scale space} \cite{LR95, LR97, FileR, AMech, ABook, A-Monopoles} 
\be 
\FrR(d, N; \lFrg(s), \Gamma)  \:=  \Rel(d, N; \lFrg(s), \Gamma)  \m .
\ee
\n{\bf Definition 4} For those $\lFrg$ that do not include a scaling transformation, 
the relational space notion specializes to {\it shape space} \cite{Kendall84, Kendall, FileR, AMech, PE16, A-Monopoles} 
\be 
\FrS(d, N; \lFrg(\emptyset), \Gamma)  \:=  \Rel(d, N; \lFrg(\emptyset), \Gamma)   \m . 
\ee
\n{\bf Remark 4} Relational Theory is thus a portmanteau of Shape Theory and Shape-and-Scale Theory.  
The distinction of whether or not scaling is among the automorphisms is significant in practise because many of the hitherto most-studied models are part of a 
{\it shape space and shape-and-scale-space pair}.
This corresponds to Shape Theories which are distinct while remaining algebraically consistent upon removal of an overall dilation generator. 

\m 

\n{\bf Example 3}  More generally however there are plenty of instances of singletons, as we shall see below.  
As a first example of singleton theory, for $\mathbb{S}^1$ 
\be 
\lFrg  =  Isom(\mathbb{S}^1) 
       =  SO(2) 
	   =  U(1)  
	   =  \mathbb{S}^1                                    \m \mbox{ as a manifold}            
\ee
can be quotiented out of $\FrQ(\mathbb{S}^1, N)$, but dilations cannot be since the dilational operator is not consistent with the sphere's periodicity, 
by which on $\mathbb{S}^1$ the similarity Killing equation has no more solutions than the Killing equation.
Also note that 
\be 
{\cal R}(\mathbb{S}^1, N)  =  \lFrr(\mathbb{S}^1, N)                                                 \m , 
\ee 
so we already have the shape-and-scale space in this case.  

\m

\n{\bf Definition 5} Suppose that the carrier space axes $A$ is to join the carrier space origin $0$ in having no meaning.
I.e.\ we quotient $\FrQ(d, N)$ by the Euclidean group $Eucl(d)$, thus forming the 
\be
(\mbox{\it relational space}) \mma \m  {\cal R}(d, N)  \:=  \frac{\FrQ(d, N)}{Eucl(d)} 
                                                       \es  \frac{\lFrr(d, N)}{Rot(d)} 
	    			                                   \es  \frac{\mathbb{R}^{n \, d}}{SO(d)}         \m  .
\ee
This corresponds to {\it Metric Shape-and-Scale Theory}, and is of relevance to the well-known Absolute versus Relational Debate's relational side in its most traditional context.

\m 

\n{\bf Remark 5} Suppose that all three of absolute scale $S$, axes $A$ and origin $0$ have no meaning.
I.e.\ we quotient $\FrQ(d, N)$ by the similarity group $Sim(d)$, thus forming Kendall's \cite{Kendall84, Kendall} 
\be
(\mbox{\it shape space}) \mma    \m  \FrS(d, N) \:=  \frac{\FrQ(d, N)}{Sim(d)} 
                                                    \es  \frac{\lFrr(d, N)}{Rot(d) \times Dil} 
 			                                        \es  \frac{\FrP(d, N)}{Rot(d)} 
 			                                        \es  \frac{\mathbb{S}^{n \, d - 1}}{SO(d)}               \m  .
\ee
This corresponds to {\it Kendall's Similarity Shape Theory} \cite{Kendall84}.   

\m 

\n{\bf Remark 6} The similarity and Euclidean theories moreover constitute a first shape and shape-and-scale pair.  

\m 

\n{\bf Remark 7} As regards invariants, in 1-$d$, we have just ratios of separations: the combined restrictions of $Tr(d)$ and $Dil$ invariants. 
In $\geq 2$-d, these continue to be invariants because separations are rotational invariants as well, 
and are joined moreover by relative angle invariants. 

\m 

\n For $Eucl(d)$, the general invariants are of the form 
\be 
f(\urho_i \cdot \urho_j)                \m , 
\ee 
of which relative separations 
\be 
||\urho_i|| = \sqrt{\urho_i\cdot\urho_i}
\ee
and relative angles 
\be 
\theta_{ij} = \mbox{arccos}
\left( 
\frac{\rho_i \cdot \rho_j}{||\urho_i|| ||\urho_j||}
\right)
\label{rel-angles}
\ee 
are subcases.

\m 

\n For $Sim(d)$, the general invariants are of the form 
\be 
f
\left(
\frac{\urho_i \cdot \urho_j}{\urho_k \cdot \urho_l} 
\right)                                                       \m , 
\ee
of which ratios of relative separations
\be
\frac{||\urho_i||}{||\urho_k||}
\ee 
and relative angles (\ref{rel-angles}) are subcases. 

\m 

\n Triangles can be parametrized using 1 of each, though there is freedom to parametrize triangles in other ways.  

\m 

\n Quadrilaterals can be parametrized with 2 of each 
(the shape space topology and geometry of Secs 11 and 12 makes clear that this is a recurring pattern for polygons: $N - 2$ of each).
%
%FFFFFFFFFFFFFFFFFFFFFFFFFFFFFFFFFFFFFFFFFFFFFFFFFFFFFF  C O O R D I N A T E   S Y S T E M S   F O R   3   P A R T I C L E S  FFFFFffFFFFFFFFFFFFFFFFFFFFFFFFFFFFFFFFFFFFFFFFFFFFFFFFFFFFF
{            \begin{figure}[!ht]
\centering
\includegraphics[width=0.8\textwidth]{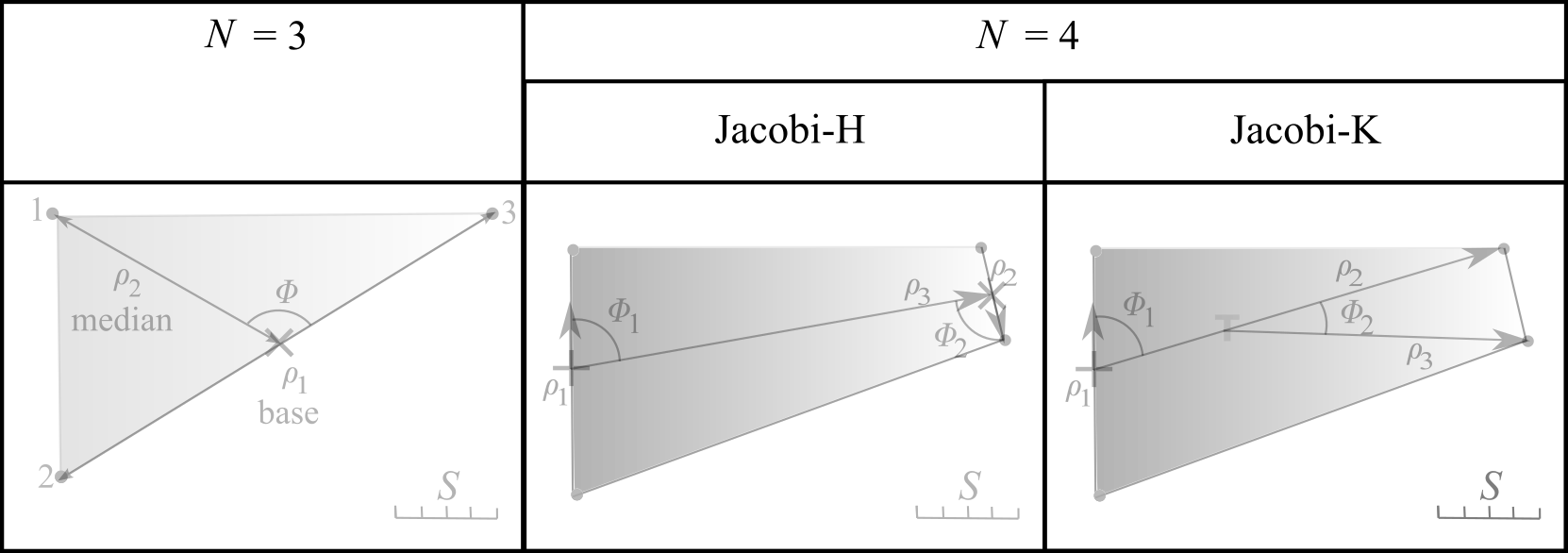}
\caption[Text der im Bilderverzeichnis auftaucht]{        \footnotesize{ Coordinates not depending on the absolute axes $A$ either: Jacobi magnitudes and the angles between them. 
To not depend on the scale, take the ratio $\rho_2/\rho_1$ of the Jacobi magnitudes alongside this angle. 
On the shape sphere, moreover, $\Phi$ plays the role of polar angle and the arctan of this ratio plays the role of azimuthal angle.
See e.g.\ \cite{QuadI} for further $\mathbb{CP}^2$ interpretation of the Jacobi H and K cases' relational coordinates. 
} }
\label{Relational-Coordinates-3}\end{figure}            }
%FFFFFFFFFFFFFFFFFFFFFFFFFFFFFFFFFFFFFFFFFFFFFFFFFFFFFFFFFFFFFFFFFFFFFFFFFFFFFFFFFFFFFFFFFFFFFFFFFFFFFFFFFFFFFFFFFFFFFFFFFFFFFFFFFFFFFFFFFFFFFFFFFFFFFFFFFFFFFFFFFFFFFFFFFFFFFFFFFFFFFFFFF

%=========================================================================================================================================================================================
\subsection{Configuration space dimension counting}\label{Dimension}
%=========================================================================================================================================================================================

\n{\bf Remark 1}
\be
\mbox{dim}(Rot(d))   =   \mbox{dim}(SO(d)) 
                    \es  \frac{d\{d - 1\}}{2}                                                                 \m .  
\ee
From this and (\ref{dim-Tr}),  
\be
\mbox{dim}(Eucl(d)) = \mbox{dim}(Tr(d) \rtimes Rot(d)) 
                    \es  d + \frac{d\{d - 1\}}{2} \es  \frac{d\{d +  1\}}{2}                                  \m  \mbox{ and }
\ee
\be
\mbox{dim}(Sim(d))   =   \mbox{dim}(Tr(d) \rtimes \{ Rot(d) \times Dil \}) 
                    \es  d + \frac{d\{d - 1\}}{2} + 1 
					\es  \frac{d\{d +  1\}}{2} + 1                                                            \m .  
\ee
\n{\bf Remark 2} The above and (\ref{A-B}) give preliminary na\"{\i}ve dimension counts of 
\be
\mbox{dim}(\FrS(d, N)) \es  \mbox{dim}\left(  \frac{\FrQ(d, N)}{Sim(d)}  \right) 
                       \es  \mbox{dim}(\mathbb{R}^{N \, d})  \, - \,  \mbox{dim}(Sim(d)) 
			           \es  N \, d - \frac{d ( d + 1 )}{2}    - 1 
		               \es  \frac{d ( 2 \, n + 1 - d )}{2} - 1                                                 \m , \m \mbox{ and }
\ee
\be
\mbox{dim}({\cal R}(d, N)) \es  \mbox{dim}\left(  \frac{\FrQ(d, N)}{Eucl(d)} \right) 
%                          \es  \frac{\lFrr(d, N)}{Rot(d)} 
						   \es  \mbox{dim}(\mathbb{R}^{N \, d}) - \mbox{dim}(Eucl(d)) 
					       \es   N \, d - \frac{d ( d + 1 ) }{2} 
						   \es  \frac{d ( 2 \, n + 1 - d ) }{2}                                                 \m  .						   
\ee
\n {\bf Remark 3} None of these dimension counts have a `simple product' form, unlike for $\FrQ(d, N)$ or $\lFrr(d, N)$.

\m

\n{\bf Remark 4} The most usual cases of these are  
\be
\mbox{dim}({\cal R}(1, N))  =  N - 1   
                            =  n               \m , 
\ee 
\be 
\mbox{dim}({\cal R}(2, N))  =  2 \, N - 3  
                            =  2 \n - 1    \m ,
\ee
\be 
\mbox{dim}({\cal R}(3, N))  =  3 \, N - 6  
                            =  3 \, n - 3  \m ,
\ee 
\be 
\mbox{dim}(\FrS(1, N))      =  N - 2   
                            =  n - 1           \m, 
\ee
\be
\mbox{dim}(\FrS(2, N))      =  2 \, N - 4  
                            =  2 \, n - 2   \m \mbox{ and }
\ee
\be
\mbox{dim}(\FrS(3, N))      =  3 \, N - 7  
                            =  3 \, n - 4   \m . 
\ee
\n{\bf Definition 1} {\it No full group action} is a triviality criterion linked to the Jacobi vectors not spanning the carrier space. 

\m 

\n{\bf Caveat 1} $\lFrg$ does not act fully on a) small enough particle number configurations or b) some nongeneric configurations; 
this is a caveat to the quotient dimension proposition (\ref{A-B}).  

\m 

\n{\bf Remark 5} The first correction to these formulae is to replace all negative dimensions by zero, 
giving the second kind of na\"{\i}ve dimension count as per Fig \ref{Shape-Triviality-Count}.  
%
%FFFFFFFFFFFFFFFFFFFFFFFFFFFFFFFFFFFFFFFFFFFFFFFFFFFFFFFFFFFFFFFFFFFFFFFFFFFFFFFFFFFFFFFFFFFFFFFFFFFFFFFFFFFFFFFFFFFFFFFFFFFFFFFFFFFFFFFFFFFFFFFFFFFFFFFFFFFFFFFFFFFFFFFFFFFFFFFFFFFFFFFFF
{            \begin{figure}[!ht]
\centering
\includegraphics[width=1.0\textwidth]{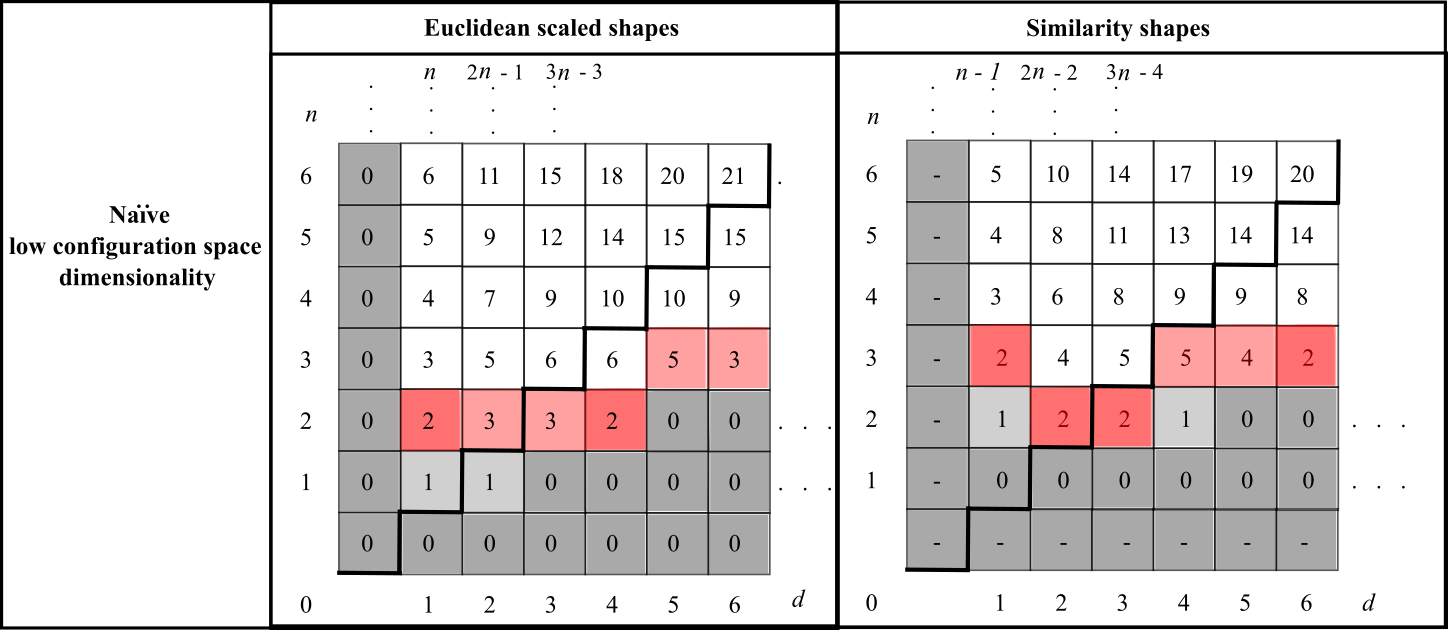}
\caption[Text der im Bilderverzeichnis auftaucht]{        \footnotesize{Tables of Euclidean scale-and-shapes and similarity shapes's na\"{\i}ve configuration space dimensions, 
presented in $(d, n)$ grid form.  
Triviality, relational triviality, relational minimality and other minimal relationally nontrivial units are marked thereupon in dark grey, pale grey, red and pink; 
these remain na\"{\i}ve allocations, subject to  the corrections made in Fig \ref{Shape-Triviality-Count-2}. 
We also mark a line of reflection symmetry in the dimensionality, albeit in this case this turns out to be an artifact of the na\"{\i}ve count.  
}} 
\label{Shape-Triviality-Count}\end{figure}            }
%FFFFFFFFFFFFFFFFFFFFFFFFFFFFFFFFFFFFFFFFFFFFFFFFFFFFFFFFFFFFFFFFFFFFFFFFFFFFFFFFFFFFFFFFFFFFFFFFFFFFFFFFFFFFFFFFFFFFFFFFFFFFFFFFFFFFFFFFFFFFFFFFFFFFFFFFFFFFFFFFFFFFFFFFFFFFFFFFFFFFFFFFF

\m

\n{\bf Remark 6} For 
\be 
d = n    \mma  \mbox{any basis of $\rho_i$ constitutes a basis of $\mathbb{R}^d$}  \m .  
\ee
This corresponds to a diagonal in the $(d, n)$ grid: a {\sl critical diagonal} separating qualitatively distinct linearly-dependent and non-spanning triangular half-grids. 

\m 

\n{\bf Remark 7} Were $n$ portrayed descending, this would be an infinite matrix diagonal separating upper and lower triangular entries.

\m 

\n{\bf Remark 8} We have however argued for $d$ to have the status of independent variable and $n$ of dependent variable, 
presented in the usual upper-right quadrant of the plane, so $n$ is portrayed ascending.  
In this presentation, one has a diagonal in the sense of a unit-gradient line through the $(d, n)$ grid's origin, 
with the upper and lower triangle statuses reversed from the previous Remark's. 

\m 
  
\n{\bf Remark 9} The basis diagonal has also been termed {\sl Casson diagonal} \cite{Kendall}, after topologist Andrew Casson who first pointed out its significance.

\m 

\n In the upper-right-quadrant presentation, the terminology `sub-Casson' and `super-Casson' for the lower and upper triangle regions respectively makes sense. 
This nomenclature has the added benefit that the (sub-Casson, Casson, super-Casson) progression is aligned with sources of increasing technical difficulty 
for the corresponding $N$-Body Problems. 

\m 

\n{\bf Definition 2} We introduce the Linear Algebra motivated term {\it basislands} $\FrB(d)$ as the conceptual name for the infinite series of models along this diagonal 
({\it Cassonlands} would be an alternative name).  
We use the notation $\FrR(\FrB(d))$ and $\FrS(\FrB(d))$ so as to distinguish between the scaled and pure-shape basislands respectively.  

\m 

\n This diagonal splits the lower wedge of {\it nonspanninglands} from the upper wedge of (linearly-){\it dependentlands}.
We term the first parallel of the diagonal in each of these wedges {\it minimal}.  
({\it Sub- and super-Cassonlands} would be alternative names.)

\m 

\n{\bf Remark 10} The first basislands are $(d, N) = (1, 2), (2, 3)$ and $(3, 4)$ i.e.\ the spaces of 1-$d$ intervals, 2-$d$ triangles and 3-$d$ tetrahaedrons respectively, 
the last of which places the 3-$d$ 4-Body Problem in this category. 
\be 
\mbox{ The general basisland is the space of $d$-simplices in dimension \m $d$}  \m .
\ee 
{\it Simplexlands} are thus a third name for these spaces.  

\m 

\n{\bf Remark 11} The corresponding minimal nonspanninglands are (1, 1), (2, 2) and (3, 3), so in particular the 3-$d$ 3-Body Problem lies in this category. 

\m 

\n{\bf Remark 12} The corresponding minimal dependantlands are (1, 3), (2, 4) and (3, 5), placing the quadrilaterals and the 3-$d$ 5-Body Problem in this class. 

\m

\n{\bf Remark 13} The basislands' dimensions are given by 
\be
\mbox{dim}(\FrS(N - 1, N)) + 1               =  
\mbox{dim}(\FrR(N - 1, N))                  \es 
\frac{1}{2}\,(N - 1)(2 \, N - (N - 1) - 1)  \es  \frac{N(N - 1)}{2} 
                                            \es  \frac{n(n + 1)}{2} 
										    \es  \frac{d(d + 1)}{2}    \m : 
\ee
the {\it triangular numbers}. 

\m 

\n{\bf Proposition 1} The corrected dimension count throughout the $(d, n)$ grid is given by 
\be 
\mbox{dim}(\FrS(d, n)) + 1   =   \mbox{dim}(\FrR(d, n))  
                            \es  \left\{ \s{        \mbox{    $\frac{  \mbox{$d$(2\, $n$ - $d$ + 1)}  }{  \mbox{2}  }     \m \m  n \geq d  $}      }
							               {        \frac{d(d - 1)}{2}                                                     \m \m  n   <  d        }         \right.   \m .  
\ee 
%
%FFFFFFFFFFFFFFFFFFFFFFFFFFFFFFFFFFFFFFFFFFFFFFFFFFFFFFFFFFFFFFFFFFFFFFFFFFFFFFFFFFFFFFFFFFFFFFFFFFFFFFFFFFFFFFFFFFFFFFFFFFFFFFFFFFFFFFFFFFFFFFFFFFFFFFFFFFFFFFFFFFFFFFFFFFFFFFFFFFFFFFFFF
{            \begin{figure}[!ht]
\centering
\includegraphics[width=1.0\textwidth]{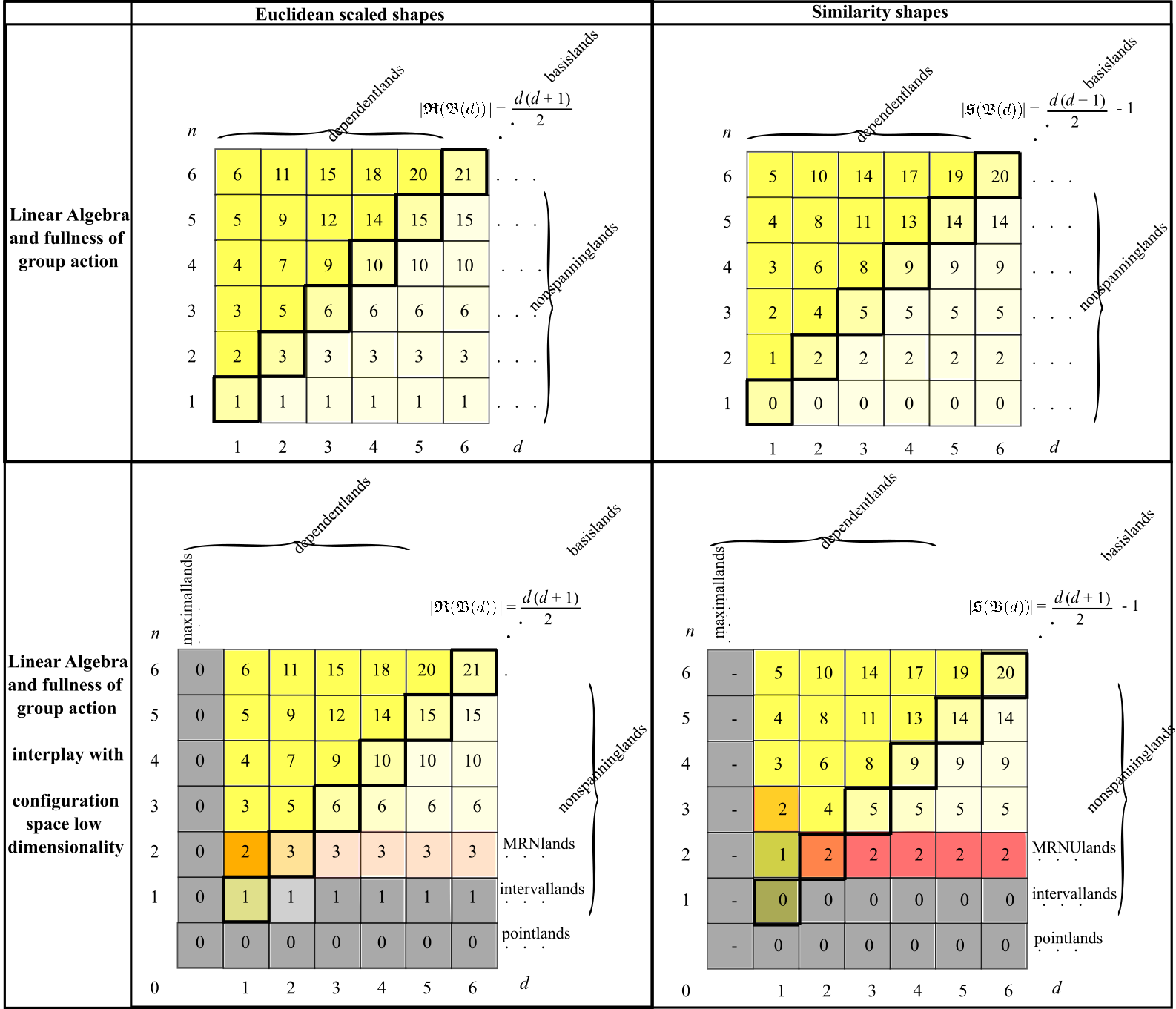}
\caption[Text der im Bilderverzeichnis auftaucht]{        \footnotesize{Overview of Euclidean scale-and-shapes and similarity shapes's configuration space dimensions, 
now taking into account partially acting groups.  
In the first row, distinction is made between the lower triangle of non-spanning and upper triangle of linear dependence as separated by the basis diagonal, 
all with reference to the number indepedent Jacobi vectors possessed by the relative space configurations.  
These are marked in ivory, butter and yellow respectively.  
The second row marks interplay between this Linear Algebra and low $k$.   
Orange denotes minimally relational and dependent, 
pale orange is minimally relational and a basis,
ochre       is relationally trivial and dependent, and 
cork        is relationally trivial and a basis. 
These colouring schemes are superposed when both apply, so e.g.\ the minimal relationally nontrivial units end up being various shades of purple.  
$d = 1$ is exceptional in that the first relationally nontrivial shape theory -- $n = 3$ i.e.\ $N = 4$ -- is {\sl two} entries up from the basis diagonal.   
Vertical, horizontal and diagonal series are labelled with vertically, horizontally and diagonally oriented script respectively.  
The MRNUland row is only valid as far left as the red and orange colouring indicates: $d \geq 2$;  
the MRNland row likewise in red pastel and orange, to $d \geq 1$, $d = 1$ being exceptionally furthermore a MRNU.
}} 
\label{Shape-Triviality-Count-2}\end{figure}            }
%FFFFFFFFFFFFFFFFFFFFFFFFFFFFFFFFFFFFFFFFFFFFFFFFFFFFFFFFFFFFFFFFFFFFFFFFFFFFFFFFFFFFFFFFFFFFFFFFFFFFFFFFFFFFFFFFFFFFFFFFFFFFFFFFFFFFFFFFFFFFFFFFFFFFFFFFFFFFFFFFFFFFFFFFFFFFFFFFFFFFFFFFF

\n{\bf Remark 14} Using the first of these counts within its domain of validity, for scaled models the trivial case gives 
\be 
d(2 \, N - d - 1) = 0  \m \Rightarrow \m   d = 0 \m \mbox{ or } \m  N = \frac{d + 1}{2} \m ; 
\ee 
the smallest cases of the latter are $(d, N) = (1, 1), (2, 3)$... 

\m 

\n The relationally trivial case returns  
\be 
d(2 \, N - d - 1) = 2  \m \Rightarrow \m  d = 1 \m \mbox{ and } \m  2 \, N - d - 1 = 2 \m \Rightarrow \m N = 2 
                       \mbox{ or }    \m  d = 2 \m \mbox{ and } \m  2 \, N - d - 1 = 1 \m \Rightarrow \m N = 2  \mbox{ } ;  
\ee 
the smallest cases of the latter are $(d, N) = (1, 2)$, (2, 2)...  
\be 
d ( 2 \, N - d - 1 ) = 4 \m  \Rightarrow \m    d = 1  \m \mbox{ and } \m  2 \, N - d - 1 = 4 \mma  \Rightarrow \m  N = 3 
\ee 
so $(d, N) = (1, 3)$ or cases we discard for being non-integer solutions.  

\m 

\n For pure-shape models, the trivial case gives 
\be 
d(2 \, N - d - 1) = 2 \m , 
\ee 
so $(d, N) = (1, 2)$ and (2, 3) arise by the above argument.

\m 

\n In the relationally trivial case,  
\be 
d(2 \, N - d - 1) = 4 \m , 
\ee 
so $(d, N) = (1, 3)$ ensues.  

\m 

\n Finally, in the minimal relationally nontrivial case 
\be 
d(2 \, N - d - 1) = 6 \m , 
\ee
giving $(d, N) = (1, 4)$, $(2, 3)$ and $(3, 3)$ as solutions in integers. 

\m 

\n{\bf Remark 15} The count conversion
\be 
\mbox{dim}(\FrR(d, N)) = \mbox{dim}(\FrS(d, N)) + 1
\ee 
includes 0 = -- 1 + 1 in passage from $\emptyset$ of undefined dimension to $\{\mbox{pt}\}$ of dimension.

%==================================================================================================================================================================
%==================================================================================================================================================================
\section{Flat Geometry study of configurations}\label{Geom}
%==================================================================================================================================================================
%==================================================================================================================================================================

\n{\bf Remark 1} We need 
\be
N \geqs 2 \m \mbox{ to have a bona fide side/path:       $\mP_2$ graph } , 
\ee
\be
N \geqs 3 \m \mbox{ to have a bona fide perimeter/cycle: $\mC_3$ graph } ,  
\ee
\n{\bf Remark 2} 2-$d$ possesses a figure area form, 3-$d$ a figure volume form and so on. 
\n In 2-$d$, area turns out to be a useful variable for triangles.

\m 

\n{\bf Remark 3}
\be 
\mbox{For $N \geq 3$, a meaningful notion of collinear configurations is exhibited by \m $d \geq 2$ \m figures.} 
\ee  
\n{\bf Remark 4} Symmetry considerations return equilateral (order 6), isosceles and uniform-collinear (both order 2) and generic classes for $N = 3$.

\m 

\n $N = 4$ has rather more nontrivial symmetry classes: square (order 8), rectangle and rhombus (order 4), uniform-collinear, parallelogram, kite and isosceles trapezium (all order 2).  
%
%FFFFFFFFFFFFFFFFFFFFFFFFFFFFFFFFFFFFFFFFFFFFFFFFFFFFFF  C O O R D I N A T E   S Y S T E M S   F O R   3   P A R T I C L E S  FFFFFffFFFFFFFFFFFFFFFFFFFFFFFFFFFFFFFFFFFFFFFFFFFFFFFFFFFFF
{            \begin{figure}[!ht]
\centering
\includegraphics[width=0.75\textwidth]{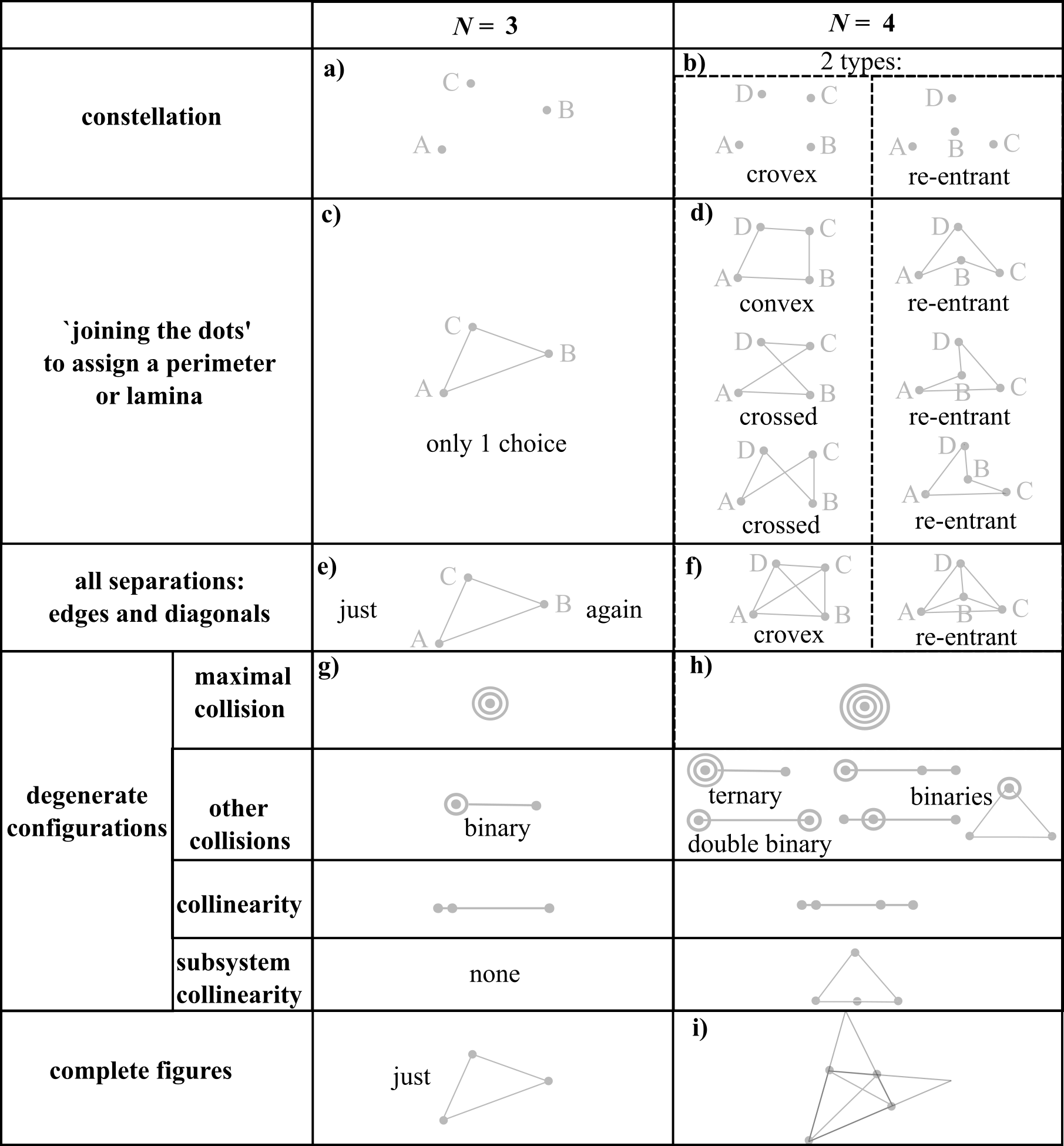}
\caption[Text der im Bilderverzeichnis auftaucht]{        \footnotesize{For each of $(d, N) = (2, 3)$ and (2, 4) in turn, we depict 
nondegenerate constellations a)-b), 
`joining the dots' to make a perimeter or lamina in c)-d), 
inclusion of all separations -- diagonals as well as edges -- in e)-f), 
and demonstration of the variety of types of degenerate configurations in g)-h). 
The onset of a complete figure is for (2, 4) as per i).  
} }
\label{Join-the-Dots}\end{figure}            }
%FFFFFFFFFFFFFFFFFFFFFFFFFFFFFFFFFFFFFFFFFFFFFFFFFFFFFFFFFFFFFFFFFFFFFFFFFFFFFFFFFFFFFFFFFFFFFFFFFFFFFFFFFFFFFFFFFFFFFFFFFFFFFFFFFFFFFFFFFFFFFFFFFFFFFFFFFFFFFFFFFFFFFFFFFFFFFFFFFFFFFFFFF

\m 

\n{\bf Remark 5} Quadrilaterals have more diversity of congruence conditions \cite{H15}, and in symmetries both realized in space and on the sets of separations and angles \cite{Forth}.  
Some key qualitative properties first manifest for quadrilaterals are listed in the subsequent Remarks. 

\m 

\n{\bf Remark 6}
\be 
N \geqs 4 \m \mbox{ for the perimeter to be distinct from the totality of relative separations} \m .  
\label{edge-diagonal}
\ee
i.e.\ quadrilaterals are the smallest figures to possess {\it diagonals} as well as edges (Fig \ref{Join-the-Dots}.f).   

\m

\n{\bf Remark 7} $N = 4$ is also minimal for the polygon to complete polygon distinction (Fig \ref{Join-the-Dots}.i).  

\m 

\n{\bf Remark 8} Coolidge's formula \cite{Coolidge} for the area of a quadrilateral in terms of separation data generalizes Heron's formula for the area of a triangle 
though involving diagonals as well as edges.  
Diagonals also feature in Ptolemy and Euler's results \cite{CG67} for quadrilaterals;  
the latter also involve the Newton interval Jacobi H-coordinate indicated in Fig 2.c). 

\m 

\n{\bf Remark 9} Not all levels of modeling's configurations are uniquely specified by constellations.  
For instance, paths only uniquely determine a figure in 1-$d$, whereas in 2-$d$, perimeter is only a unique specification for $N = 3$'s triangles.  
Also, whereas specifying three side lengths fully determines a triangle (if no length exceeds the sum of the other two),
giving four side lengths at most only specifies a continuum family of quadrilaterals.

\m 

\n{\bf Remark 10} A consequence of (\ref{edge-diagonal}) is that (Fig \ref{Join-the-Dots}.d).  
\be 
\mbox{$N = 4$ \m is minimal for the re-entrant, convex and crossed figure distinction} \m .
\ee 
\n {\it Sylvester's problem} \cite{Sylvester, Pfiefer} of what is Prob(convex) for a quadrilateral 
is thus but the $N$-minimal case of a question that can be posed of any subsequent $N$-a-gon as well.  

\m 

\n{\it Carroll's problem} \cite{Pillow, III} of what is Prob(obtuse) for a triangle is an analogue of this supported for $N = 3$. 
This analogy is via obtuseness being of generic measure among triangles and convexity being of generic measure among quadrilaterals. 
At the level of shape spaces, this means that these quantities are regions of the shape space of the same dimensionality as the shape space.  

\m 

\n At the level of the constellations itself, a re-entrant versus `crovex' (crossed or convex) remains.  

\m  

\n{\bf Remark 11} 
\be
\mbox{$N = 4$ \m is minimal for non-collinear configurations with collinear subsystems} \m .  
\ee 
These are moreover intermediate configurations between the crovex and re-entrant ones.

\m 

\n{\bf Remark 12} 
\be 
\mbox{$N = 4$ \m is minimal to have multiple types of partial (non-maximal) coincidences-or-collisions} \m .  
\ee
\n{\bf Remark 13} 
\be
\mbox{$(d, N) = (2, 4)$ \m is minimal to have non-collinear coincidences-or-collisions} \m . 
\ee 
\n{\bf Remark 14}
\be 
\mbox{(2, 4) \m is minimal as regards exhibiting parallelism which is not just collinearity}  \m .
\ee
\n{\bf Remark 15} 
\be 
\mbox{$N = 3$ \m exihibits total rigidity of centre of mass positions, as per Fig \ref{Rigid}.a, b)}  \m . 
\ee
\be 
\mbox{$N = 4$ \m is minimal for some centre of mass position flexibility, as per Fig \ref{Rigid}.c, d) }  \m . 
\ee
%
%FFFFFFFFFFFFFFFFFFFFFFFFFFFFFFFFFFFFFFFFFFFFFFFFFFFFFFFFFFFFFFFFFFFFFFFFFFFFFFFFFFFFFFFFFFFFFFFFFFFFFFFFFFFFFFFFFFFFFFFFFFFFFFFFFFFFFFFFFFFFFFFFFFFFFFFFFFFFFFFFFFFFFFFFFFFFFFFFFFFFFFFFF
{\begin{figure}[ht]
\centering
\includegraphics[width=0.75\textwidth]{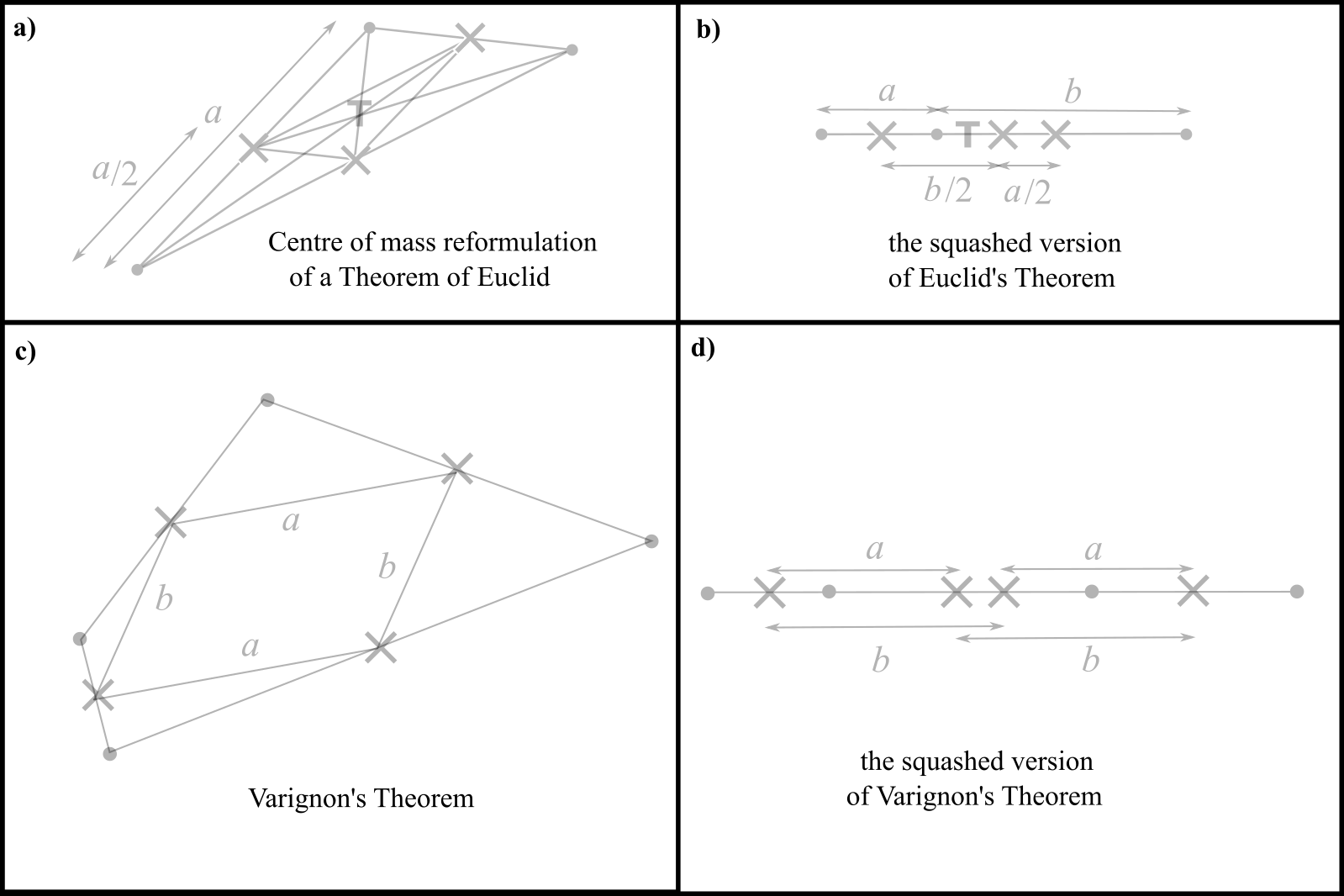}
\caption[Text der im Bilderverzeichnis auftaucht]{\footnotesize{Loss of cluster hierarchy rigidity.  
At the geometrical level, in 2-$d$ the $N = 3$ case is recognizable as the Median Concurrence Theorem (dating back to Euclid), 
and the $N = 4$ case is Varignon's Theorem \cite{Varignon, CG67} that the pairwise centres of mass form a parallelogram.
This result moreover admits both arbitrary-$N$ and arbitrary-$d$ generalizations.}} 
\label{Rigid}\end{figure} } 
%FFFFFFFFFFFFFFFFFFFFFFFFFFFFFFFFFFFFFFFFFFFFFFFFFFFFFFFFFFFFFFFFFFFFFFFFFFFFFFFFFFFFFFFFFFFFFFFFFFFFFFFFFFFFFFFFFFFFFFFFFFFFFFFFFFFFFFFFFFFFFFFFFFFFFFFFFFFFFFFFFFFFFFFFFFFFFFFFFFFFFFFFFF

\m 
 
\n{\bf Remark 16} Much as the Jacobi H and K both enter quadrilateral study, we expect all of the Jacobi coordinate systems in rows 3 and 4 of Fig \ref{Jacobi-Trees-3-to-6}  
to enter the detailed study of pentagons and hexagons respectively.

\m 

\n{\bf Remark 17} $N$-a-gon theory complexity does keep on increasing with $N$, 
for all that $N = 3$ is the first first nontrivial polygon and pasage to $N = 4$ is the first big jump in complexity. 
\be 
\mbox{Hexagons -- (2, 6) -- are minimal for many projective results} \m : 
\ee 
Desargues', Pappus' and Pascal's Theorems \cite{HCV}; see also \cite{CG67} for hexagonal configurations more generally.

\m 

\n{\bf Remark 18} Tetrahaedrons -- $(3, 4)$ -- are minimal to have nonplanar figures. 
These have the general della Francesca--Tartaglia formula for volume in terms of separation data.
Tetrahaedrons admit symmetry groups of order 24 (regular), 8, 6, 4, 2 and 1.   

\m 

\n Nonplanar tetrahaedrons however have no diagonals and are all convex. 

\m 

\n{\bf Remark 19} We thus need $(3, 5)$ to have diagonals -- separations that are not edges -- nonplanar non-convex figures, and coplanar subsystems that are not just collinear.

\m 

\n{\bf Remark 20} In fact, the whole Casson line $(d, d + 1)$ is minimal to have $d$-dimensional figures. 
These admit Cayley--Menger formulae for $d$-volume in terms of separation data, of which Heron and della Francesca--Tartaglia are the first nontrivial two.  

\m 

\n{\bf Remark 21} Minimal dependentlands $(d, d + 2)$ are required to have $d$-dimensional non-convex figures and thus a generalization of Sylvester's Problem, 
as well as nontrivially $(d - 1)$-dimensional subsystems.  

\m 

\n{\bf Remark 22} On the other hand, the basislands $(d, d + 1)$ admit a generalization of Carroll's problem along the lines of the size of the maximal angle present.

%==================================================================================================================================================================
%==================================================================================================================================================================
\section{Topological shape spaces}\label{TSS}
%==================================================================================================================================================================
%==================================================================================================================================================================

\n{\bf Structure 1} Topological adjacency graphs parallel submanifold structure gluing. 
\n In any one given step, one can carry out as many fissions as one pleases, or as many fusions as one pleases, but not a mixture of both.  
This step specification is moreover topologically well-defined.    

\m 

\n{\bf Structure 2} Topological shape spaces are more than the partitions due to their edge structure.
This is moreover in excess of the edge structure in Partition Theory's natural lattice of partition refinements, $\lattice_{\sP}$ \cite{Top-Shapes-2}.  
What $\lattice_{\sP}$ encodes, rather, is the the dimensional descent ladder 1 dimension at a time.
The two are already clearly not the same for $N = 3$ since corners can be adjacent to faces as well as to edges.

\m
 
\n{\bf Structure 3} We are to next comment on which $N$ are minimal as regards the following graph-theoretic nontrivialities; see Fig \ref{Graph-MN}for the graphs in question.  

\m 

\n{\bf Criterion 0} The empty graph -- no edges or vertices -- is graph-theoretically trivial. 

\m

\n{\bf Criterion 1} The disconnected graphs $\mD_n$ are graph-theoretically trivial. 
This is because these have no edge structure, so the definition of a graph in practise collapses to just the definition of a point set. 

\m

\n{\bf Criterion 2} Paths -- $\mP_n$ graphs -- are graph-theoretically trivial. 
This is because ordered point sets suffice to describe these.  

\m 

\n{\bf Criterion 3} Cycles -- $\mC_n$ graphs -- are graph-theoretically trivial. 
This is because a cyclic `joining of the dots' order, alias an order modulo periodicity and choice of starting point, suffices to describe these. 

\m 

\n{\bf Criterion 4} Disjoint unions $\coprod$ of trivial graphs are themselves trivial. 
This is because they can be treated component by component, with each component itself requiring less mathematics than Graph Theory to treat. 

\m  

\n{\bf Criterion 5} A graph $\mH$ is trivial if its complement $\overline{\mH}$ is; 
this is complementation in the sense that all edges become non-edges and vice versa.  

\m 

\n{\bf Remark 1} Topological relational spaces are cones, mostly double cones in fact, with some triple cones as well.
Such graphs are further studied in the recent article \cite{Top-Shapes-2}. 
Stars, fans and wheels -- as per Fig \ref{Graph-MN} -- are simple examples of cones.  

\m 

\n{\bf Criterion 6} We moreover regard cones as graph-theoretically trivial.  
%
%FFFFFFFFFFFFFFFFFFFFFFFFFFFFFFFFFFFFFFFFFFFFFFFFFFFFFFFFFFFFFFFFFFFFFFFFFFFFFFFFFFFFFFFFFFFFFFFFFFFFFFFFFFFFFFFFFFFFFFFFFFFFFFFFFFFFFFFFFFFFFFFFFFFFFFFFFFFFFFFFFFFFFFFFFFFFFFFFFFFFFFFFF
{            \begin{figure}[!ht]
\centering
\includegraphics[width=0.68\textwidth]{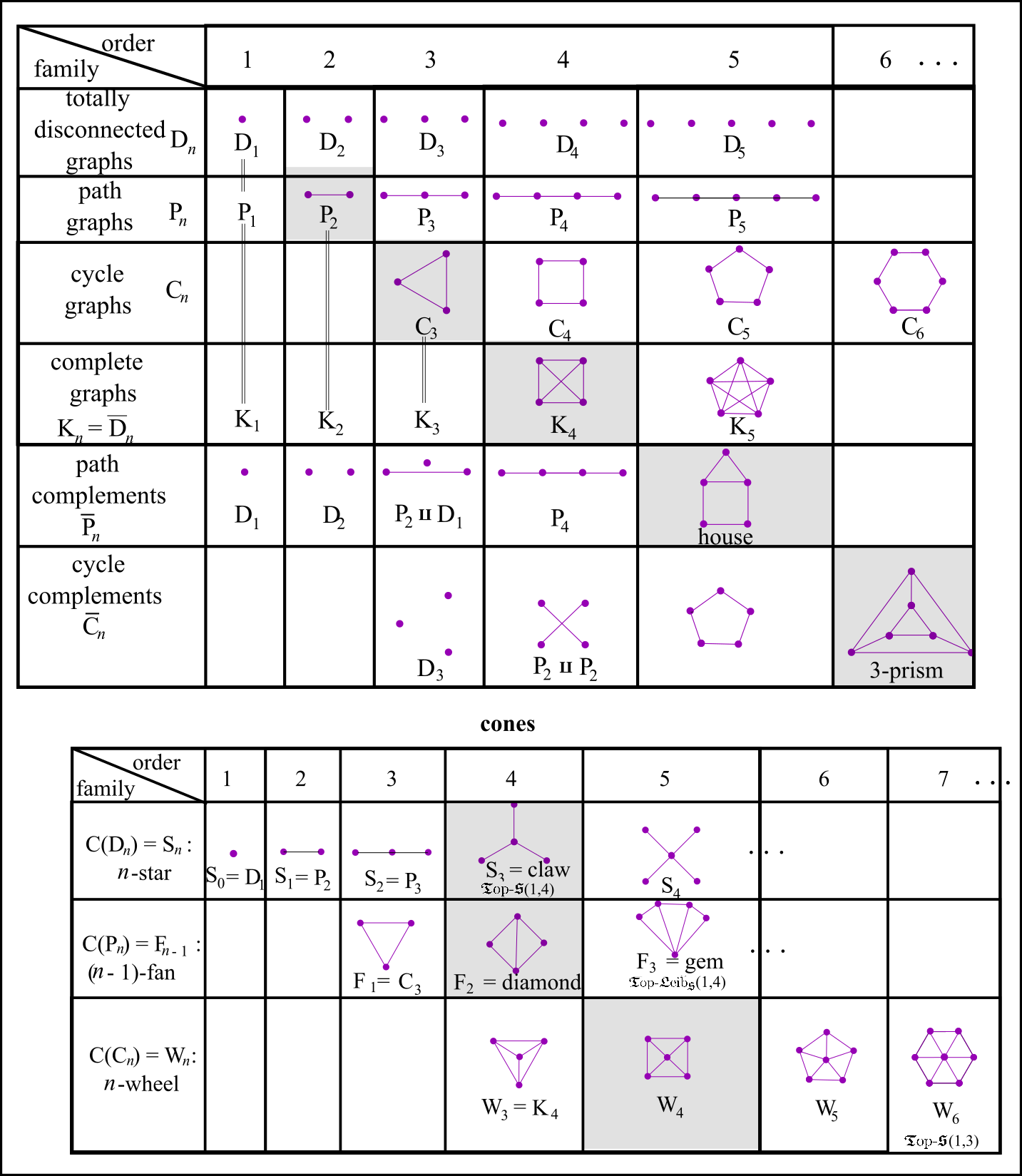}
\caption[Text der im Bilderverzeichnis auftaucht]{        \footnotesize{Totally disconnected, path, cycle and complete graphs, their complements (where distinct) and their cones.
Grey shading denotes first distinctive members. 
} }
\label{Graph-MN} \end{figure}          }
%FFFFFFFFFFFFFFFFFFFFFFFFFFFFFFFFFFFFFFFFFFFFFFFFFFFFFFFFFFFFFFFFFFFFFFFFFFFFFFFFFFFFFFFFFFFFFFFFFFFFFFFFFFFFFFFFFFFFFFFFFFFFFFFFFFFFFFFFFFFFFFFFFFFFFFFFFFFFFFFFFFFFFFFFFFFFFFFFFFFFFFFFF

\m 

\n{\bf Definition 1} $\mG_{\sD}$, the  {\it deconing} of $\mG$ is the graph obtained by sequentially removing all cone points present.  

\m 

\n{\bf Definition 2} Let us define the {\it residue} of a graph as the end-point of deconing it. 
For rubber Leibniz space, the complement graphs of the residues are simpler than the graphs themselves.  
Fig \ref{Top-2-Front} contains the first few examples of Shape-Theoretic Leibniz graphs' resides.

\m 

\n{\bf Criterion 7} Planarity \cite{Graphs-1, Graphs-2} is another triviality criteria for graphs.  

\m 

\n{\bf Criterion 8)} Planarity-or-coplanarity (planarity of the complement). 

\m 

\n{\bf Criterion 9)} Planarity {\sl and} coplanarity. 

\m

\n{\bf Remark 2} We moreover apply criteria 7, 8 and 9 to rubber Leibniz space's residue graphs.  

\m

\n{\bf Criterion 10} That the graph is `{\it modular}':  a finite sequence of $\mK_p$ strung out in a line with further edges only between adjacent $\mK_p$ 
along this line.  
See the next version of \cite{Top-Shapes-2} for a Linear Algebra characterization of this.  

\m

\n{\bf Remark 3} (1, 3)'s configuration spaces are all trivial; even the intervening lattice of shape and relational spaces are all trivial \cite{Top-Shapes}.  

\m 

\n{\bf Remark 4} $\FrS(\FrC^{\geq 2}, 3)$ is the claw graph, though this is the cone over $\mD_3$ and thus trivial according to criterion 5.   

\m

\n{\bf Remark 5} (1, 4) has considerably nontrivial $\Top\mbox{-}\FrS(1, 4)$, though its rubber Leibniz space is still trivial by criterion 5 due to being the cone over a path.

\m 

\n{\bf Remark 6}  
\be
\mbox{$N = 5$ \m is minimal to have a nontrivial rubber Leibniz space for $\mathbb{R}$}, 
\ee 
and 
\be
\mbox{$N \geq 6$ \m for $\mathbb{S}^1$ \m and \m $\FrC^{d \geq 2}$}  \m .  
\ee 
\n{\bf Remark 7} 
\be 
\mbox{ $N = 6$ \m is minimal in 1-$d$ to have nonplanarity and noncoplanarity as well as modularity}   \m , 
\ee
collectively `non-modplanarity', and
\be 
\mbox{$N \geq 8$ \m for $\mathbb{S}^1$ \m and \m $\FrC^{d \geq 2}$}    \m .  
\ee 
\n See Fig \ref{Top-2-Front} and \cite{Top-Shapes-2} for further details, and \ref{Shape-Triviality-Count-4} for the updated $(d, n)$ grid.
%
%FFFFFFFFFFFFFFFFFFFFFFFFFFFFFFFFFFFFFFFFFFFFFFFFFFFFFFFFFFFFFFFFFFFFFFFFFFFFFFFFFFFFFFFFFFFFFFFFFFFFFFFFFFFFFFFFFFFFFFFFFFFFFFFFFFFFFFFFFFFFFFFFFFFFFFFFFFFFFFFFFFFFFFFFFFFFFFFFFFFFFFFFF
{            \begin{figure}[!ht]
\centering
\includegraphics[width=0.7\textwidth]{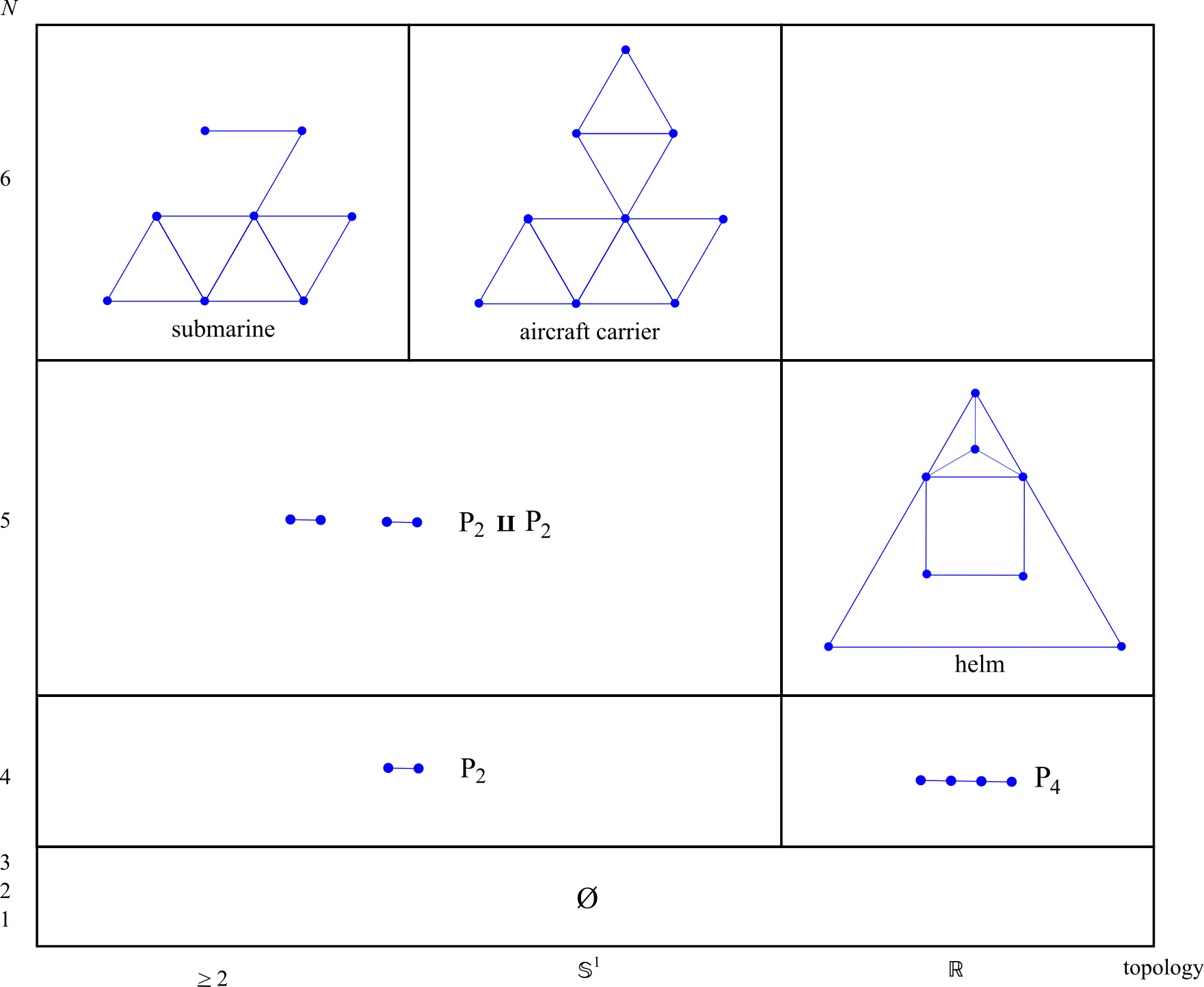}
\caption[Text der im Bilderverzeichnis auftaucht]{        \footnotesize{Rubber Leibniz space graph residue complements. 
Note that in the case of rubber, three topologies takes the place of $\mathbb{N}_0$ of dimensions.
} }
\label{Top-2-Front} \end{figure}          }
%FFFFFFFFFFFFFFFFFFFFFFFFFFFFFFFFFFFFFFFFFFFFFFFFFFFFFFFFFFFFFFFFFFFFFFFFFFFFFFFFFFFFFFFFFFFFFFFFFFFFFFFFFFFFFFFFFFFFFFFFFFFFFFFFFFFFFFFFFFFFFFFFFFFFFFFFFFFFFFFFFFFFFFFFFFFFFFFFFFFFFFFFF
%
%FFFFFFFFFFFFFFFFFFFFFFFFFFFFFFFFFFFFFFFFFFFFFFFFFFFFFFFFFFFFFFFFFFFFFFFFFFFFFFFFFFFFFFFFFFFFFFFFFFFFFFFFFFFFFFFFFFFFFFFFFFFFFFFFFFFFFFFFFFFFFFFFFFFFFFFFFFFFFFFFFFFFFFFFFFFFFFFFFFFFFFFFF
{            \begin{figure}[!ht]
\centering
\includegraphics[width=1.0\textwidth]{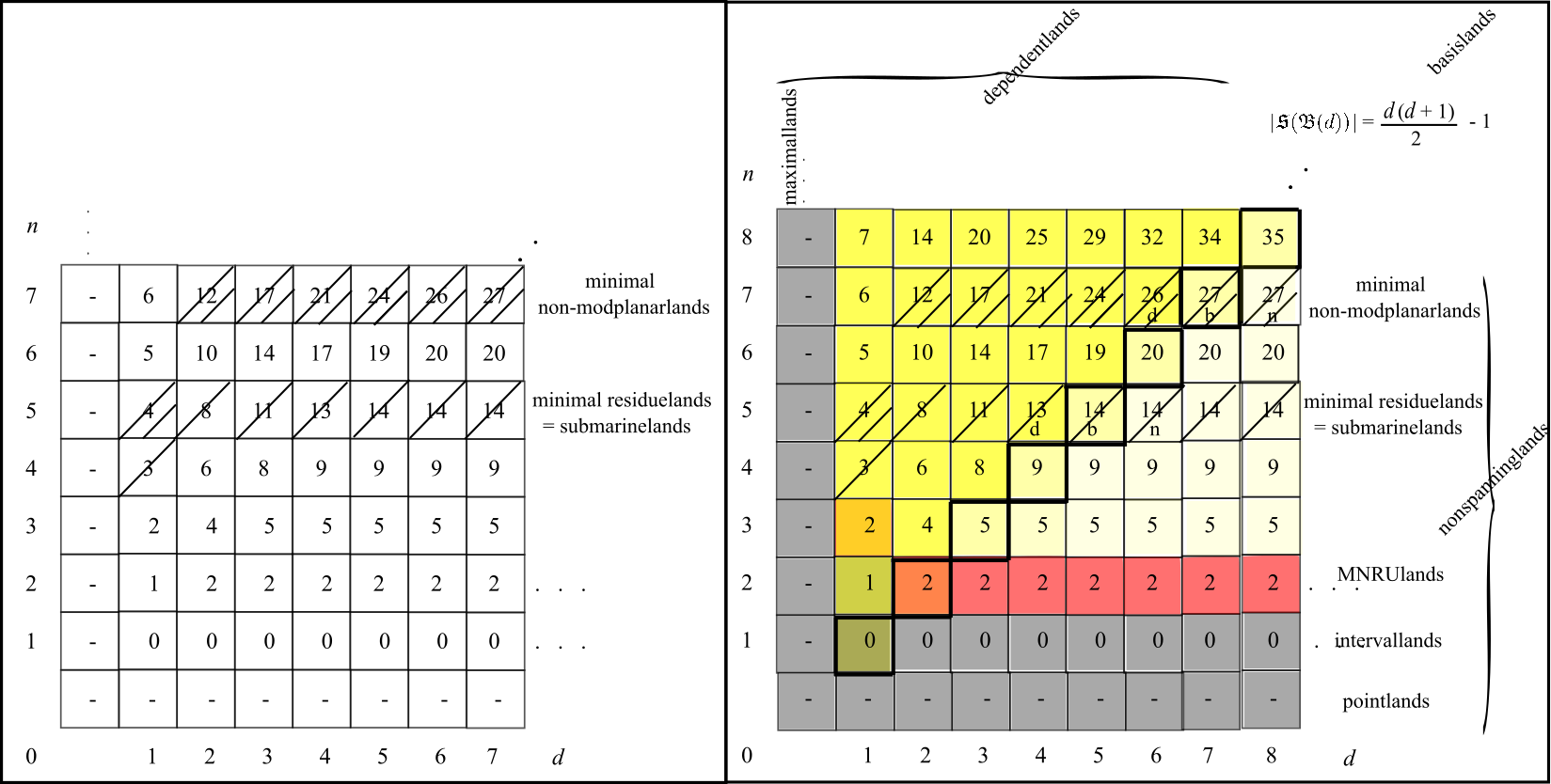}
\caption[Text der im Bilderverzeichnis auftaucht]{        \footnotesize{The current article's rubber nontrivialities and their interplay with triviality counts and Linear Algebra.
The first requires expanding our previous picture to have the $n = 7$ row, whereas the second requires a $d = 8$ column to exhibit the first non-modplanar nonspanningland.
The marked letters are `d' for the minimal dependentland, `b' for basisland and `n' for minimal nonspanningland cases to exhibit residue nontriviality and non-modplanarity.   
}} 
\label{Shape-Triviality-Count-4}\end{figure}            }
%FFFFFFFFFFFFFFFFFFFFFFFFFFFFFFFFFFFFFFFFFFFFFFFFFFFFFFFFFFFFFFFFFFFFFFFFFFFFFFFFFFFFFFFFFFFFFFFFFFFFFFFFFFFFFFFFFFFFFFFFFFFFFFFFFFFFFFFFFFFFFFFFFFFFFFFFFFFFFFFFFFFFFFFFFFFFFFFFFFFFFFFFF

\m 

\n{\bf Remark 8} For the distinguishably labelled mirror images distinct case, $\Top\mbox{-}\FrS(1, 3)$ is the 6-wheel graph with generic G at the centre and B's round the rim, 
$\Top\mbox{-}\FrS(1, 4)$ is a 74-vertex decoration of the cubic net \cite{II} and $\Top\mbox{-}\FrS(2, 3)$ is the claw graph with generic G at the centre and B's on each talon. 
The first and third of these are marked on Fig \ref{Graph-MN}.

%==================================================================================================================================================================
%==================================================================================================================================================================
\section{Relational space topology}\label{RST}
%==================================================================================================================================================================
%==================================================================================================================================================================

\n{\bf Structure 1} Since 
\be 
\FrS(0, N)  =  \FrP(0, N)                 \m \mbox{ and } 
\ee
\be 
\FrS(1, N)  =  \FrP(1, N)
\ee
due to absense of continuous rotations in 0- and 1-$d$, the first two columns -- maximallands and metrolands -- cases' topological-level structures have already been covered: 
\be 
\FrS(0, N)  =  \emptyset    \m \mbox{ and } 
\ee 
\be 
\FrS(1, N)  =  \mathbb{S}^{n - 1}
\ee 
\n The first two rows -- pointlands and intervallands -- topological level structures have also been covered: 
\be
\FrS(d, 1)  =  \FrP(d, 1) = \emptyset                       \m , 
\ee 
\be
\FrS(d, 2)  =  \{\mbox{pt}\} \m \mbox{ for } \m   d \geq 2  \m , 
\ee
with the exception that 
\be
\FrS(1, 2)  =  C_2                                          \m . 
\ee
In Fig \ref{Shape-Triviality-Top-5}.a), these rows and columns are marked in grey, bar the less trivial metrolands, which are accorded cyan. 

\m 
 
\n{\bf Structure 2} The $N$-a-gonland shape spaces also constitute a series at the topological level \cite{Smale70, Kendall84}: 
\be 
\FrS(2, N) = \mathbb{CP}^{n - 1}                            \m . 
\ee 
The column these form is indicated in green in Fig \ref{Shape-Triviality-Top-5}.a). 

\m 

\n{\bf Structure 3} The basislands constitute a third, now diagonal, nontrivial topological-level series: 
\be 
\FrB(\FrS(d))  =  \FrS(d, d + 1) 
               =  \mathbb{S}^q
\label{Basislands-Top}
\ee
for 
\be
q  \es  \frac{d(d + 1)}{2} - 1 
   \es  \frac{(d - 1)(d + 2)}{2}  
   \es  \frac{(N - 2)(N + 1)}{2}          \m .   
\ee 
This is also an observation of Casson's \cite{Kendall}, adding significance to this diagonal.   
These are depicted in butter in Fig \ref{Shape-Triviality-Top-5}.a).  

\m 

\n{\bf Remark 1} Dynamical and relational nontriviality already covered why pure-shape (1, 3) and (1, 4) [orange and purple in Fig \ref{Shape-Triviality-Top-5}.a) respectively] 
                                                                                                          are special minimal cases.

\m

\n{\bf Remark 2} The basislands and the $N$-a-gonlands moreover share a common member: triangleland.
At the level of the shape spaces, this is consistent because of the topological coincidence 
\be 
\mathbb{CP}^1 = \mathbb{S}^2                                                                                            \m . 
\ee 
So, on the one hand, for triangleland, the number of methods available is `doubled': one has spherical methods as well as projective space methods.  

\m 

\n On the other hand, 
\be 
\mbox{$N \geq 4$ \m is required for topological-level typicality and specifically complex-projective mathematics}       \m . 
\ee
Quadrilateralland is thus minimal in this sense; as this carries over at the geometrical level as well, it is marked by the letter `G' on subsequent figures.  

\m

\n{\bf Smale's simplicity condition} \cite{Smale70} is that increasing $N$-a-gonlands' $N$ from 4 upward involves moving along a series of nontrivially $\mathbb{CP}^{N - 2}$ spaces. 

\m 

\n{\bf Remark 3} This results in there being many ways in which $N$-a-gonlands can be treated systematically at the topological level. 

\m 

\n{\bf Remark 4} A counterpoint to this is the various other levels of structure's distinctions and minimalities for $N$-a-gons 
as laid out in the rest of the current article and \cite{Minimal-N-2}.

\m 

\n{\bf Remark 5} $(d, N) = (3, 4)$ -- {\it tetrahaedronland} -- is the next basisland: minimal as regards not also being an $N$-a-gonland. 

\m 

\n{\bf Structure 4} Topologically, tetrahaedronland is, as a subcase of (\ref{Basislands-Top})
\be 
\FrB(\FrS(3)) = \FrS(3, 4)  =  \mathbb{S}^5  \mbox{ } .  
\ee  
\n{\bf Structure 5} The linearly-dependent sector can be viewed as all of a diagonal infinite series, horizontal equalities and vertical terminating series. 
This sector also has a systematically-known topological form,  
\be 
\FrS(d, < d + 1)  \es  \frac{\mathbb{S}^{(d - 1)(d + 2)/2}}{C_2}  \m . 
\ee 
Note that this is not however $\mathbb{RP}^q$, 
as the (3, 3) case's closed hemisphere already indicates by B being clearly distinct from its antipode U (correcting some previous literature).  
They are, rather, hemispheres:
\be 
\FrS(d, < d + 1)  \es  \mathbb{S}^{(d - 1)(d + 2)/2}_0    \m . 
\ee
This corresponds to mirror image identification being obligatory in this sector, due to rotation through the extra dimension(s).  
%
%FFFFFFFFFFFFFFFFFFFFFFFFFFFFFFFFFFFFFFFFFFFFFFFFFFFFFFFFFFFFFFFFFFFFFFFFFFFFFFFFFFFFFFFFFFFFFFFFFFFFFFFFFFFFFFFFFFFFFFFFFFFFFFFFFFFFFFFFFFFFFFFFFFFFFFFFFFFFFFFFFFFFFFFFFFFFFFFFFFFFFFFFF
{            \begin{figure}[!ht]
\centering
\includegraphics[width=1.0\textwidth]{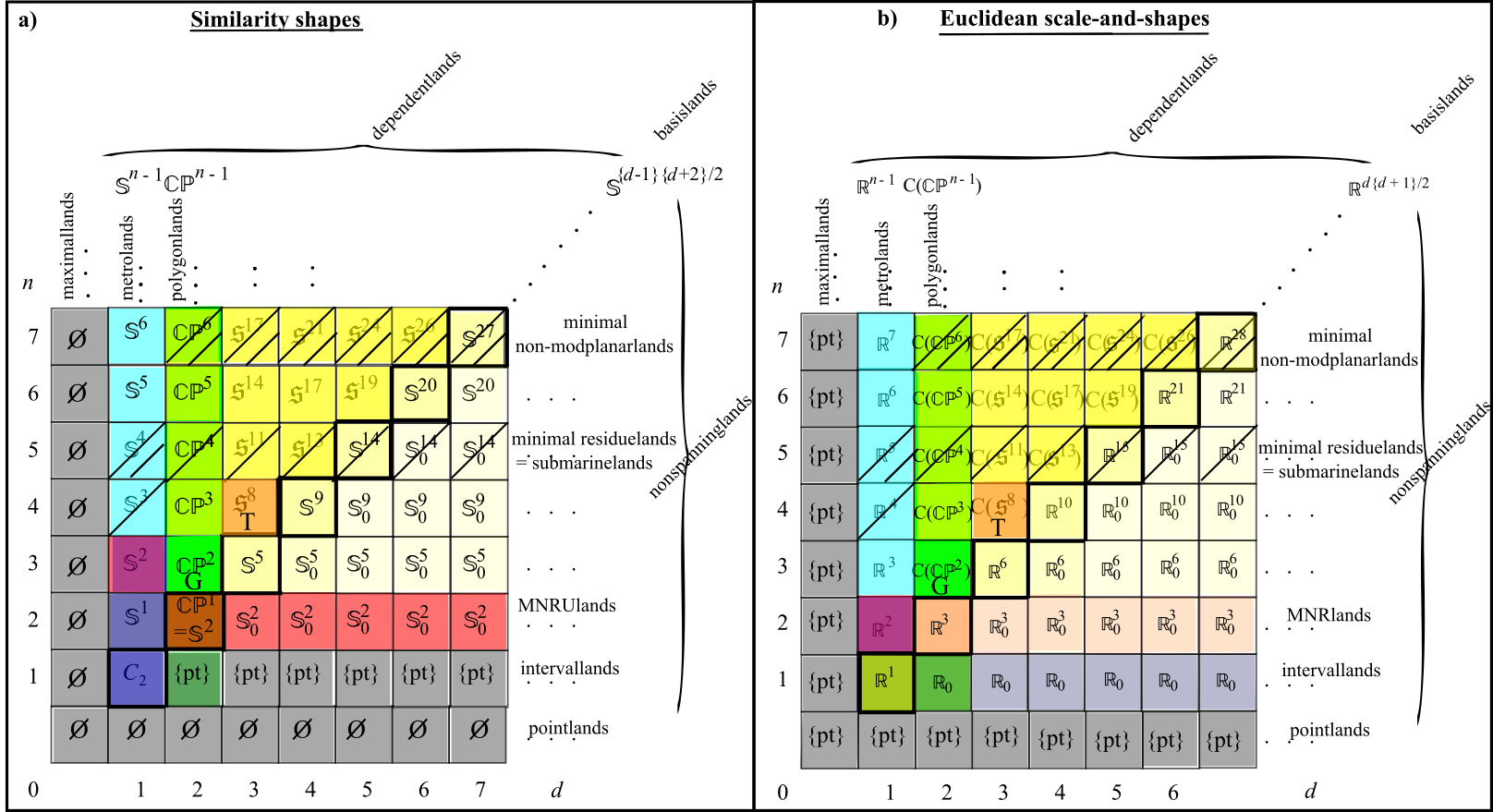}
\caption[Text der im Bilderverzeichnis auftaucht]{        \footnotesize{a) and b) are pure-shape and shape-and-scale topological manifolds 
(see \cite{FileR} for a summary of further topological results about shape and relational configuration spaces).  
This gives 3 nontrivial tractable series per table.    
Note that in the scaled case the intervallands become distinct from the pointlands and maximallands. 
Also note that the $\mathbb{S}^3$ and $\mathbb{S}^4$ of 5- and 6-stop metroland are the first rubber nontrivial cases, 
which criteria single out likewise the $\mathbb{CP}^4$ and $\mathbb{CP}^6$ of hexagonland and octagonland. 
}   }
\label{Shape-Triviality-Top-5} \end{figure}         } 
%FFFFFFFFFFFFFFFFFFFFFFFFFFFFFFFFFFFFFFFFFFFFFFFFFFFFFFFFFFFFFFFFFFFFFFFFFFFFFFFFFFFFFFFFFFFFFFFFFFFFFFFFFFFFFFFFFFFFFFFFFFFFFFFFFFFFFFFFFFFFFFFFFFFFFFFFFFFFFFFFFFFFFFFFFFFFFFFFFFFFFFFFF

\m 

\n{\bf Remark 6} In 1-$d$, 
\be 
\mathbb{S}^0    =  C_2 \m \mbox{ and }  \m 
\mathbb{S}^0_0  =  \{\mbox{pt}\}        \m . 
\ee 
\n{\bf Remark 7} That 3-$d$ is not such a topological series means that the $N \geq 5$ have further surprises in store in 3-$d$, 
rather than the $d = 2$ situation whose main leap in complexity is from $N = 3$ to $N = 4$. 
In particular, to have a 3-$d$ model with the generic feature of linear dependence -- and consequently of not having a merely spherical shape space -- 
we must turn to (3, 5) (indicated in orange).  
This singles out 
\be
\mbox{$(d, N) = (3, 5)$ \m as the {\sl bottom corner} T of the topologically hard wedge} \m ; 
\ee  
this offers a {\sl partial} explanation -- for many aspects of (3, 5) being more complex than (3, 4).  

\m 

\n{\bf Remark 8} The general case of similarity shape space is 
\be 
\FrS(d, N)  \es  \frac{\mathbb{S}^{n \, \d - 1}}{SO(d)}  
            \es  \frac{SO(n \, \d)}{SO(n \, d - 1) \times SO(d)}  
			\es  A(n \, d; n \, d - 1, d)  \m , 
\ee
where the second equality uses the Stiefel space result and the fourth brings in Appendix C.3's A-space. 
A-spaces are useful in that yet further open problems in Shape Theory can be phrased in terms of these (see e.g.\ Sec \ref{Orbits}).

\m 

\n{\bf Proposition 1} Shape-and-scale spaces are the corresponding topological cones over shape space \cite{LR97, Cones}, 
\be
{\cal R}(d, N) = \mC(\FrS(d, N))   \m .  
\ee  
\n This requires the following definition.  

\m 

\n{\bf Definition 1} A {\it topological cone} $\mC(\FrX)$ over a topological space $\FrX$ is     
\beq
\mC(\FrX) = \FrX \times [0, \infty)/\mbox{ } \widetilde{\mbox{ }} \mbox{ } . 
\eeq
$\widetilde{\mbox{ }}$ \mbox{ } here means that all points of the form \{p $\in \FrX$, 0 $\in [0, \infty)$\} 
are `squashed' or identified to a single point termed the {\it cone point} 0. 

\m
 
\n{\bf Structure 6}  Maximalland shape-and-scale spaces are 
\be 
{\cal R}(0, N)  =  \lFrr(0, N)    
                =  \mC(\emptyset) 
                =  \{\mbox{pt}\}                   \m ,   
\ee
indicated in grey in Fig \ref{Shape-Triviality-Top-5}.b). 

\m 

\n Metroland shape-and-scale spaces are 
\be 
{\cal R}(1, N)  =  \lFrr(1, N) 
                =  \mC(\mathbb{S}^{n - 1})
			    =  \mathbb{R}^{n}                  \m , 
\ee
which are depicted in cyan in Fig \ref{Shape-Triviality-Top-5}.b), with dynamical and relational nontriviality singling out (1, 2) and (1, 3) models respectively.  

\m 

\n Also pointland shape-and-scale spaces are 
\be
{\cal R}(d, 1)  =  \lFrr(d, 1)  
                =  \mC(\emptyset) 
				=  \{\mbox{pt}\}                   \m ; 
\ee 
These are also indicated in grey in Fig \ref{Shape-Triviality-Top-5}.b). 

\m 

\n Finally intervalland shape-and-scale spaces are 
\be
{\cal R}(d, 2)  =  \mathbb{R}_+ \m \m d \geq 2     \m , 
\ee
with the exception that 
\be
\FrS(1, 2)  =  \mC(C_2)  
            =  \mathbb{R}                          \m . 
\ee
These are indicated in pale grey in Fig \ref{Shape-Triviality-Top-5}.b);   
the scaled (1, 2) model (purple) is furtherly special through being the sole intersection of the metroland and the basisland series.  
$\mC(\emptyset) = \{\mbox{pt}\}$ acounts for coning's +1 count, working out as $0 = -1 +1$.  

\m  

\n{\bf Structure 7} The $N$-a-gonland shape-and-scale spaces also form a series at the topological level 
\be 
\FrR(2, N)  =  \mC(\FrS(2, 3)) 
            =  \mC(\mathbb{CP}^{n - 1})                                                                                              \m .
\ee 
This does not further simplify, other than in triangleland's excseptional case, for which 
\be 
\FrR(3, 2)  =  \mC(\mathbb{S}^2)
            =  \mathbb{R}^3                                                                                                          \m .    
\ee
So 
\be 
\mbox{in 2-$d$, $N \geq 4$ \m is required for topological-level typicality: specifically cones over complex-projective spaces}       \m . 
\ee
The shape-and-scale space of the first basisland which is not also an $N$-a-gonland -- tetrahaedronland -- is 
\be 
\FrB({\cal R}(3))  =  {\cal R}(3, 4)  
                   =  \mC(\FrS(3, 4)) 
				   =  \mC(\mathbb{S}^5) 
				   =  \mathbb{R}^6                                                                                                  \m .  
\ee  
More generally, the basislands are topologically 
\be 
\FrB({\cal R}(d))  =  {\cal R}(d, d + 1) = \mC(\FrS(d, d + 1)) = \mC(\mathbb{S}^{d(d + 1)/2 - 1}) = \mathbb{R}^{d(d + 1)/2}         \m .  
\ee
\n{\bf Structure 8} The linearly-dependent sector can also be viewed as diagonal infinite series as well as horizontal equalities and vertical terminating series. 
This sector also has a systematically-known topological form,  
\be 
{\cal R}(d, < d + 1)  =  \mC(\mathbb{S}^{d(d + 1)/2 - 1}_0)  
                      =  \mathbb{R}^{d(d + 1)/2}_0                                                                                 \m : 
\ee 
half-spaces.  

\m 

\n{\bf Remark 9} The collinear configurations form $\mathbb{RP}^{n - 1}$ submanifolds of the $\mathbb{CP}^{n - 1}$; 
this is an Aufbau result since these are the mirro-image-identified versions of the 1-$d$ shape spaces;   
this was already known to Kendall \cite{Kendall84} and even to Smale \cite{Smale70}.

%==================================================================================================================================================================
%==================================================================================================================================================================
\section{Relational space metric geometry}\label{RSG}
%==================================================================================================================================================================
%==================================================================================================================================================================

\n{\bf Remark 1} The $\mathbb{S}^{N - 2}$, $\mathbb{RP}^{N - 2}$ and $\mathbb{CP}^{N - 2}$ series are moreover geometrical series.  
The first two are as per Secs \ref{Preshape} and \ref{Gamma}.  
The following holds for the third. 

\m 

\n{\bf Kendall's Simplicity postulate} \cite{Kendall84} The $N$-a-gonland $\mathbb{CP}^{N - 2}$'s are moreover naturally equipped as a series with Fubini--Study metrics   
\beq
\d s^2_{\sF\sS}  =  \frac{\{1 + ||\mbox{\boldmath$Z$}||_{\sC}^2\} ||\d\mbox{\boldmath$Z$}||_{\sC}^2 -  |(\mbox{\boldmath$Z$} \cdot \d \mbox{\boldmath$Z$})_{\sC}|^2} 
                         {\{1 + ||\mbox{\boldmath$Z$}||_{\sC}^2\}^2} \m  .
\label{FS-2}
\eeq
The $\sC$ suffix here denotes the $\mathbb{C}^{n - 1}$ version of inner product and norm, with $Z_{\bar{p}}$'s indices running over $n - 1$ copies of $\mathbb{C}$.
\be
Z_{i} = {\cal R}_{i} \, \mbox{exp}(i\slPhi_{i})  \m :
\label{Zpolar}
\ee 
the `multiple copies of $\mathbb{C}$ plane-polar coordinates' version of ratios of the $\u{\rho}_i$.
Therein, the $\slPhi_{i}$ are an independent set of $n$ relative angles $\theta_{ij}$ between $\u{\rho}_i$'s, 
whereas the ${\cal R}_{i}$ are the corresponding set of $n$ ratios of magnitudes $||\u{\rho}_i||$ \cite{FileR}.
Thus in 2-$d$ relative angles and ratios of magnitudes occur in 1 : 1 proportion as modulus--phase pairs.  

\m 

\n{\bf Remark 2} Writing the 1-$d$ case in Beltrami coordinates brings it into close analogy with the second.  
\beq
\d\fs_{N-\sss\st\so\sp\,\sS\sR\sP\sM}^{\sr\se\sd}\mbox{}^2  \es  
\frac{  ||\mbox{\boldmath${\rho}$}||^2||\d{\mbox{\boldmath${\rho}$}}||^2 - (\mbox{\boldmath${\rho}$}, \d{\mbox{\boldmath${\rho}$}})^2}
     {  \{ 1 + ||\mbox{\boldmath${\rho}$}||^2\}^2  }                                                                                                  \m ,  
\eeq
\n The $\mathbb{S}^q$ basisland topological series is not known to form a metric series; consequently the $\mathbb{S}^q/C_2$ are not either. 

\m 

\n Because of this, the metrical version of the non-systematic wedge is 1 taller (past this point, $\mathbb{S}^q$ to $\mathbb{S}^q/C_2$ metric inheritance applies).  
%
%FFFFFFFFFFFFFFFFFFFFFFFFFFFFFFFFFFFFFFFFFFFFFFFFFFFFFFFFFFFFFFFFFFFFFFFFFFFFFFFFFFFFFFFFFFFFFFFFFFFFFFFFFFFFFFFFFFFFFFFFFFFFFFFFFFFFFFFFFFFFFFFFFFFFFFFFFFFFFFFFFFFFFFFFFFFFFFFFFFFFFFFFF
{            \begin{figure}[!ht]
\centering
\includegraphics[width=0.87\textwidth]{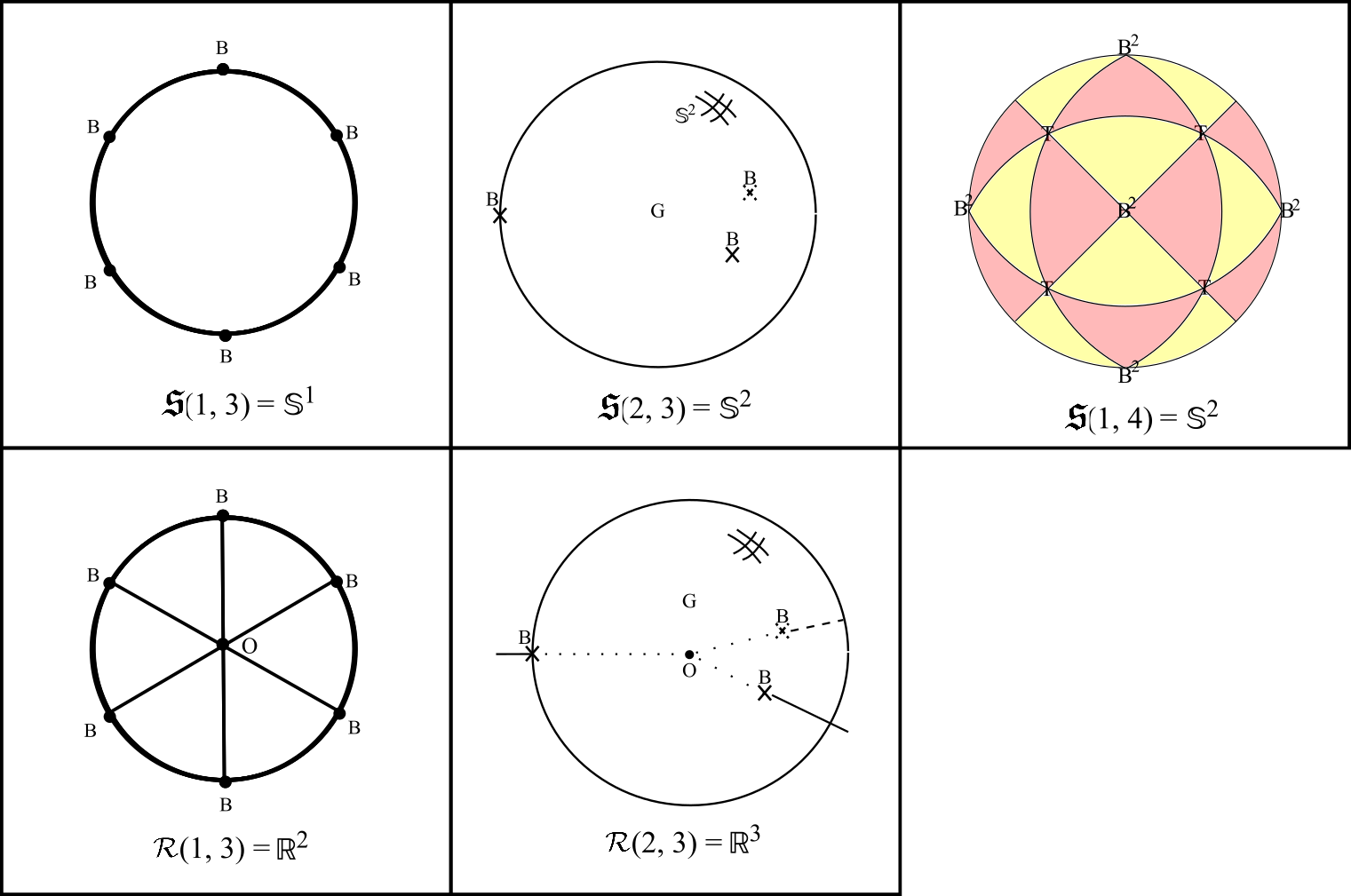}
\caption[Text der im Bilderverzeichnis auftaucht]{        \footnotesize{Rubber relational space features that recur in metric relational spaces.
}   }
\label{Rubber-Recurrs} \end{figure}         } 
%FFFFFFFFFFFFFFFFFFFFFFFFFFFFFFFFFFFFFFFFFFFFFFFFFFFFFFFFFFFFFFFFFFFFFFFFFFFFFFFFFFFFFFFFFFFFFFFFFFFFFFFFFFFFFFFFFFFFFFFFFFFFFFFFFFFFFFFFFFFFFFFFFFFFFFFFFFFFFFFFFFFFFFFFFFFFFFFFFFFFFFFFF

\m 

\n{\bf Remark 3} As examples of relevance of the topological shape-and-scale space, the (1, 3) cycle-and-wheel graphs, (2, 3) claw graph, and (1, 4) cubic net 
recurs in metric-level relational spaces as per columns 1, 2 and 3 of Fig \ref{Rubber-Recurrs} respectively.  

\m 

\n{\bf Remark 4} $N$-a-gons have an element of systematicness to them, 
which is underlied by the $\mathbb{CP}^{n - 1}$ series of shape space geometries whose unificatory power and simplicity has no 3- or higher-$d$ counterpart.
Shape theoretic studies have moreover been shown to be capable of yielding interesting geometrical results in their own right 
\cite{Kendall84, Kendall89, Small, Kendall, MIT, III, Forth}.

%==================================================================================================================================================================
%==================================================================================================================================================================
\section{Symmetry, uniformity, Lagrangian and Jacobian structure}\label{Lag}
%==================================================================================================================================================================
%==================================================================================================================================================================

\n The symmetry structure in space is required to have all the boundaries of Leibniz space and the other quotients; see \cite{I} for the (1, 3) case, 
                                                                                                                        \cite{II} for the (1, 4) case 
																												    and \cite{III} for the (2, 3) case.
This exceeds the Flat Geometry section's consideration by ascertaining where in the relational shape spaces the various configurations with elements of spatial symmetry reside.

\m 

\n We next provide some definitions; see \cite{I, II, III} for further discussion of these concepts.  
%
% including their entrenchment on notions of invariant.

\m

\n{\bf Definition 1} {\it Lagrangian structure} is based on relative Lagrange separation vectors $\u{r}_{IJ}$ having zero magnitudes, 
                                                                                                                   equal nonzero magnitudes, 
																												   zero angles, 
																												and equal nonzero angles. 

\m 

\n{\bf Definition 2} The {\it Lagrangian uniformity structure} consists of the second and fourth of these. 

\m 

\n{\bf Definition 3} The {\it Jacobian (uniformity) structure} are defined likewise, except now in terms relative Jacobi separation vectors $\urho_i$.

\m 

\n{\bf Definition 4} {\it Mergers} M are zero $\urho_i$ which are not also zero $\u{r}_{IJ}$: coincidences-or-collisions; 
mergers are rather the analogues thereof involving at least one nontrivial centre of mass. 

\m 

\n{\bf Definition 5} The 1-$d$ `diameters' are mass-weighted diameters per square root of moment of inertia, 
and the `areas' in 2-$d$ are mass-weighted areas per unit moment of inertia.
We refer to these, and further higher-$d$ analogues as {\it normalized notions of size}, and consider in particular which shapes extremize these notions. 

\m 

\n{\bf Remark 1} A number of extremal properties are furthermore realized by the Leibniz space `corner or side' points, as per Fig \ref{Extrema}. 
%
%FFFFFFFFFFFFFFFFFFFFFFFFFFFFFFFFFFFFFFFFFFFFFFFFFFFFFFFFFFFFFFFFFFFFFFFFFFFFFFFFFFFFFFFFFFFFFFFFFFFFFFFFFFFFFFFFFFFFFFFFFFFFFFFFFFFFFFFFFFFFFFFFFFFFFFFFFFFFFFFFFFFFFFFFFFFFFFFFFFFFFFFFF
{\begin{figure}[ht]
\centering
\includegraphics[width=1.0\textwidth]{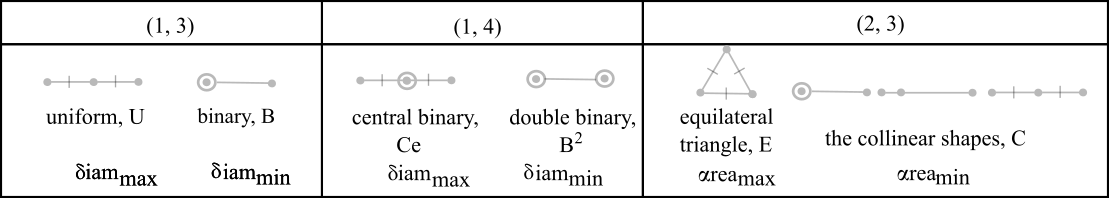}
\caption[Text der im Bilderverzeichnis auftaucht]{\footnotesize{a) Maxima and minima of diameter, and area, in each case per unit moment of inertia.}} 
\label{Extrema}\end{figure} } 
%FFFFFFFFFFFFFFFFFFFFFFFFFFFFFFFFFFFFFFFFFFFFFFFFFFFFFFFFFFFFFFFFFFFFFFFFFFFFFFFFFFFFFFFFFFFFFFFFFFFFFFFFFFFFFFFFFFFFFFFFFFFFFFFFFFFFFFFFFFFFFFFFFFFFFFFFFFFFFFFFFFFFFFFFFFFFFFFFFFFFFFFFFF

\m 

\n For (1, 3), there is a single Lagrange-uniform shape and a single symmetric shape (reflection symmetry, Ref) and these moreover coincide with the maximal diameter, 
\be
\mU = \mbox{Ref} = \diam_{\sm\sa\sx} = \mM   \m .
\label{MUSym}
\ee 
On the other hand, the binary shape and the minimal diameter coincide, 
\be 
\mB =  \diam_{\sm\si\sn}
\ee
Furthermore,  
\be 
\mbox{$N = 3$ \m is minimal as regards supporting nontrivial mergers}   \m . 
\ee
\n{\bf Remark 2} For (1, 3), the U shape depicted in Fig \ref{Extrema} is the only shape with any element of uniformity.  

\m 

\n To have multiple points in Leibniz space with elements of Lagrangian uniformity exhibiting moreover various distinct strengths of uniformity, 
$(d, N) = (1, 4)$ or (2, 3) are required.  

\m 

\n For (2, 3), the most uniform shape is of course the equilateral triangle, E.   
This generalizes to the regular $N$-a-gon for $(2, N)$. 
For $d \geq 3$, the most uniform states are much more complicated to describe \cite{Forth}.  

\m 

\n{\bf Remark 3} To have distinctly realized notions of symmetry, uniformity, and extremal normalized size, $(d, N) = (1, 4)$ and (2, 3) are minimal.  

\m 

\n{\bf Remark 4} \cite{II} bears further witness to $(d, N) = (1, 4)$ similarity shape space being far more complex than the (1, 3) one.
This is partly through $k = 2$ -- the minimal relationally nontrivial number -- supporting submanifolds, 
and partly through (1, 4) supporting a wider range of notions of merger, which moreover form an intricate pattern on the shape space.

%====================================================================================================================================================================================
\section{Isometry groups of relational spaces}
%====================================================================================================================================================================================

{\bf Structure 1} From Sec 12's metric-level geometry results, 
\be 
Isom(\FrS(1, N)) = Isom(\mathbb{S}^{n - 1}) = SO(n)   \m , 
\ee 
\be 
Isom(\FrS(2, N)) = Isom(\mathbb{CP}^{n - 1}) \es \frac{SU(n)}{C_n}   \m .  
\ee 
For triangleland's (2, 3), 
\be 
Isom(\mathbb{CP}^{1}) \es \frac{SU(2)}{C_2} \es SO(3) =     Isom(\mathbb{S}^{2})  \m .  
\ee 
Also 
\be 
Isom({\cal R}(1, N)) = Isom(\mathbb{R}^{n}) = Eucl(n) = \mathbb{R}^n \rtimes SO(n)   \m , 
\ee 
\be 
Isom({\cal R}(2, N)) = Isom(\mC(\mathbb{CP}^{n - 1})) \es \frac{SU(n)}{C_n}   \m , 
\ee 
with an extra radial similarity. 
This includes triangleland's (2, 3) case, since its origin is privileged by a curvature singularity, breaking the 
\be 
\mC(\mathbb{CP}^{1})) = \mC(\mathbb{S}^{2})) = \mathbb{R}^{3}
\ee
translation isometries. 

\m 

\n{\bf Remark 1} For triangleland, there is the further coincidence that all of the embedding space, the space of separations and the isometry group are 3-$d$. 
This coincidence underlies various further results, such as that the Hopf quantities associated with the isometry group constitute embedding space Cartesian coordinates, 
and that both Kendall's Theorem and the Hopf map can be derived from Heron's formula in this case \cite{III}. 

\m 

\n However, for the general $N$-a-gon, the dimensions of these three spaces are
\be 
\mbox{dim(embedding space)} = \mbox{dim}({\cal R}(2, N) = 2 \, N - 3  = 2 \n - 1          \m ,
\ee 
\be 
\mbox{dim(separation space)} = C(2, N)  \es  \frac{N(N - 1)}{2}  
                                        \es  \frac{n(n + 1)}{2}                           \m , 
\ee 
and 
\be 
\mbox{dim(isometry group)} = \mbox{dim}(Isom(\FrS(2, N))) = \mbox{dim}(SU(n)) = n^2 - 1   \m .
\ee 
These only coincide pairwise if, firstly, 
\be 
2 \, n - 1   \es  \frac{n(n + 1)}{2}  \m \mbox{ i.e.\ } \m (n - 1)(n - 2) = 0    \m , 
\ee 
so $n = 1$: intervalland, or $n = 2$: triangleland. 

\m 

\n Secondly, 
\be 
2 \, n - 1   \es n^2 - 1  \m \mbox{ i.e.\ } \m n(n - 2) = 0  \m, 
\ee 
so $n = 0$: pointland, or $n = 2$: triangleland.  

\m 

\n Thirdly, 
\be 
\frac{n(n + 1)}{2}  \es  n^2 - 1  \m \mbox{ i.e.\ } \m (n + 1)(n - 2) = 0  \m, 
\ee 
so $n = -1$: an empty model, or $n = 2$: triangleland.  

\m 

\n Thus no further such coincidences occur for $n \geq 3$, leaving quadrilateralland as minimal to have these distinctions. 

\m 

\n{\bf Remark 2} The $N$-a-gonlands continue to enjoy a {\it generalized} Hopf map for all $N$, 
\be 
\mathbb{S}^{2n - 1} \longrightarrow \mathbb{CP}^{n - 1} \m , 
\ee 
but this ceases to be aligned with representing the shape space isotropy group. 
While quadrilaterals continue to enjoy an area in terms of separation formula -- the Coolidge formula mentioned in Sec \ref{Geom} --,
neither Kendall's Theorem nor the generalized Hopf map can be derived from this \cite{Forth}.  

\m

\n{\bf Remark 3} Triangleland and quadrilateralland are sequentially simpler than subsequent $N$-a-gonlands in the following further ways. 

\m 

\n Another generalization of triangleland's Hopf map -- in its aspect as an embedding into a linear space -- is to quadrilateralland's Veronese embedding \cite{Veronese, Kuiper, QuadI}.  
This does moreover further generalize to the arbitrary $N$-a-gons via the Veronese--Whitney embeddings, 
a point already well-known in the Shape Statistics literature \cite{Bhatta, DM16, PE16}.  

\m 

\n The two hemi-spaces formed by the equator of collinearity are metrically and topologically hemispheres for triangleland, 
                                                                            and topologically $\mathbb{S}^4$ for quadrilateralland \cite{Kuiper}. 
While such a splitting into hemi-spaces is generally realized for $d \geq 2$, the topological nature of the two identical halves thus produced is less straightforward, 
both for subsequent $N$-a-gonlands and for $d \geq 3$
																			
\m 

\n Triangleland's symmetry-adapted coordinates are just spherical coordinates, while quadrilateralland's are Euler-angle-adapted coordinates \cite{GP78, Page-Instanton, MacFarlane03}. 
While the latter $SU(n)$-generalize, this does place some limitations on subsequent applications 
(compare \cite{MacFarlane03}'s quadrilateralland interpretation \cite{QuadI} with its general-$N$ counterpart in \cite{MacFarlane03b}).  

\m

\n{\bf Remark 4} Let us finally mention that one result which is known for both 1- and 2-$d$ with arbitrary $N$ is the eigenspectrum of the shape space Laplacian, 
following from the geometrical series status of the 1- and 2-$d$ shape spaces \cite{Berger}.  

\vspace{10in}

%==================================================================================================================================================================
%==================================================================================================================================================================
\section{Strata}\label{Strata}
%==================================================================================================================================================================
%==================================================================================================================================================================
%
%
%FFFFFFFFFFFFFFFFFFFFFFFFFFFFFFFFFFFFFFFFFFFFFFFFFFFFFFFFFFFFFFFFFFFFFFFFFFFFFFFFFFFFFFFFFFFFFFFFFFFFFFFFFFFFFFFFFFFFFFFFFFFFFFFFFFFFFFFFFFFFFFFFFFFFFFFFFFFFFFFFFFFFFFFFFFFFFFFFFFFFFFFFF
{            \begin{figure}[!ht]
\centering
\includegraphics[width=1.0\textwidth]{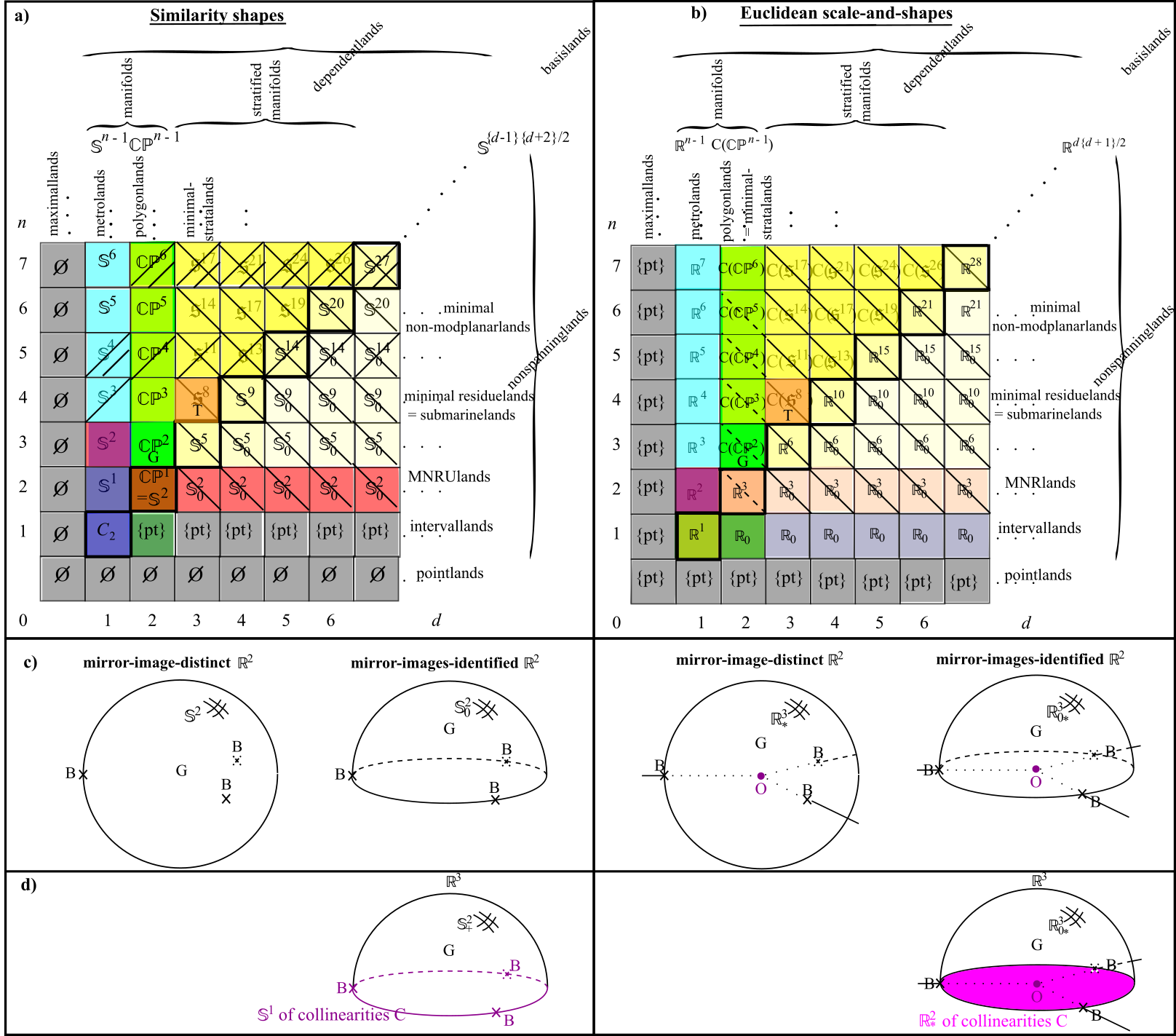}
\caption[Text der im Bilderverzeichnis auftaucht]{        \footnotesize{a), b) Further refinement by the stratificational structure, without and with scale. 
In the scaled case of {\it minimal stratalands} -- 2-$d$: polygonlands -- the nontrivial strata are just the maximal coincidences-or-collisions O. 
c) and d) contrast (2, 3) and (3, 3) from a stratificational point of view; non-principal strata are picked out in purple and magenta. 
* denotes an excluded point and + denotes the open-edged version of a space. }   }
\label{Shape-Triviality-Top-6} \end{figure}         } 
%FFFFFFFFFFFFFFFFFFFFFFFFFFFFFFFFFFFFFFFFFFFFFFFFFFFFFFFFFFFFFFFFFFFFFFFFFFFFFFFFFFFFFFFFFFFFFFFFFFFFFFFFFFFFFFFFFFFFFFFFFFFFFFFFFFFFFFFFFFFFFFFFFFFFFFFFFFFFFFFFFFFFFFFFFFFFFFFFFFFFFFFFF

\n{\bf Caveat 2} 
\be
\frac{A}{B} \m  \mbox{ need not be of a single dimension, even if $A$ and $B$ are of a single dimension }  \m .
\label{A-B-2}
\ee
This is because quotienting a manifold by a group in general produces a stratified manifold, 
which no longer obeys the local Euclideanness axiom of manifolds whereby manifold dimension is unique.
See \cite{ABook, Generic} for outlines or \cite{Pflaum, BanaglBook, Kreck} for detailed references on stratified manifolds, 
and also \cite{Fischer70, FM96, Giu09} for stratification in General Relativity.

\m 

\n Thus (\ref{A-B})'s dimension count may just refer to the {\it top} stratum.

\m

\n{\bf Definition 1} For $\lFrg$ a group acting on a set $\FrX$, and $x \in \FrX$, the {\it group orbit}  
\be 
\mbox{Orb}(x) := \{ g \, x \, | \, g \in \lFrg\}  \m :
\ee
the set of images of $x$.  

\m 

\n{\bf Definition 2} The set of orbits for a given group and group action constitute the {\it orbit space}, $\FrO$.  

\m 

\n{\bf Remark 1} Relational spaces are thus orbit spaces, which are moreover one of the more habitual settings in which stratification arises. 

\m 

\n{\bf Remark 2} Orbits are furthermore associated with stabilizers, which are defined as follows. 

\m 

\n{\bf Definition 3} A {\it stabilizer} alias {\it isotropy group} (and {\it little group} in Theoretical/Particle Physics: Wigner's useage \cite{Wigner39}) is 
\be 
\mbox{Stab}(x) := \{ g \in \lFrg \, | \, g \, x = x \}  \m :
\ee    
the set of $g \in \lFrg$ that fix $x$.  

\m 

\n{\bf Remark 3} The Orbit--Stabilizer Theorem -- that 
\be
|\lFrg| = |\mbox{Orb}(x)| \, |\mbox{Stab}(x)|
\ee 
for finite $\lFrg$ and $\FrX$ -- gives some indication of such a link.
Differences in isotropy group leading to multiplicity of orbits, including to stratification of orbit spaces, is a further phenomenon which moreover transcends finiteness.  

\m 

\n{\bf Example 0} None of $\FrR(1, N)$, $\FrS(1, N)$, $\FrS(2, N)$ or $\FrR(\mathbb{S}^1, N)$ are stratified; manifold theory suffices for all of these.  

\m 

\n{\bf Example 1} $SO(2)$ acts differently on O and all other configurations: the rotation is inactive in the first case and active in the second. 
Thus in $\FrR(2, N)$, O and the rest constitute distinct orbits and distinct strata.

\m  

\n{\bf Example 2} $SO(3)$ has 3 distinct actions: full, $SO(2)$ and $id$.
For $(3, N)$, the $id$ action applies to the maximal coincidence-or-collision $\mO$, and the $SO(2)$ action to all other collinear configurations. 
Because of this, the (2, 3) and (3, 3) models are not the same. 
This rests in part on Remark 10 of Sec \ref{RST}'s mirror-image identification obligatoriness applyng to (3, 3) but not to (2, 3).  
However, even if mirror-image identification obligatoriness is considered in each case, 
$\FrS(3, 3)$ is a hemisphere with edge whose edge is a distinct stratum $\mathbb{S}^1$         (Fig \ref{Shape-Triviality-Top-6}.e)
to $\FrS(3, 2)$ being a hemisphere with edge whose edge is not stratifictionally distinguished (Fig \ref{Shape-Triviality-Top-6}.d).
$\FrR(3, 3)$ is a half space with edge plane, whose punctured edge plane $\mathbb{R}^2_*$ and puncture point $\mO$ are distinct strata (Fig \ref{Shape-Triviality-Top-6}.e), 
whereas $\FrR(2, 3)$ is a half space with edge plane with a puncture point O in the edge plane alone constituting a distinct stratum   (Fig \ref{Shape-Triviality-Top-6}.d).  

\m 

\n{\bf Remark 4} The above-mentioned strata are moreover contiguous-as-manifolds in their relative placing. 
The maximal collision has some problematic features, by which it is more often excised as regards mathematical treatment than the collinear configurations.
Note that $\FrS(2, N)$ study avoids the need of either excision, and $\FrS(3, N)$ of the more mathematically problematic one. 

\m 

\n{\bf Remark 5} Whereas non-normalizability of the maximal $\mO$, and removal of one rubber relational graph vertex affects all dimensions, 
many mathematical difficulties with including the maximal $\mO$ stem rather from stratificational differences, which require $d \geq 2$ to materialize.  
                                                                                                                                                                                         
\m 

\n{\bf Remark 6} Collinearity in $\geq 3$-$d$ has the knock-on effect of inertia quadric noninvertibility, amounting to configuration space geometry singularness. 

\m

\n{\bf Remark 7} In 3-$d$, $N = 3$ and $N \geq 4$ are moreover qualitatively different \cite{Gergely} as regards configuration space Ricci curvature singularities on the strata.  
On the one hand, 
\be 
\mbox{for $N = 3$ the Ricci curvature scalar is regular at collinear configurations C}    \m . 
\ee 
On the other hand, 
\be 
\mbox{for $N \geq 4$ the Ricci curvature scalar is singular at C}                         \m .  
\ee 
Moreover, even $N = 3$ has a Ricci curvature singularity at $\mO$ \cite{Iwai87}.  

\m 

\n{\bf Remark 8} Finally see \cite{Generic} for a conceptual classification of stratified spaces by complexity, 
and Paper II for realization of a harder case than that in the current Sections' specific examples.

\vspace{10in}

%==================================================================================================================================================================
%==================================================================================================================================================================
\section{Orbit space decompositions}\label{Orbits}
%==================================================================================================================================================================
%==================================================================================================================================================================

\n{\bf Definition 1} The {\it kinematic group} of the $N$-Body Problem in $\mathbb{R}^d$ is the `internal' $SO(n)$ rotations 
acting on whichever basis choice of $n$ mass-weighted Jacobi vectors $\u{\rho}_i$ in the natural manner. 
(This treats the components of each $\u{\rho}_i$ together as a package.) 

\m 

\n Let us denote the arbitrary {\it kinematic rotation matrix} alias {\it internal rotation matrix} by $\bK$. 

\m

\n{\bf Remark 1}  
\be 
\mbox{$N = 1$ \m has \m $n = 0$ \m and so no \m $\urho_i$ \m for any kinematical rotations to act upon} \m .
\ee    
\be 
\mbox{$N \geq 2$ \m is required for kinematical rotation matrices to be defined}  \m{ } , 
\ee 
whereas 
\be 
\mbox{$N \geq 3$ \m is required for their kinematical rotation group to be nontrivial}  \m{ } .   
\ee 

\m 

\n{\bf Lemma 1} \cite{ML00}
\be
[\bK, \, \bL] = 0 
\ee 
for $\bL$ the arbitrary {\it external} alias {\it spatial rotation}.

\m 

\n{\bf Remark 2} $SO(n)$ consequently has a well-defined action on the relational space 
\be 
\FrR(d, N) = \frac{  \mathbb{R}^{n \, d}  }{  SO(d)  }                   \m . 
\ee 
\n{\bf Definition 2} The {\it kinematic orbit} through a specific shape-and-scale configuration $\mR$ is 
\be 
\FrO(\mR) :=  \{ \, \bK \, \mR \, | \, \bK \in SO(n) \, \}                                          \m .
\ee
\n{\bf Remark 3} Furthermore, 
\be 
\FrO(\mR) \m \s{\mbox{diffeo}}{=} \m  \frac{SO(n)}{Isot(\mR)}                                         \m .  
\ee 
\n{\bf Definition 3} The kinematical action on $\mR$'s {\it isotropy group}  
\be 
Isot(\mR)  :=  \{ \, \bK \, \in \, SO(n) \, | \, \mR \m \mbox{ is invariant under } \m K \, \} 
            =  \{ \, \bK \, \in \, SO(n) \, | \, \mR \, \bK^{\sT} = \bL \,  \mR \, \}               \m , 
\ee 
since $\mR \, \bK^{T}$ and $\mR$ have the same shape-and-scale if they are related by a spatial rotation $\bL$ \cite{ML00}.  

\m

\n{\bf Definition 4} We term a model {\it C-generic} if it has a full count of distinctly realized isotropy groups.  

\m  

\n{\bf Proposition 1} i) $d = 1$ possesses 2               isotropy groups     
                                       and 2 corresponding kinematical orbits, all as per Fig \ref{7-Isot} and \ref{7-Orb}'s first columns. 
\be 
\mbox{$N = 3$ \m is minimal to distinctly realize these}                                            \m .   
\ee 

\m 

\n ii) $d = 2$ possesses 4               isotropy groups     
                     and 2 corresponding kinematical orbits, as per Fig \ref{7-Isot} and \ref{7-Orb}'s second columns. 
\be 
\mbox{$N = 4$ \m is minimal to distinctly realize these}                                           \m .    
\ee 
\n{\bf Proposition 2} $d = 3$ possesses 7               isotropy groups     
                                                                          and 7 corresponding kinematical orbits, all as per Fig \ref{7-Isot} and and \ref{7-Orb}'s third columns.
\be 
\mbox{$N = 5$ \m is minimal to distinctly realize all 7 isotropy groups and corresponding kinematical orbits that $\mathbb{R}^3$ possesses}  \m . 
\ee 
\n{\u{Derivations}} In 3-$d$, this was worked by Mitchell and Littlejohn \cite{ML00}, building on earlier work with Reinsch, Aquilanti and Cavalli \cite{LMRAC98} for the $N = 4$ case.  
See \cite{ML-1-2} for detailed consideration of the 1-and-2-$d$ counterparts. 

\m 

\n In brief, this follows from the following series of coincidences and Lie group accidents.

\m 

\n $SO(p)$ undefined for $p \leq 0$ and $SO(0) = \emptyset$ removes all isotropy groups for $N = 1$ and all but the top one for $N = 2$.

\m 

\n $N = 3$ already has the $C$-generic number of distinct isotropy groups in 1-$d$, as $SO(2)$ and $C_2$.  
In 2-$d$, the first and third are conflated by $SO(0) \times SO(2)$ collapsing to  $SO(2)$, 
whereas the fourth is knocked out.  
In 3-$d$, the first four survive, as $SO(2)$, $C_2$, $O(2)$ and $V_4$.  

\m 

\n $N = 4$ has the $C$-generic number of full isotropy groups in 2-$d$, as $SO(3)$, $O(2)$, $SO(2) \times SO(2)$ and $C_2$.  
For 3-$d$, the first and fifth and second and sixth coincide.

\m 

\n $N = 5$ has the $C$-generic number of full isotropy groups in 3-$d$.                                   $\Box$

\m 

\n{\bf Proposition 3} As regards the continuous parts of the isotropy groups, i) 
\be 
\mbox{$N = 3$ \m is minimal to distinctly realize these in 1-$d$}                                           \m ,  
\ee 
\be 
\mbox{$N = 5$ \m is minimal to distinctly realize these in 2-$d$}                                           \m ,  
\ee 
                                                                         and 7 corresponding kinematical orbits, all as per Fig \ref{7-Isot} and and \ref{7-Orb}'s third columns.\be 
\mbox{$N = 8$ \m is minimal to distinctly realize these in 3-$d$}                                           \m .  
\ee 
\n{\bf Derivation} For $N = 4$, the continuous parts of the first and third coincide in 2-$d$, but all are distinct for $N = 5$. 

\m 

\n For $N = 5$, the continuous parts of the second and fifth coincide, as do the third and sixth.  

\m  

\n $N = 6$'s continuous part realizes $SO(3) \times SO(2)$ twice.

\m

\n $N = 7$'s continuous part realizes $SO(4)$ twice, once directly and once via the accidental Lie group relation 
\be
SO(3) \times SO(3) = SO(4) 
\label{SO4}                                                                                  \m .  
\ee 
\n $N = 8$ distinctly realizes the continuous part of the isotropy group in 3-$d$.                           $\Box$  
%
%FFFFFFFFFFFFFFFFFFFFFFFFFFFFFFFFFFFFFFFFFFFFFFFFFFFFFFFFFFFFFFFFFFFFFFFFFFFFFFFFFFFFFFFFFFFFFFFFFFFFFFFFFFFFFFFFFFFFFFFFFFFFFFFFFFFFFFFFFFFFFFFFFFFFFFFFFFFFFFFFFFFFFFFFFFFFFFFFFFFFFFFFF
{\begin{figure}[ht]
\centering
\includegraphics[width=0.6\textwidth]{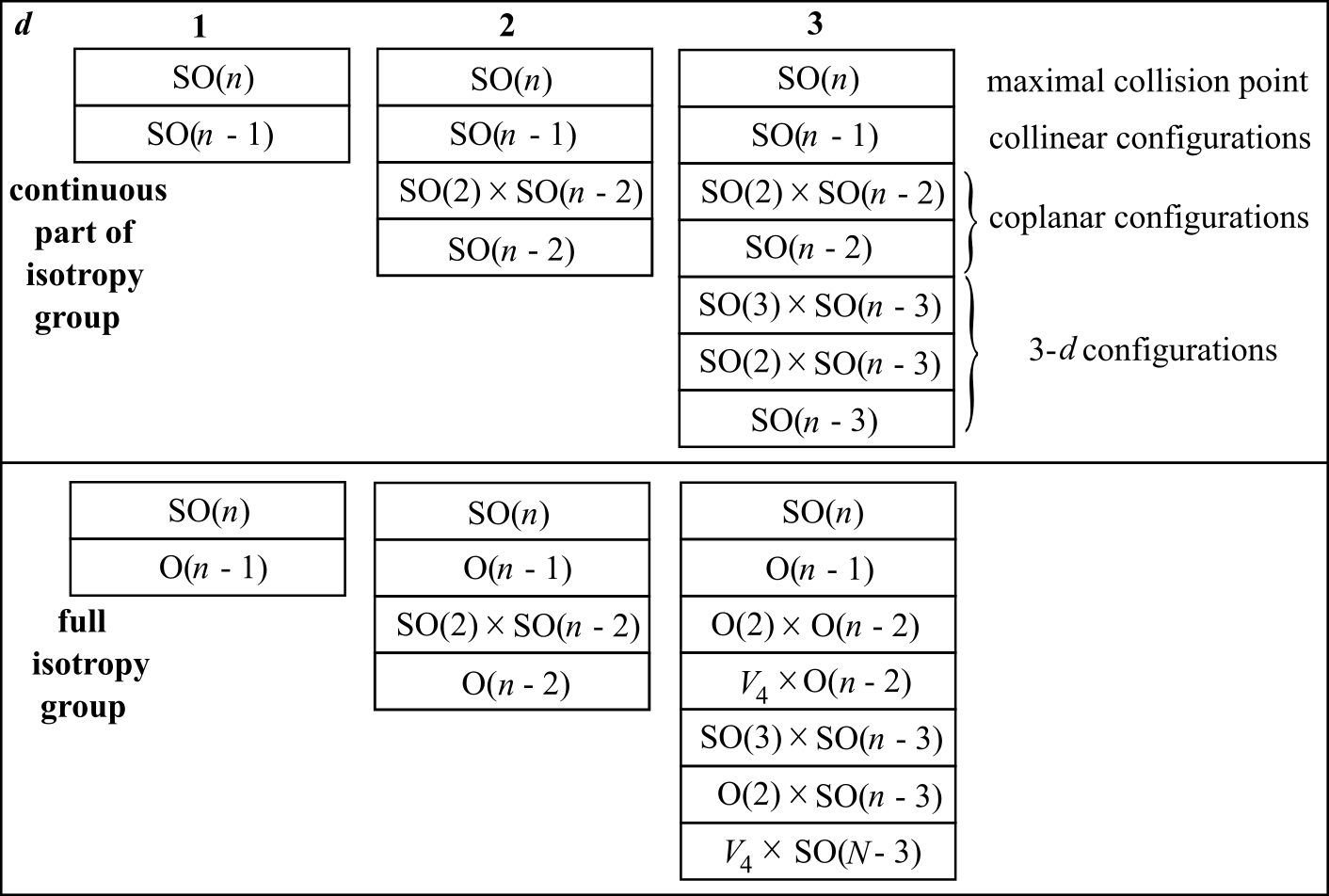}
\caption[Text der im Bilderverzeichnis auftaucht]{\footnotesize{Isotropy groups in a)1-$d$, b) 2-$d$, and c) in 3-$d$. 
The first row are the continuous parts of the group, whereas the second row are the full group: discrete parts included.
$V_4 := C_2 \times C_2$: the Klein 4-group.   
The Aufbau Principle is once again in evidence.
}} 
\label{7-Isot}\end{figure} } 
%FFFFFFFFFFFFFFFFFFFFFFFFFFFFFFFFFFFFFFFFFFFFFFFFFFFFFFFFFFFFFFFFFFFFFFFFFFFFFFFFFFFFFFFFFFFFFFFFFFFFFFFFFFFFFFFFFFFFFFFFFFFFFFFFFFFFFFFFFFFFFFFFFFFFFFFFFFFFFFFFFFFFFFFFFFFFFFFFFFFFFFFFFF
%
%FFFFFFFFFFFFFFFFFFFFFFFFFFFFFFFFFFFFFFFFFFFFFFFFFFFFFFFFFFFFFFFFFFFFFFFFFFFFFFFFFFFFFFFFFFFFFFFFFFFFFFFFFFFFFFFFFFFFFFFFFFFFFFFFFFFFFFFFFFFFFFFFFFFFFFFFFFFFFFFFFFFFFFFFFFFFFFFFFFFFFFFFF
{\begin{figure}[ht]
\centering
\includegraphics[width=0.85\textwidth]{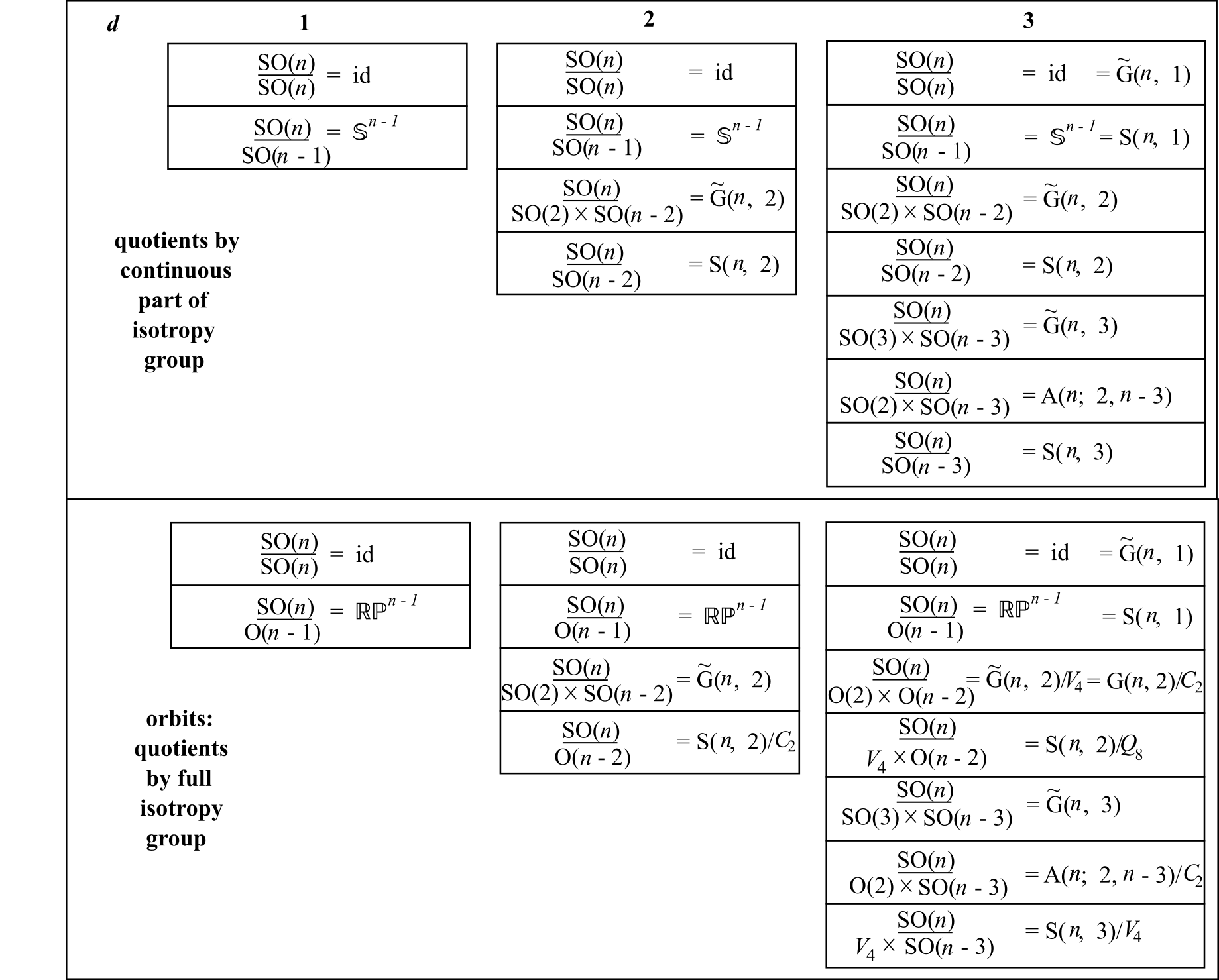}
\caption[Text der im Bilderverzeichnis auftaucht]{\footnotesize{The corresponding orbit space topologies.
}} 
\label{7-Orb}\end{figure} } 
%FFFFFFFFFFFFFFFFFFFFFFFFFFFFFFFFFFFFFFFFFFFFFFFFFFFFFFFFFFFFFFFFFFFFFFFFFFFFFFFFFFFFFFFFFFFFFFFFFFFFFFFFFFFFFFFFFFFFFFFFFFFFFFFFFFFFFFFFFFFFFFFFFFFFFFFFFFFFFFFFFFFFFFFFFFFFFFFFFFFFFFFFFF

\m 

\n{\bf Proposition 4} The general result is that 
\be 
\mbox{$N - 1 = n = d + 1$ \m is minimal for C-genericity}  \m .   
\ee 
Thus 
\be 
\mbox{(minimal linear dependence)} \m = \m \mbox{(minimal C-genericity)}  \m . 
\ee 
\n{\u{Derivation}} This follows by generalizing Mitchell and Littlejohn's point about the smallest-dimension generic orbit, which in their $d = 3$ case has dimension 
\be 
3 \, N - 12  =  3 (N - 4)
\ee 
thus requiring $N \geq 5$ to realize.
For such dimension counting, the continuous part of the orbit suffices. 

\m 

\n I moreover identify this orbit's continuous part in arbitrary-$d$ to be the oriented Grassmannian  
\be 
\w{G}(n, d)  \es  \frac{SO(n)}{SO(d) \times SO(n - d)}                                                              \m .  
\label{Grass}
\ee
See Appendix C.2 for more on this, including evaluating its dimension to be $d(n - d)$.  
From the second factor in this, these are all zero-dimensional along the basis diagonal. 
Inclusion of one more point-or-particle than the basis diagonal however suffices for this to attain a positive-integer value.  $\Box$

\m 

\n{\bf Remark 4} So, once variable dimension is incorporated, 
Mitchell and Littlejohn's condition is not a bound on $N$ but rather a further ladder of unit slope in the $(d, N)$ grid. 

\m 

\n On the one hand, the quadrilaterals in the plane           -- $(d, N) = (2, 4)$ -- are revealed to be a meaningful model arena for 
the notoriously hard and interesting 5-Body Problem in 3-$d$:    $(d, N) = (3, 5)$, 
with the                          step-up in complexity from the tetrahaedrons (3, 4) to (3, 5) sharing some conceptual features 
with the {\sl much more familiar} step-up in complexity from the triangles (2, 3) to the quadrilaterals (2, 4). 

\m 

\n On the other hand, $(d, N) = (4, 6)$ is revealed to be substantially more of a sequel to $(d, N) = (3, 5)$ than $(d, N) = (3, 6)$ is.
Such sequels are moreover never-ending, for all the $(d, N)$ pairs satisfying 
\be 
(d, N)          =  (d + 2, d)  
       \m [ \m  =  (n + 1, n - 1) 
                =  (N, N - 2)      \m ]   \m . 
\ee
These are the first parallel above the basisland diagonal, i.e.\ the minimal dependentlands.  

\m 

\n{\bf Remark 5} This completes realization of the {\it qualitatively-distinct triplets} of $N$-Body Problems for values of $N$ 
(1, 2, 3) in 1-$d$, 
(2, 3, 4) in 2-$d$, 
(3, 4, 5) in 3-$d$, 
(4, 5, 6) in 4-$d$ ... and 
\be
(d, d + 1, d + 2) \m 
\ee 
in general dimension $d$.  
%
% 3-$d$: is the first nontrivial supercritical case: mirror image identification obligatoriness apart, this reduces to 2-$d$ case in many ways (but not all). 
%
In particular, in 2-$d$ this now amounts to familiar increases in complexity in passing sequentially from intervals to triangles to quadrilaterals.

\m 

\n{\bf Remark 6} I furthermore observe a sense in which Mitchell and Littlejohn's condition for $N = 5$ is not generic.  
This is based on considering the {\sl bounded lattice formed by} the isotropy subgroups; 
the continuous parts for this are presented for $d = 2$ and 3 in Figs \ref{Isot-Latt}.a) and \ref{Isot-Latt}.b).\footnote{This might in general be just a bounded poset, 
%OOOOOOOOOOOOOOOOOOOOOOOOOOOOOOOOOOOOOOOOOOOOOOOOOOOOOOOOOOOOOOOOOOOOOOOOOOOOOOOOOOOOOOOOOOOOOOOOOOOOOOOOOOOOOOOOOOOOOOOOOOOOOOOOOOOOOOOOOOOOOOOOOOOOOOOOOOOOOOOOOOOOOOOOOOOOOOOOOOO
but in all cases featuring in the current article, it is a fortiori a bounded lattice.} 
%OOOOOOOOOOOOOOOOOOOOOOOOOOOOOOOOOOOOOOOOOOOOOOOOOOOOOOOOOOOOOOOOOOOOOOOOOOOOOOOOOOOOOOOOOOOOOOOOOOOOOOOOOOOOOOOOOOOOOOOOOOOOOOOOOOOOOOOOOOOOOOOOOOOOOOOOOOOOOOOOOOOOOOOOOOOOOOOOOOO
%
%FFFFFFFFFFFFFFFFFFFFFFFFFFFFFFFFFFFFFFFFFFFFFFFFFFFFFFFFFFFFFFFFFFFFFFFFFFFFFFFFFFFFFFFFFFFFFFFFFFFFFFFFFFFFFFFFFFFFFFFFFFFFFFFFFFFFFFFFFFFFFFFFFFFFFFFFFFFFFFFFFFFFFFFFFFFFFFFFFFFFFFFFF
{\begin{figure}[ht]
\centering
\includegraphics[width=0.7\textwidth]{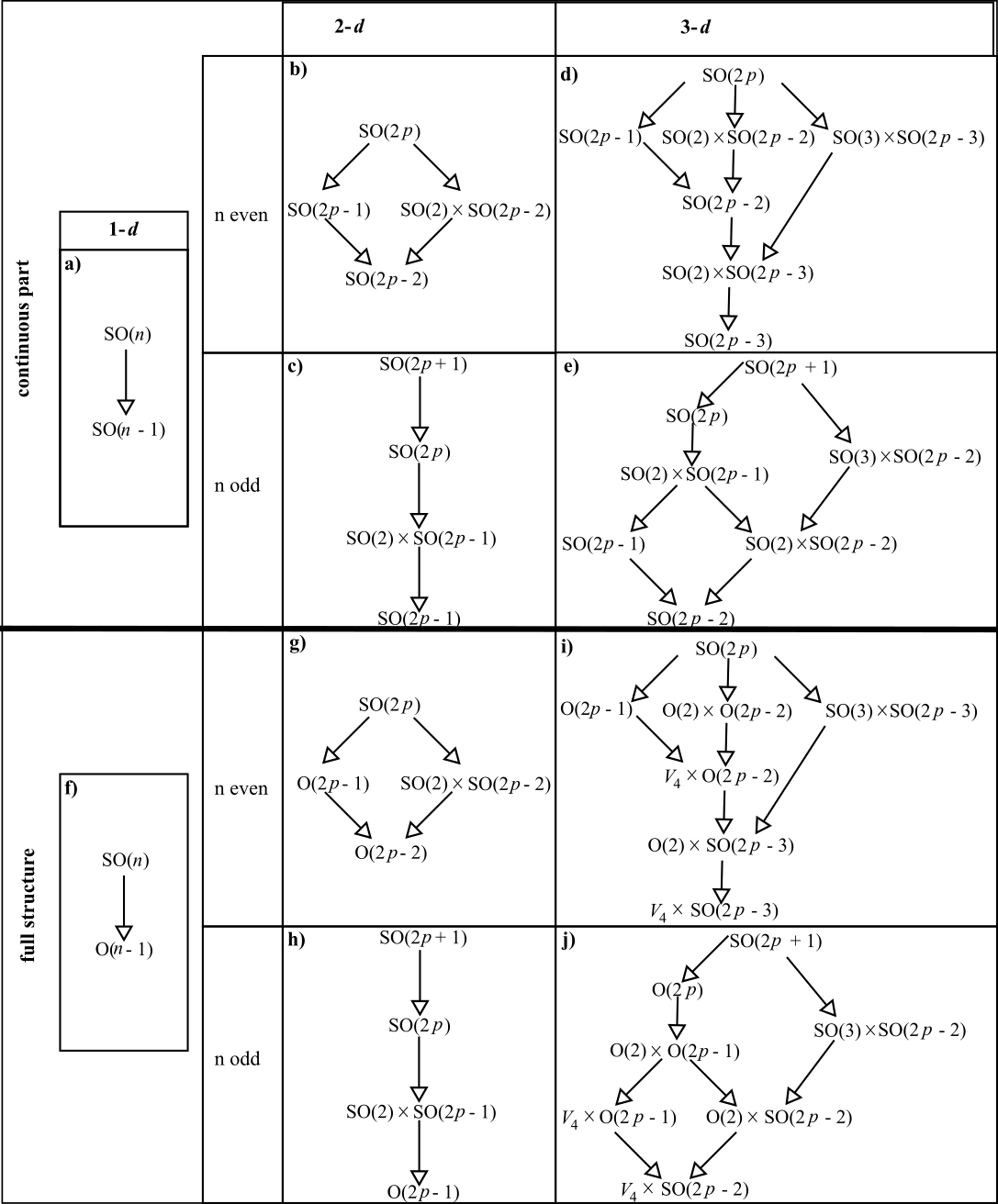}
\caption[Text der im Bilderverzeichnis auftaucht]{\footnotesize{The general lattices.
}} 
\label{Target-Lattices}\end{figure} } 
%FFFFFFFFFFFFFFFFFFFFFFFFFFFFFFFFFFFFFFFFFFFFFFFFFFFFFFFFFFFFFFFFFFFFFFFFFFFFFFFFFFFFFFFFFFFFFFFFFFFFFFFFFFFFFFFFFFFFFFFFFFFFFFFFFFFFFFFFFFFFFFFFFFFFFFFFFFFFFFFFFFFFFFFFFFFFFFFFFFFFFFFFFF
%
%FFFFFFFFFFFFFFFFFFFFFFFFFFFFFFFFFFFFFFFFFFFFFFFFFFFFFFFFFFFFFFFFFFFFFFFFFFFFFFFFFFFFFFFFFFFFFFFFFFFFFFFFFFFFFFFFFFFFFFFFFFFFFFFFFFFFFFFFFFFFFFFFFFFFFFFFFFFFFFFFFFFFFFFFFFFFFFFFFFFFFFFFF
{\begin{figure}[ht]
\centering
\includegraphics[width=1.0\textwidth]{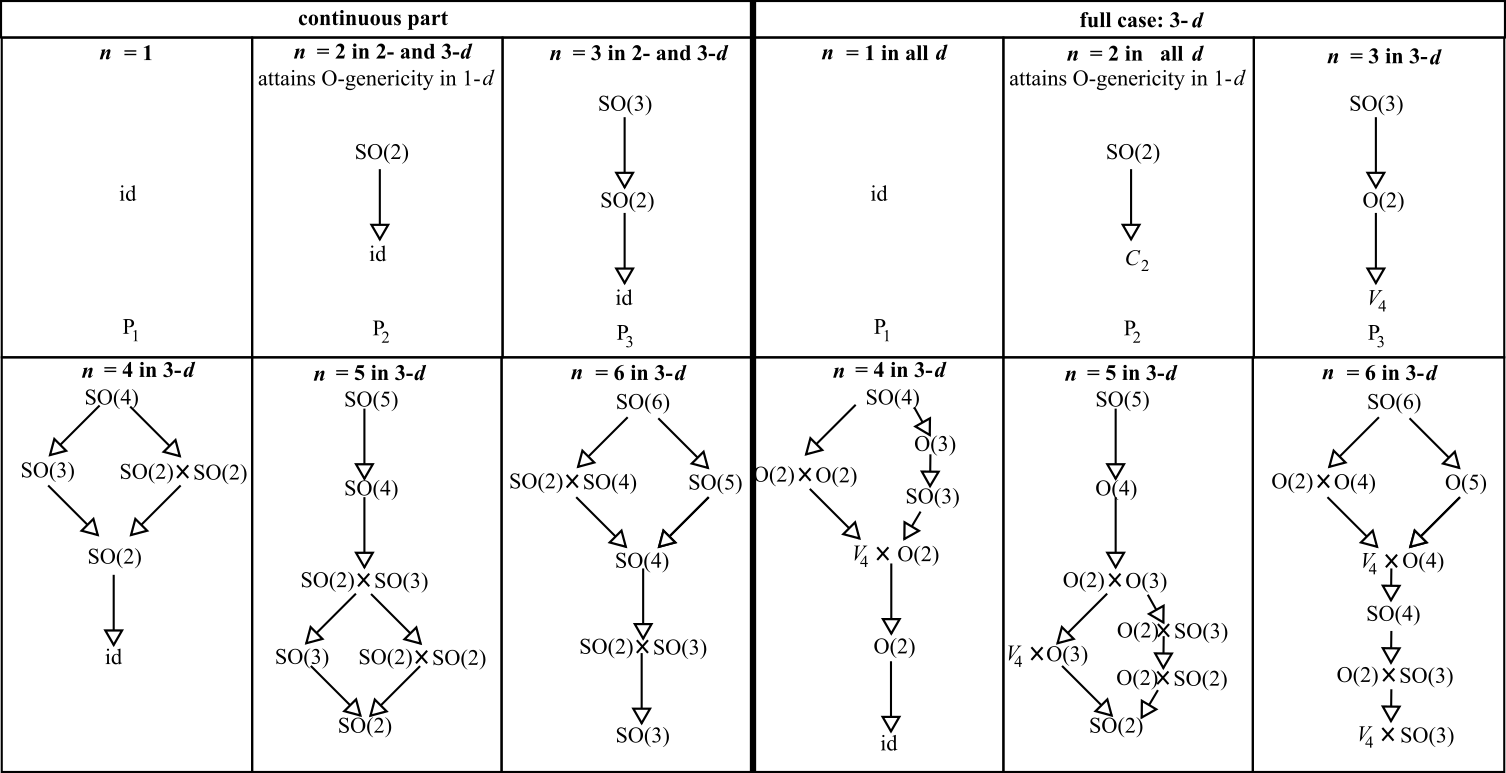}
\caption[Text der im Bilderverzeichnis auftaucht]{\footnotesize{The exceptional cases' isotropy subgroups lattices.}} 
\label{Isot-Latt}\end{figure} } 
%FFFFFFFFFFFFFFFFFFFFFFFFFFFFFFFFFFFFFFFFFFFFFFFFFFFFFFFFFFFFFFFFFFFFFFFFFFFFFFFFFFFFFFFFFFFFFFFFFFFFFFFFFFFFFFFFFFFFFFFFFFFFFFFFFFFFFFFFFFFFFFFFFFFFFFFFFFFFFFFFFFFFFFFFFFFFFFFFFFFFFFFFFF

\m 

\n{\bf Remark 7} The even--odd distinction of these lattices in 2- and 3-$d$ follows from the number of Casimirs going up by one for every even $SO(n)$ 
                                                                                                                   but not at all for every odd $SO(n)$.  
Thus 
\be 
SO(2\,p - 1) \m \leq \m  SO(2 \, p) 
\ee 
leaves one Casimir unused, which can be used to generate an extra $SO(2)$, so 
\be 
SO(2) \times SO(2\,p - 1) \m \leq \m  SO(2 \, p)                                  \m .  
\ee 
On the other hand, 
\be 
SO(2\,p - 2) \m \leq \m SO(2 \, p - 1) 
\ee
uses up all of the Casimirs, so an extra $SO(2)$ subgroup cannot be included.  

\m

\n{\bf Remark 8} We can place a sequence of qualitative criteria in terms of increasing complexity of the bounded lattice of isotropy subgroups as follows.   

\m 

\n For arbitarary $d$, 
\be 
\mbox{$N = 2$'s isotropy subgroup lattice is only a point}                                                     \m , 
\ee
\be 
\mbox{$N = 3$'s isotropy subgroup lattice is the first with a distinct top and bottom but has no middle}       \m .   
\ee 
In 1-$d$, this attains genericity.  
\be 
\mbox{$N = 4$'s isotropy subgroup lattice is the first to have a middle but is still just a chain}             \m ,
\ee 
and 
\be  
\mbox{$N = 5$'s isotropy subgroup lattice is the first with a nontrivial -- rather than just chain -- middle}  \m .  
\ee
\n{\bf Proposition 5} Realizing the generic lattice of the continuous parts of the isotropy subgroups requires 
\be 
N = 4 \m \mbox{ in 2-$d$ }
\ee 
and 
\be 
N = 8 \m \mbox{ in 3-$d$ }         \m .  
\ee 
\n{\bf Proposition 6} 
\n i)   For $d \neq 3$, 
\be 
n = 2 \, d \m \mbox{ ( i.e. \ $N = 2 \, d + 1$ ) }  
\ee
is an upper bound (`B-genericity', with `B' standing for bounding) on O-genericity.  

\m 

\n ii)  For $d = 3$, $n = 7$ is required. 

\m 

\n{\bf Remark 9} The $n = 2 \, d$ formulation of i) is a line in the $(d, n)$ grid with twice the slope of Casson's basis diagonal, also through the origin (Fig \ref{Slope-2}.a).  
ii)'s exception follows from the Lie group accidental relation (\ref{SO4}). 
%
%FFFFFFFFFFFFFFFFFFFFFFFFFFFFFFFFFFFFFFFFFFFFFFFFFFFFFFFFFFFFFFFFFFFFFFFFFFFFFFFFFFFFFFFFFFFFFFFFFFFFFFFFFFFFFFFFFFFFFFFFFFFFFFFFFFFFFFFFFFFFFFFFFFFFFFFFFFFFFFFFFFFFFFFFFFFFFFFFFFFFFFFFF
{\begin{figure}[ht]
\centering
\includegraphics[width=1.0\textwidth]{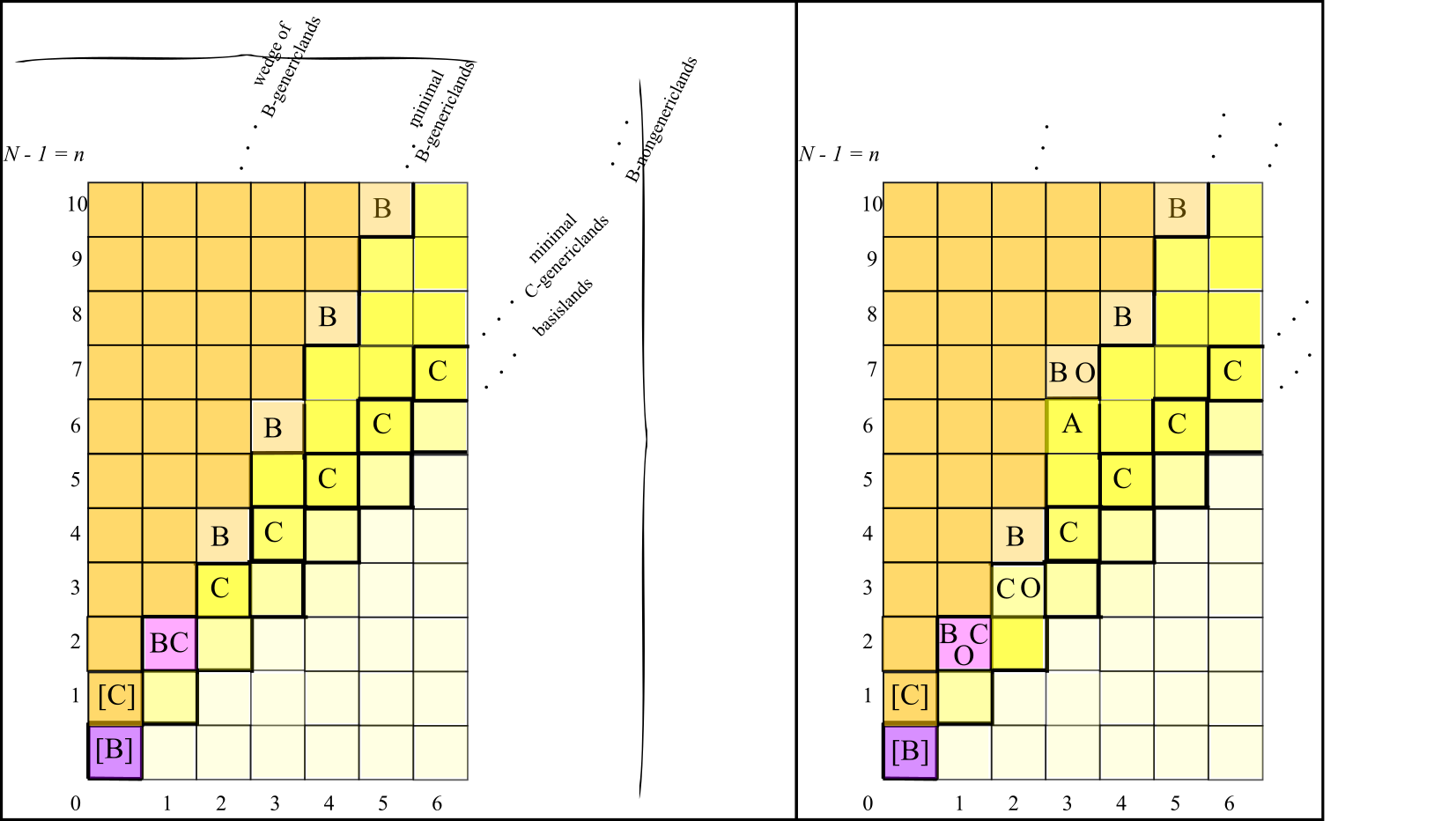} 
\caption[Text der im Bilderverzeichnis auftaucht]{\footnotesize{Realization of the generic isotropy subgroup lattice  
a) systematically, with C standing for C-generic, and B for upper bound on O-generic.  
[] denotes cases included to see the intersection point, but for which the properties themselves are not fully defined. 

\m

\n b) Making allowance for the Lie group accident based exception (marked with an A), as well as entering known O-generic values.  

\m

\n Our final figure shall suppress square-by-square C and B labels in all systematically-distributed cases.  
}} 
\label{Slope-2}\end{figure} } 
%FFFFFFFFFFFFFFFFFFFFFFFFFFFFFFFFFFFFFFFFFFFFFFFFFFFFFFFFFFFFFFFFFFFFFFFFFFFFFFFFFFFFFFFFFFFFFFFFFFFFFFFFFFFFFFFFFFFFFFFFFFFFFFFFFFFFFFFFFFFFFFFFFFFFFFFFFFFFFFFFFFFFFFFFFFFFFFFFFFFFFFFFFF

\m 

\n{\bf Remark 10} We thus have the overall pattern of Fig \ref{Slope-2}.b) in the $(d, n)$ grid.  

\m 

\n{\bf Remark 11} Further examination of the continuous part of the subgroup lattice reveals an Aufbau Principle: 
the split of Fig \ref{7-Isot} into the 3-$d$, 2-$d$ and 1-$d$ chains. 
A generic chain of length $d$ is moreover present in arbitrary dimension.
Aside from the top element of this chain being the Grassmann space of eq (\ref{Grass}), the bottom such is the Stiefel space 
\be 
\FrS(p, q) \:= \frac{SO(p)}{SO(q)}                                      \m .  
\ee 
The middle elements of the continuous-part chain are moreover more general than Grassmannians as per (\ref{A-Spaces}) -- of the specific form (\ref{Middles}).  
More knowledge about such `A-spaces' would be very desirable for Shape(-and-Scale) Theory, since this is but one of many places where such spaces occur in this subject.
%
%FFFFFFFFFFFFFFFFFFFFFFFFFFFFFFFFFFFFFFFFFFFFFFFFFFFFFFFFFFFFFFFFFFFFFFFFFFFFFFFFFFFFFFFFFFFFFFFFFFFFFFFFFFFFFFFFFFFFFFFFFFFFFFFFFFFFFFFFFFFFFFFFFFFFFFFFFFFFFFFFFFFFFFFFFFFFFFFFFFFFFFFFF
{\begin{figure}[ht]
\centering
\includegraphics[width=0.85\textwidth]{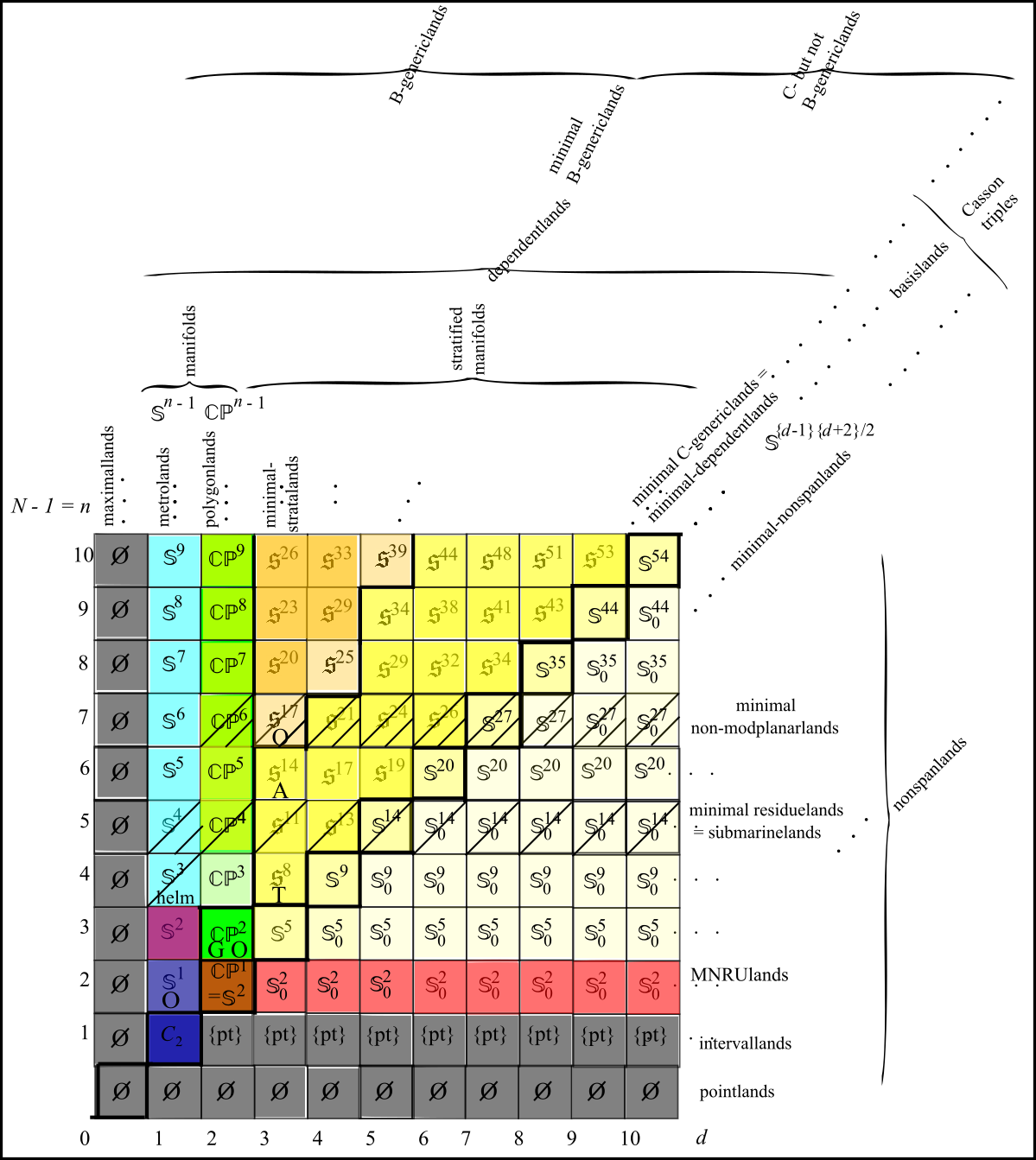}
\caption[Text der im Bilderverzeichnis auftaucht]{\footnotesize{Combined summary figure for the article on the $(n, d)$ grid for $n:= N - 1$. 
}} 
\label{N-Body-Summary}\end{figure} } 
%FFFFFFFFFFFFFFFFFFFFFFFFFFFFFFFFFFFFFFFFFFFFFFFFFFFFFFFFFFFFFFFFFFFFFFFFFFFFFFFFFFFFFFFFFFFFFFFFFFFFFFFFFFFFFFFFFFFFFFFFFFFFFFFFFFFFFFFFFFFFFFFFFFFFFFFFFFFFFFFFFFFFFFFFFFFFFFFFFFFFFFFFFF

\vspace{10in}

%==================================================================================================================================================================
%==================================================================================================================================================================
\section{Conclusion}\label{Conclusion}
%==================================================================================================================================================================
%==================================================================================================================================================================

%==================================================================================================================================================================
\subsection{Summary of results}
%==================================================================================================================================================================

\n In the current article, we made systematic use of Discrete Mathematics equations in considering minimal counts and dimensional coincidences of properties.  

\m 

\n We also provided a structural analysis level by level of what relational theory -- Shape Theory and Shape-and-Scale Theory -- is. 
Shape Theory consists of Topology, Geometry, Group Theory, Linear Algebra, Graph Theory and Order Theory.  
In this manner we have gone from the existing literature's qualitative distinctions for $N$ = 3, 4 and 5 to finding some further such which apply to $N = 6$ and $N = 8$. 

\m 

\n In particular, rubber shapes' shape spaces are graphs.
Metric shapes' shape spaces are in general stratified manifolds, which arise by topology--group theory interaction.  
Graphs are clearly much simpler, and yet still encode some useful information.  

\m 

\n The rubber level of structure brings emphasis to $\mathbb{S}^1$ absolute space. 
This admits a compactified version of Jacobi coordinates.  

\m 

\n Rubber moreover picks out $N = 5$ and 6 in 1-$d$ and 6 and 8 for $d > 1$. 
The first of each of these have nontrivial residues after deconing, while the second of these are all of nonplanar, non-coplanar and non-modular.  

\m 

\n  This leads to hexagons and octagons having further significance as minimal cases, 
as well as (3, 6) and (3, 8) models, which are in excess of the usually largest considered and sometimes purportedly generic (3, 5) model.

\m 

\n We moreover observe regularities for specific $d$'s, as well as $N(d)$ relations by which the distribution of minimal properties on the $(n, d)$ grid becomes pertinent.  

\m 

\n The current article furthermore clarifies various previously decribed concepts in Linear Algebra terms. 
In particular, the `Casson diagonal' in the $(d, n)$ grid consists of basislands, 
which moreover split the $(n, d)$ grid into an upper (linear) dependentland wedge and a lower nonspanningland wedge.  
Both basislands and minimal dependentlands pick up further significance as the study progresses. 

\m

\n For metric shapes, the corresponding shape spaces are in general stratified manifolds, though these are just manifolds in 1 and 2-$d$.

\m 

\n The current article also gives an isotropy groups and kinematical orbits treatment that is more extensive than in the previous literature.
This is firstly by considering spatial dimensions other than 3. 
Secondly, by identifying Mitchell and Littlejohn's genericity condition as a counting `C'-genericity, which picks out the minimal dependentlands. 
Thirdly, by use of a further Order-Theoretic `O'-genericity criterion, for which the $n = 2 \, d$ line provides an upper bound: a line twice as steep as the basislands diagonal. 
$d = 3$ is a sole exception to this, due to the Lie group accident $SO(4) \cong SO(3) \times SO(3)$ pushing $N$ up by 1 to 8.    
Fourthly, we point to Grassmann and Stiefel spaces playing top and bottom element roles within this ordering structure.   
Between these, an A-space generalization of Grassmann spaces occurs; 
this generalization is further warranted by another instance of A-spaces entering Relational Theory at the level of Euclidean Relational Spaces. 

\m 
 
\n The current article's systematic structural build-up is summarized in Fig \ref{N-Body-Summary}.
A fairly high proportion of this information is new to the current article and to \cite{I, II, III}.  
In particular, applying Graph Theory to Shape Theory is new to the current program (now including also \cite{Top-Shapes}).  

\m 

\n Dimension by dimension, for $d = 1$, the 
$n = 1$ basisland case's shape space is exceptionally $C_2$ rather than $\{\mbox{pt}\}$.

\m 

\n $n = 2$ is the minimal dependentland, supports the first relationally nontrivial shape-and-scale theory \cite{I}, and is both C- and O-generic. 
Its shape-and-scale supports the first nontrivial stratum: its maximal coincidence-or-collision.  

\m 

\n $n = 3$ supports the first relationally nontrivial shape space \cite{AF, II}.

\m

\n $n = 4$'s rubber theory has the first nontrivial residue: the helm.

\m 

\n $n = 5$ has the first nonplanar, non-coplanar and non-modular residue graph rubber theory.  

\m 

\n For $d = 2$,   
$n = 2$ -- triangleland -- is the only $N$-a-gonland that is also a basisland and that simplifies from complex-projective mathematics to spherical mathematics.
Many further distinctive features of triangleland can be found in the current article and \cite{III}.  

\m 

\n $n = 3$ -- quadrilateralland -- is all of a minimal dependentland,the first instance of complex projective mathematics that is not reducible to spherical mathematics, 
and both C- and O-generic.
Many further distinctive features of quadrilateralland can be found in the current article and \cite{QuadI, Forth}.  

\m 

\n $n = 4$ -- pentagonland -- is B-generic.

\m 

\n $n = 5$ -- hexagonland -- is the first whose rubber theory has a nontrivial residue: the submarine. 

\m 

\n $n = 7$ -- octagonland -- has the first nonplanar, non-coplanar and non-modular residue graph rubber theory.  

\m 

\n For $d = 3$, 
$n = 2$ -- the 3-Body Problem -- is the first nonspanningland, with further stratificational complexity than its 2-$d$ counterpart \cite{A-Monopoles}.  

\m 

\n $n = 3$ -- the 4-Body Problem's tetrahaedronland -- is the first basisland that is not also an $N$-a-gonland; its relational spaces are moreover still topologically standard. 

\m 

\n $n = 4$ -- the 5-Body Problem -- is the first dependentland, the extreme tip of the wedge of topologically nonstandard shape spaces \cite{Kendall} and C-generic.  

\m 

\n $n = 5$ -- the 6-Body Problem's octahaedronland -- has the first rubber theory with nontrivial residue: the submarine.  

\m 

\n $n = 7$ -- the 8-Body Problem's cubeland -- has the first nonplanar, non-coplanar and non-modular residue graph rubber theory and is O-generic, 
the 7-Body Problem being the case that is exceptionally unable to attain O-genericity. 

\m 

\n Finally returning to the Introduction's opening and closing adages, 
we no longer say that the 3-Body Problem is hard, the 4-body problem harder and the 5-body problem even harder, and we have scarcely looked beyond. 
We say, rather, that {\sl the first non-spanning parallel is hard, the basis diagonal is harder and the first linear dependence parallel harder still, 
with some sources of complexity then stabilizing, in all spatial dimensions}.  
We also no longer say that 3-body problem intuitions are necessary for the study of Background Independence. 
We say, rather, for now, {\sl that 3 to 8 body problem intuitions are necessary 
in the study of Background Independence, depending on the level of structure under consideration}.
This second upgrade is further supported by article II's affine, projective and conformal analysis.

%===================================================================================================================================================================================
\subsection{Follow-up Projects}
%===================================================================================================================================================================================

\n{\bf Project 1} In view of the current article, a detailed study of 5 and 6 points on a line,  
                                                                      pentagons, 
                                                                      hexagons 
				             and nontrivialities first arising in yet larger $N$-a-gons 
							 as a useful preliminary to the $N = 5$ to $8$ Body Problems in 3-$d$, 
							 for all that each of points on the line and polygons have a topological and geometrical series simplicity absent from 3-$d$.  
These series, and familiarity with polygons through humanity's past interest renders this a rather more tractable problem than in 3-$d$.  
If even this were to badly misbehave with increasing $N$, caution about 3-$d$ $N$-body problems attaining further impasses beyond $N = 5$ would increase. 
It is moreover valuable to split those effects which are serial from those which are not.  

\m 

\n{\bf Project 2} Consider $(d, n) = (4, 3)$, (4, 4) and (4, 5): the minimal nonspanningland, basisland and minimal dependentland in 4-$d$, 
including the analogue of Mitchell and Littlejohn \cite{ML00}'s analysis of isotropy groups and kinematical orbits.  
Consider also O-genericity in this 4-$d$ setting; the current article's B-generic upper bound for this being (4, 8).

\m 

\n To Project 1 having planar geometry as a source of of shape-theoretic detail and cautions, 
      Project 2's main source of insights is the $N$-Body Problem and Molecular Physics literatures.
	  
\m 

\n{\bf Project 3 (maximally uniform states)} Foor $d = 3$, $N = 4$, 6, 8, 12 and 20 have platonic solid maximally uniform states. 
These play a significant role in the structure and interpretation of shape spaces \cite{+Tri, ABook, II, III}. 
Platoonic solids moreover dominate other $N$'s maximally uniform states by appending points-or-particles to the centre, or, once there are enough, equably among the edges or faces.
6 and 8 coincide with rubber nontrivialities, and 8 with kinematical orbit nontrivialities as well. 

\m 

\n $d = 4$ has these for $N = 5$, 8, 16, 24, 120 and 600.

\m 

\n $d \geq 5$ however has just 3, for $N = d + 1$, $2 \, d$ and $2^d$.  
The first of these is yet another realization of Casson's diagonal of simplices-and-bases, which is thus imbued with further uniformity significance.
The second of these is another slope-2 property.  
The third forms an exponential curve above that in the $(d, n)$ grid.  
We moreover have uniform quantifiers to test out \cite{I, II, III}, and these yield highest fractional values in the case of the Casson diagonal.

\m 

\n{\bf Acknowledgments} I thank Chris Isham and Don Page for concrete discussions about configuration space topology, geometry and background independence. 
The Introduction's final adage about the 3-body problem, that the current series of articles replaces with an extended precise statement about when further $N(d)$ are required,  
featured in discussions with Julian Barbour in the 2000's.  
I also thank Jeremy Butterfield, Christopher Small, Bryce DeWitt, Jurgen Ehlers and Jimmy York Jr for encouragement over the years, 
Malcolm MacCallum for some mathematical advice in the 2015-2017 period, and A.C. for a useful comment.     
I thank Don, Jeremy, Malcolm, Enrique Alvarez and Reza Tavakol for support with my career.  

\vspace{10in}

\begin{appendices}

%===================================================================================================================================================================================
\section{Critical \biN \, \, in Dynamics} 
%===================================================================================================================================================================================

\n Dynamical consequences of the current article's considerations include maximal coincidence-or-collision mathematical intractability, 
                                                                         inertia tensor noninvertibility,                                and
                                                                         mathematical intractability on the reduced configuration space of collinearities: 
																		 all global obstructions to modelling motion as a geodesic on configuration space.  
Many of the article's variables are among the more straightforward simplifying variables in such studies.  
Some $N$ and $d$ dependent highlights from Dynamics with Newtonian gravitational potentials are moreover as follows.  

\m 

\n{\bf Definition 1} The standard {\it first integrals} for the $N$-body problem in $\mathbb{R}^d$ are those resulting from centre of mass motion, 
the energy and the $d(d - 1)/2$ angular momenta.

\m 

\n{\bf Bruns' Theorem} is that no further independent first integrals that are algebraic in position--momentum Cartesian coordinates 
$(\u{q}^I, \u{p}_I)$ for $N \geq 3$ in $\mathbb{R}^3$, or more generally, on $d \leq N$, which we recognize as the dependentlands.  
See e.g.\ \cite{J00} including for reference to previous literature. 

\m

\n{\bf Definition 2} {\it Restricted} $N${\it-Body Problems} are ones in which the bodies are split up into massive ones and infinitesimal ones; 
e.g.\ the model for the motion of the Moon in which the Sun and Earth are massive. 
Within such models, {\it Hill's regions} \cite{Marchal} are those which are accessible by the `infinitesimal moons'.

\m 

\n{\bf Remark 1} $N = 3$ admits disconnected Hill regions, whereas their $N \geq 4$ analogues are connected, 
so $N = 4$ -- the tetrahaedron -- is minimal for the latter qualitatively-distinct behaviour. 

\m 

\n{\bf Remark 2} 1- and 2-$d$ moreover support some simpler arbitrary-$N$ results \cite{Moulton, Smale70}.  
Some such results are underpinned by the geometrical series of shape spaces; 
in particular, Smale derived his result in awareness of the 2-$d$ series of $\mathbb{CP}^{n - 1}$ shape spaces.

\m 

\n{\bf Painlev\'{e}'s conjecture} \cite{Pain} is that for $N \geq 4$ the Newtonian gravitational $N$-Body Problem admits singularities that are not collisions.

\m 

\n{\bf Remark 3} Specific results about 1-$d$ collisions were obtained for $N = 3$ by McGehee \cite{McGehee} 
                                                                       and $N = 4$ by Mather and McGehee \cite{MM75}; the latter includes a non-collision singularity. 

\m

\n{\bf Remark 4} The Painlev\'{e} conjecture was established by Xia \cite{Xia} for $N \geqs 5$. 

\m 

\n It moreover remains an open question for $N = 4$ in $d \geq 2$.  

\m 

\n{\bf Definition 3} A {\it central configuration} is one in which each $\ddot{\u{q}}_I$ is aligned with the corresponding force vector and with a common proportionality. 

\m 

\n{\bf Remark 5} Under these circumstances, the dynamics of each particle takes central force problem form, considerably simplifying matters.  
Central configurations are moreover significant as regards the structure of more general dynamical solutions.  
\n Central configurations lead to numerous results on qualitatively critical values of $N$ and $d$. 

\m 

\n Moulton \cite{Moulton} gave a general-$N$ counting result for central configurations in 1-$d$ (there are $N!/2$ in the mirror-image-identified case). 

\m 

\n Away from 1-$d$ counts are more involved -- $N$-dependent -- and the shapes formed by central configurations are less straightforward. 
See e.g.\ \cite{Albouy} for central configurations for $N = 5$ in 2-$d$, and \cite{M94, Saari, M14} for general reviews.  

\vspace{10in}

%===================================================================================================================================================================================
\section{Larger \biN}
%===================================================================================================================================================================================

\n Given we have increased $N$ from 5 to 8 in 3-$d$, and further up the basis and isotropy subgroup lattice genericity diagonals, 
limitations on exact $N$-body study from larger $N$ asymptotic formulae are pertinent. 

\m

\n Firstly, the Hardy-Ramanujan asymptotic formula \cite{HR18} for number of partitions is pertinent to $d \geq 2$ rubber shapes (and as a bound on 1-$d$ counterparts), 
\be  
p(N)  \m \sim \m  \frac{1}{4\sqrt{3} \, N} \, \mbox{exp}
\left(\pi \sqrt{    \frac{2 \, N}{3}    } 
\right)                                                    \m 
\mbox{ as } \m  N \longrightarrow \infty                   \m .
\label{HR}
\ee 
This furthermore gives a crude bound on the number of Leibniz space edges using the semi-saturation bound on edge number for graphs modulo complementation,   
\be 
e(N)  \m \approx \m  \frac{1}{192 \,  \, N^2} \, \mbox{exp}
\left(2 \, \pi \sqrt{\frac{2 \, N}{3}    } 
\right)                                                    \m 
\mbox{ as } \m  N \longrightarrow \infty                   \m .  
\ee
%
% For $N = 30$, 100, 300, 1000 and 3000, this returns $10^7$, $10^{15}$, $10^{31}$, $10^{62}$ and $10^{111}$.
%
This estimate could furthermore be improved by obtaining a residue-complement-graph-specific bound on unsaturation.  

\m 

\n Secondly, Otter \cite{O48} showed that the number of unlabelled trees is 
\be 
t(N) \sim \frac{C}{N^{5/2}}\mbox{exp}(k N) \m \mbox{ as } \m N \longrightarrow \infty \m , 
\ee 
for $C \approx 0.53494961$ and 
    $k \approx 1.08375760$.
This is relevant to the variety of clustering alias Jacobi coordinates involved for each $N$. 
With smaller-$N$ studies' substantially benefitting from probing with all Jacobi coordinates (and subsequent relational and shape space coordinates anchored upon each such choice), 
it is important to point out that this very quickly becomes an impossible feat with increasing $N$.

\m 

\n See Fig \ref{Large-N} for a table of order-of-magnitude values.  
%
%FFFFFFFFFFFFFFFFFFFFFFFFFFFFFFFFFFFFFFFFFFFFFFFFFFFFFFFFFFFFFFFFFFFFFFFFFFFFFFFFFFFFFFFFFFFFFFFFFFFFFFFFFFFFFFFFFFFFFFFFFFFFFFFFFFFFFFFFFFFFFFFFFFFFFFFFFFFFFFFFFFFFFFFFFFFFFFFFFFFFFFFFF
{\begin{figure}[ht]
\centering
\includegraphics[width=0.55\textwidth]{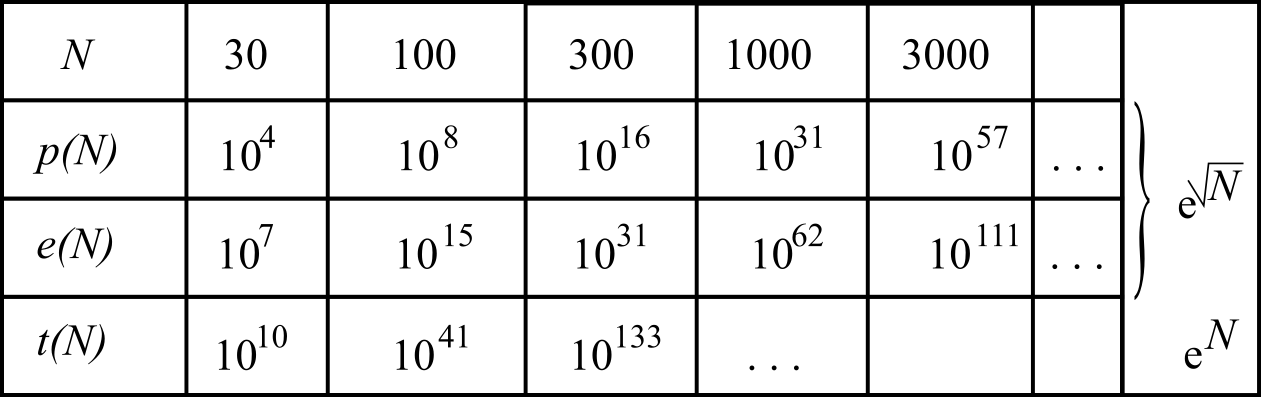}
\caption[Text der im Bilderverzeichnis auftaucht]{\footnotesize{Order of magnitude estimates for numbers of partitions, Leibniz space edges and trees for moderately large $N$.}} 
\label{Large-N}\end{figure} } 
%FFFFFFFFFFFFFFFFFFFFFFFFFFFFFFFFFFFFFFFFFFFFFFFFFFFFFFFFFFFFFFFFFFFFFFFFFFFFFFFFFFFFFFFFFFFFFFFFFFFFFFFFFFFFFFFFFFFFFFFFFFFFFFFFFFFFFFFFFFFFFFFFFFFFFFFFFFFFFFFFFFFFFFFFFFFFFFFFFFFFFFFFFF

\m 

\n There is of course a distinct -- and vast -- literature on approximate approaches to $N$-body problems for $N$ large; see e.g.\ \cite{Aar, Million}, for reviews.  

\vspace{10in}

%==================================================================================================================================================================================
%==================================================================================================================================================================================
\section{Some useful spaces for Shape Theory and the \biN-Body Problem}
%==================================================================================================================================================================================
%==================================================================================================================================================================================

%==================================================================================================================================================================================
\subsection{Stiefel spaces}
%==================================================================================================================================================================================

\n{\bf Definition 1} {\it Stiefel spaces} 
\be 
S(a, b) \:= \frac{O(a)}{O(a - b)}  \mma  b < a   \m . 
\ee 
\n{\bf Remark 1} See \cite{NSBook, Nakahara, AMP2, Frankel} for introductory accounts, or \cite{Husemoller} for a more detailed account.  

\m 

\n{\bf Remark 2} Stiefel spaces are manifolds. 

\m 

\n{\bf Remark 3} Stiefel spaces are a fortiori homogeneous spaces \cite{Helgason} due to $O(a - b)$'s {\sl natural transitive smooth action} on $O(a)$, 
which is the defining condition for homogeneous spaces.  

\m 

\n{\bf Remark 4} Stiefel spaces are compact, since the $O(a)$ \cite{Armstrong} are and compactness is quotientative (preserved under quotients). 

\m 

\n{\bf Remark 5} Stiefel spaces $S(a, b)$ can be interpreted as the topology and geometry of the set of all orthonormal $b$-frames in $\mathbb{R}^a$. 

\m 

\n{\bf Remark 6} Stiefel manifolds are furthermore a projective notion; by the Pl\"{u}cker embedding \cite{Plucker}, they can be seen to reside within projective spaces themselves.

\m 

\n{\bf Remark 7} Stiefel space play further roles \cite{Husemoller} in the theory of fibre bundles          -- universal bundle and classifying spaces -- 
                                                             and of characteristic classes: in particular of Stiefel--Whitney classes; 
the current article points to their occurrence in Shape Theory as well.  

\m 

\n{\bf Remark 8} We can furthermore write 
\be 
S(a, b) \es \frac{SO(a)}{SO(a - b)}  
\ee  
by an elementary cancellation.  

\m 

\n{\bf Remark 9} 
\be 
\mbox{dim}(S(a, b)) \es  \frac{a(a - 1)}{2} - \frac{(a - b)(a - b - 1)}{2}  
                    \es  \frac{1}{2}(a^2  - a - a^2 - b^2 + 2 \, a \, b + a - b)  
					\es  \frac{b}{2}( 2 \, a  - b - 1)                                \m . 
\ee
\n{\bf Remark 10} Some special cases are as follows. 
The bottom two $b$ for a given $a$ are 
\be 
S(a, 0) = id 
\ee 
by trivial cancellation, and \cite{Husemoller}
\be 
S(a, 1) = \mathbb{S}^{a - 1}  \m .  
\label{Stie-Sphere}
\ee 
On the other hand, the top two are 
\be 
S(a, a - 1) = SO(a) \m ,  
\ee
\be 
S(a, a) = O(a)  \m .
\ee 
 
\vspace{10in}

%==================================================================================================================================================================================
\subsection{Grassmann spaces}
%==================================================================================================================================================================================

\n{\bf Definition 1} {\it Grassmann spaces} \cite{NSBook, Nakahara, AMP2, Frankel, Husemoller} are 
\be 
G(a, b) \:= \frac{O(a)}{O(b) \times O(a - b)}  \mma                b < a  \m .  
\ee 
We are in fact more interested in {\it oriented Grassmann spaces}, 
\be 
\w{G}(a, b) \:= \frac{SO(a)}{SO(b) \times SO(a - b)}                       \m . 
\ee 
\n{\bf Remark 1}
$$
\mbox{dim}(\w{G}(a, b))  =   \mbox{dim}(G(a, b))  
                        \es  \frac{a(a - 1)}{2} -  \frac{b(b - 1)}{2} - \frac{(a - b)(a - b - 1)}{2}  
$$
\be 
                        \es  \frac{1}{2}(a^2  - a - b^2 + b - a^2 - b^2 + 2 \, a \, b + a - b)  
					    \es  b (a  - b)                                                                                  \m . 
\ee
\n{\bf Remark 2} Grassmann spaces (oriented or not) are manifolds.

\m 

\n{\bf Remark 3} They are a fortiori homogeneous spaces \cite{Helgason} due to $O(b)$ as well as $O(a - b)$, and indeed $O(b) \times O(a - b)$ having natural transitive smooth actions on $O(a)$. 

\m 

\n{\bf Remark 4} They are compact since the $O(a)$ are, and compactness is both productive [so $O(b) \times O(a - b)$ is compact] and quotientative. 

\m 

\n{\bf Remark 5} The $G(a, b)$ can be interpreted as the topology and geometry of the set of all $a$-dimensional oriented subspaces in $\mathbb{R}^b$.  

\m

\n{\bf Remark 6} They furthermore admit a projective interpretation, as a further projection of $S(a, b)$ that is both locally trivial and a fibre map \cite{Husemoller}, 
and as projective varieties; they participate in further Algebraic Geometry results.

\m 

\n{\bf Remark 7} The symmetries  
\be 
G(a, b) = G(a, a - b) \m \mma \m \w{G}(a, b) = \w{G}(a, a - b)
\ee 
-- manifest from the definition -- can moreover be further interpreted as manifestation of projective duality. 

\m 
 
\n{\bf Remark 8} Grassmanians space play further roles \cite{Husemoller} in the theory of fibre bundles and characteristic classes -- Chern classes and K-theoretic classifying spaces -- 
                                                                  and in Affine Shape Theory \cite{Sparr, GT09, Bhatta, PE16}. 
The current article points to their occurrence in Euclidean Shape-and-Scale Theory and Similarity Shape Theory as well.  

\m

\n{\bf Remark 9} See e.g. \cite{Helgason} for a brief outline of the topology of Grassmann spaces.  
 
\m

\n{\bf Remark 10} We can furthermore write 
\be 
\w{G}(a, b)  \es  \frac{S(a, b)}{SO(b)}  
             \es  \frac{S(a, a - b)}{SO(a - b)}                                                                          \m ,
\ee  
so Grassmann spaces are furthermore quotients of Stiefel spaces.  

\m 

\n{\bf Remark 11} Some special cases are as follows. 
\be 
\w{G}(n, 1) = \mathbb{S}^{n - 1}  \m ,   
\ee 
\be 
\w{G}(3, 2) = \mathbb{S}^2  \m , 
\ee 
\be 
\w{G}(4, 2) = \frac{\mathbb{S}^2 \times \mathbb{S}^2}{C_2}                                                                   \m . 
\ee 
The other cases -- not these or those equal to these by projective duality symmetry -- are all `nontrivially Grassmannian'.  

\vspace{10in}

%==================================================================================================================================================================================
\subsection{A-spaces}
%==================================================================================================================================================================================

\n{\bf Definition 1} {\it A-spaces} are 
\be 
A(a; b, c) \:= \frac{SO(a)}{SO(b) \times SO(c)}  \m ; 
\label{A-Spaces}
\ee 
this definition is new to the current article.  

\m 

\n{\bf Remark 1} For a given value of $a$, A-spaces are only defined for certain $b, c \in \mathbb{N}_0$. 
\be 
b, c \m \leq \m a
\ee 
provides a first bound.  

\m 

\n{\bf Remark 2}
\be 
\mbox{dim}(A(a; b, c)) \es  \frac{a(a - 1)}{2} -  \frac{b(b - 1)}{2} - \frac{c(c - 1)}{2}  
                       \es  \frac{1}{2}(a^2  - b^2 - c^2 - a + b + c)                                                           \m . 
\ee 
This permits full characterization of acceptable values of $b$ and $c$ for a given $a$, 
\be 
b(b - 1) + c(c - 1) \m \leq \m  a(a - 1)                                                                                        \m . 
\ee 
\n{\bf Remark 3} We still immediately have a symmetry immediately from the definition, 
\be 
A(a; b, c) = A(a; c, b) \m ; 
\ee 
This however does not in general admit a projective interpretation.

\m 

\n{\bf Remark 4} Both Stiefel spaces and oriented Grassmann spaces occur as subcases of A-spaces, according to 
\be 
A(a; b, 1) =  S(a, b)         \m , 
\ee 
\be 
A(a; b, a - b) =  \w{G}(a, b) \m .  
\ee 
I.e. if 
\be 
b + c = a \m , 
\ee 
then $A(a; b, c)$ reduces to an oriented Grassmannian.  
We use `nontrivial A spaces' of A-spaces that do not have elsewise-known alter egos in topology and geometry, Stiefel and Grassmannian spaces included.  

\m

\n{\bf Lemma 1} A-spaces are compact.

\m

\n{\u{Proof}} The previous subappendix's compactness argument carries over.  $\Box$

\m 

\n{\bf Definition 2} An A-space is {\it sub-Grassmannian} if 
\be 
b + c  \m < \m  a  \m . 
\ee
It is {\it super-Grassmannian} if 
\be 
b + c  \m > \m  a  \m . 
\ee
\n{\bf Lemma 2} Sub-Grassmannian A-spaces are manifolds and a fortiori homogeneous spaces.

\m 

\n{\u{Proof}} A standard proof that Grassmannians are homogeneous spaces rests on being able to pick bases for      a $b$-dimensional subspace 
                                                                                                        and its $(a - b)$-dimensional complement. 
This extends to the case in which a $c < a - b$-dimensional subspace of the complement is evoked instead of the whole complement.   
We finally use that all homogeneous spaces are manifolds. $\Box$

\m 

\n{\bf Remark 5}
\be 
A(n; n - 1, 1) = \mathbb{S}^{n - 1}
\ee 
is both inherited from (\ref{Stie-Sphere}) and the content of mirror-images distinct 1-$d$ Similarity Shape Theory.
\be 
A(2 \, n; 2 \, n - 1, 2) = \mathbb{CP}^{n - 1}
\ee 
is the content of mirror-images distinct 2-$d$ Similarity Shape Theory, is based on the generalized Hopf map \cite{Hopf},  
and first entered the $N$-Body Problem literature in Smale's work \cite{Smale70} and the Shape Theory literature through Kendall's \cite{Kendall84} work. 
More generally, 
\be 
A(d \, n; d \, n - 1, d) = \FrS(d, N) \m : \m \m \mbox{Kendall's similarity shape space} \m .  
\ee 
Among these, the 
\be 
A(3 \, n; 3 \, n - 1, 3) = \FrS(3, N) 
\ee
are of particular interest in the study of 3-$d$ Shape Theory and the 3-$d$ $N$-body problem.  

\m 

\n{\bf Remark 6} Aside from clarity in the mathematical posing of $N$-body problems in terms of quotients of special orthogonal groups, 
the introduction of the A-space concept is useful since further A-spaces occur in Shape Theory's or $N$-Body Problem's associated kinematical orbit problem. 
These further A-spaces occur in chains of the form 
\be 
A(n; b, n - r) \mma  b = 1 \m \mbox{ to } \m r  \m . 
\label{Middles}
\ee 
$r$ is here the rank of the configuration, ranging from 0 to the spatial dimension $d$, 
while $n$ is the relative space dimension, i.e. the number of independent relative Jacobi vectors.  
In these chains, the bottom element $b = 1$ is Stiefel,  
\be 
A(n; 1, n - r ) =  A(n; n - r, 1) = S(n, r)      \m , 
\ee   
whereas the top element $b = r$ is oriented Grassmann, 
\be 
A(n; r, n - r) = \w{G}(n, r) \m , 
\ee 
but the middle of the chain consists of nontrivial A-spaces.
$r = 3$ is the first case for which there is a nontrivial middle; this is first supported by $d = 3$. 
So one other nontrvial A-space enters the 3-$d$ $N$-Body Problem, though further such do so in extending to $d \geq 4$ dimensional $N$-body problems, 
a development that the current article elsewise strongly argues for. 

\m

\n{\bf Remark 7} From $\FrS(d, N)$ being stratified for $d \geq 3$, 
we know that super-Grassmannian A-spaces are not necessarily manifolds, and thus not necessarily homogeneous spaces eihter.
On the other hand, the above middles of chains consist of sub-Grassmannian A-spaces, and thus are homogeneous spaces and manifolds.  

\end{appendices}

% \vspace{10in}

%======================================================================  BIBLIOGRAPHY  ============================================================================================

\end{document}